%% file: frdma.tex
\documentclass[format=sigconf,usenames,dvipsnames]{acmart}

\usepackage{epsfig,endnotes}
\usepackage[english]{babel}
\usepackage{blindtext}
\usepackage{subfig}
\usepackage[inline]{enumitem}
\usepackage{booktabs}
\usepackage{amssymb}

\usepackage{subfig}
\usepackage{xcolor}
\usepackage[inline]{enumitem}

\usepackage{times}
\usepackage[small, compact]{titlesec}
\usepackage[textfont={it}, font={small,bf}, skip=1pt]{caption}

\renewcommand\footnotetextcopyrightpermission[1]{} 
\setcopyright{none}

\acmDOI{}

\acmISBN{}

\acmPrice{}

\usepackage{hyperref}
\hypersetup{
  colorlinks=true,      
  linkcolor=blue,       
  citecolor=magenta,    
  filecolor=cyan,       
  urlcolor=red          
}

\usepackage{algorithm, algorithmicx}
\usepackage{algpseudocode}
\algrenewcommand\algorithmicindent{0.5em}%
\algnewcommand\LogicAnd{\textbf{and} }

\newenvironment{denseitemize}{
\begin{itemize}[topsep=2pt, partopsep=0pt, leftmargin=1.5em]
  \setlength{\itemsep}{2pt}
  \setlength{\parskip}{0pt}
  \setlength{\parsep}{0pt}
}{\end{itemize}}

\newenvironment{denseenum}{
\begin{enumerate}[topsep=2pt, partopsep=0pt, leftmargin=1.5em]
  \setlength{\itemsep}{2pt}
  \setlength{\parskip}{0pt}
  \setlength{\parsep}{0pt}
}{\end{enumerate}}

\def\name{Justitia}

\def\ie{{i.e.}}
\def\eg{{e.g.}}

\def\safeutil{$\mathit{SafeUtil}$\xspace}
\def\targettail{$\mathit{Target_{99}}$\xspace}

\def\curtail{$\mathit{Current_{99}}$\xspace}
\def\tokenbytes{$\mathit{Token_{Bytes}}$\xspace}
\def\tokenops{$\mathit{Token_{Ops}}$\xspace}

\def\refcount{$\mathit{RefCount}$\xspace}

\def\maxrate{$\mathit{MaxRate}$\xspace}

\def\maxtput{$\mathit{MaxTput}$\xspace}
\def\tokengentime{$\tau$\xspace}
\def\refperiod{$\mathit{RefPeriod}$\xspace}
\def\refcount{$\mathit{RefCount}$\xspace}

\begin{document}
\title{RDMA Performance Isolation With {\name}}
\author{Yiwen Zhang}
\affiliation{%
  \institution{University of Michigan}
}
\author{Yue Tan}
\affiliation{%
  \institution{Princeton University}
}
\author{Brent Stephens}
\affiliation{%
  \institution{University of Illinois at Chicago}
}
\author{Mosharaf Chowdhury}
\affiliation{%
  \institution{University of Michigan}
}


\begin{sloppypar}
\input{abstract}
\maketitle
\input{intro}
\input{anomalies}
\input{properties}
\input{design}
\input{implementation}

\input{evaluation}
\input{related}
\input{outro}
\label{EndOfPaper}

\newpage
{
\bibliographystyle{ACM-Reference-Format}
\bibliography{frdma}
}

\clearpage
\input{appendix}
\input{open}
\end{sloppypar}

\end{document}

%% file: abstract.tex
\begin{abstract}
Despite its increasing popularity, most of RDMA's benefits such as ultra-low latency can be achieved only when running an application in isolation.  
Using microbenchmarks and real open-source RDMA applications, we identify a series of performance anomalies when multiple applications coexist and show that such anomalies are pervasive across InfiniBand, RoCEv2, and iWARP. 
They arise due to a fundamental tradeoff between performance isolation and work conservation, which the state-of-the-art RDMA congestion control protocols such as DCQCN cannot resolve.

We present {\name} to address these performance anomalies. 
{\name} is a software-only, host-based, and easy-to-deploy solution that maximizes RNIC utilization while guaranteeing performance isolation via shaping, rate limiting, and pacing at senders.
Our evaluation of {\name} on multiple RDMA implementations show that {\name} effectively isolates different types of traffic and significantly improves latency (by up to $56.9\times$) and throughput (by up to $9.7\times$) of real-world RDMA-based applications without compromising low CPU usage or modifying the applications.



\end{abstract}

%% file: intro.tex
\section{Introduction}
To deal with the growing application demands of ultra-low latency \cite{herd, farm, infiniswap, app-performanace-disagg-dc}, high throughput \cite{herd, farm, pilaf}, and high bandwidth \cite{azure-erasure, tachyon, tensorflow, cntk, grappa}, modern datacenters are aggressively deploying RDMA \cite{msr-rdma-15, msr-rdma-16, timely}.
The intuition is simple: RDMA can provide low latency, high throughput (measured in messages/second), and high bandwidth (measured in bytes/second) with low CPU overhead. 
Indeed, RDMA-based applications experience orders-of-magnitude improvements in latency ($<10 \mu$s) and message throughput (10s of millions operations/second) \cite{herd, farm}. 
Similarly, bandwidth-sensitive applications have been scaled to many users without CPU becoming the bottleneck \cite{azure-erasure, msr-rdma-15, msr-rdma-16}. 

Unfortunately, modern RDMA usages are often limited to optimizing individual applications with careful tuning of RDMA verbs and transport types -- each combination with its own advantages and drawbacks \cite{rdma-design-guideline, herd, fasst-osdi16, pilaf, farm}.
However, even in a private datacenter, it is reasonable to assume that diverse RDMA-enabled applications will coexist~\cite{msr-rdma-15, msr-rdma-16}. 
In this paper, we answer the question: \emph{What happens when multiple RDMA-enabled applications coexist?}

To this end, we performed a series of experiments using two state-of-the-art RDMA-based systems, FaSST \cite{fasst-osdi16} and eRPC \cite{erpc}, and three different commercial RDMA implementations: InfiniBand, RoCEv2, and iWARP (\S\ref{sec:anomalies}).  
From our measurements, we conclude that RDMA's low latency, high throughput, and high bandwidth are not guaranteed when multiple applications compete.
In fact, the throughput of FaSST and eRPC drops by 74\% and 93\%, respectively,  and eRPC's median (99th percentile) latency increases by $67\times$ ($40\times$) when competing with an RDMA-based storage application.
Those highly optimized systems have their Achilles' heel that only in fully isolated environments does the performance stay very good -- which they rarely are in practice \cite{msr-rdma-15, msr-rdma-16}.
 
Our flow-level analyses further justify our conclusion. The median (99th percentile) latency of a latency-sensitive flow -- one that sends 16B messages -- increases by $1.85\times$ ($2.23\times$) in InfiniBand, $3.82\times$ ($4\times$) in RoCEv2, and $1.11\times$ ($95\times$) in iWARP when running alongside a single 1MB bandwidth-sensitive flow. 
Similarly, throughput-sensitive flows also get throttled with throughput loss of 69.5\% in InfiniBand, and worse in RoCEv2 and in iWARP.
Surprisingly, even bandwidth-sensitive flows sending different sizes of messages do not compete fairly against each other, even though both can independently saturate line-rate.

Unfortunately, RDMA NICs (RNICs) have not been designed for multi-tenant use cases, and their isolation mechanisms are not sufficient. 
Although RDMA standards support up to 15 hardware virtual lanes \cite{ib-arch-spec-1} for separating traffic classes, such a small number of hardware shapers and/or priority queues are rarely sufficient in shared environments \cite{virtualized-shapers-hotcloud, pfabric}.
We have also confirmed that the state-of-the-art congestion control protocols such as DCQCN \cite{msr-rdma-15} do not mitigate these latency and throughput anomalies either.
 
RDMA performance isolation is further complicated by the multi-resource nature of RNICs. 
Each RNIC has two primary resources: \emph{link bandwidth} (\ie, the number of bytes it can transfer each second) and \emph{execution unit throughput} (\ie, the number of messages it can process each second). 
Bandwidth-sensitive flows send large volumes of data, throughput-sensitive ones send a large number of messages, and latency-sensitive ones care about individual message latencies -- all three need both resources in different amounts. 

Given that current RNIC implementations cannot provide performance isolation, we aim to answer the following simple yet fundamental question: \emph{Can we isolate applications and flows sharing an RNIC purely in software without compromising RDMA's performance benefits?}

An ideal solution should provide performance isolation without sacrificing RNIC utilization; it should do so in a scalable manner, with low CPU overhead, and without any hardware changes (\S\ref{sec:props}).
Note that we focus on \emph{cooperative} datacenters in this paper, where the aforementioned RDMA performance anomalies arise due to RNIC implementations and not from users/tenants gaming the system.

However, simultaneously achieving performance isolation and work conservation has a well-known tradeoff even in cooperative environments \cite{faircloud,hug}.
We address this by presenting {\name} (\S\ref{sec:design}), a pragmatic alternative that guarantees \emph{sharing incentive} \cite{jaffe-maxmin, hug}, wherein each of the $n$ flows competing on an RNIC receives at least $\frac{1}{n}$th of one of its two resources. 
We then maximize utilization as long as latency-sensitive flows are well isolated. 
To minimize application-level overhead, {\name} monitors system-wide latency characteristics by maintaining a reference flow on its own, and it arbitrates among throughput-and bandwidth-sensitive flows via multi-resource shaping.
At the possibility of slightly decreasing utilization, {\name} can effectively isolate latency-sensitive flows and ensure that throughput- and bandwidth-sensitive ones are not unfairly penalized either.
The proposed solution requires no hardware changes, provides a non-invasive service interface, and is applicable to different RDMA implementations.

We have implemented (\S\ref{sec:implementation}) and evaluated (\S\ref{sec:eval}) {\name} on InfiniBand and RoCEv2.
It mitigates the performance isolation anomalies between different types of flows while guaranteeing sharing incentive within the confines of the tradeoff space without compromising low CPU usage, introducing additional overhead, or modifying application codes.
Furthermore, it complements RDMA congestion control protocols such as
DCQCN~\cite{msr-rdma-15} and hardware virtual lanes~\cite{Qaz, dcb} (when available). 
In a large-scale experiment, {\name} improved the median and 99th percentile latencies of latency-sensitive flows by $48.8\times$ and $16.4\times$, respectively, when competing against large bandwidth-sensitive flows. 
It scales well, effectively handles remote READs, and works well in simple incast scenarios.
{\name} also isolates the performance of real-world RDMA applications.
Using {\name}, eRPC's throughput and latency improve by $9.7\times$ and $56.9\times$ when sharing RNIC resources with another storage service application.

Our paper makes the following contributions:
\begin{denseitemize}
  \item To the best of our knowledge, we are the first to perform a comprehensive analysis on RDMA sharing characteristics across all three RDMA implementations. 
  
  \item We design and implement {\name}, a software-only, host-based, and easy-to-deploy performance isolation solution that supports a wide range of RNICs.

  \item We demonstrate {\name}'s benefits on both microbenchmarks and using real-world RDMA applications.
\end{denseitemize}

%% file: anomalies.tex
\section{Performance Isolation Anomalies in RDMA}
\label{sec:anomalies}

This section establishes a baseline understanding of RDMA sharing  characteristics and identify common anomalies across different RDMA implementations (\S\ref{sec3:flow-level-analyses}),
followed by performance isolation analyses of highly optimized, state-of-the-art RDMA-based applications (\S\ref{sec3:iso-in-apps}).
We then discuss the impact of RDMA congestion control on these anomalies (\S\ref{sec3:cc-no-good}) and provide our hypothesis on the source of the anomalies (\S\ref{sec3:anomalies-source}).

\begin{figure}[!t]
	\centering
		\subfloat[][Latency Flow (Med)]{%
			\label{fig:MOTI-EvL-median}%
			\includegraphics[width=1.1in]{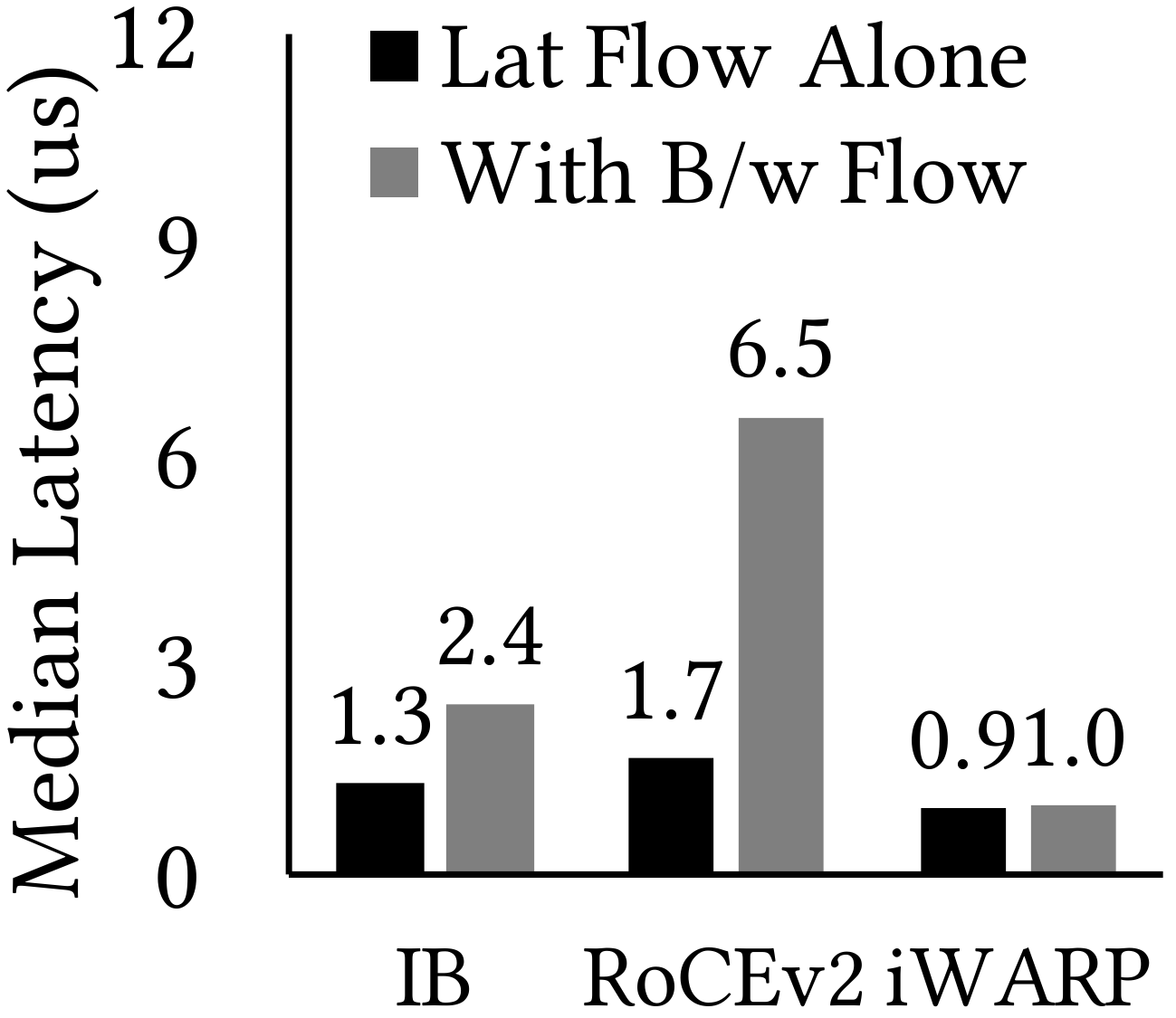}%
		}
		\hfill
		\subfloat[][{Latency Flow (99th)}]{%
			\label{fig:MOTI-EvL-tail}%
			\includegraphics[width=1.1in]{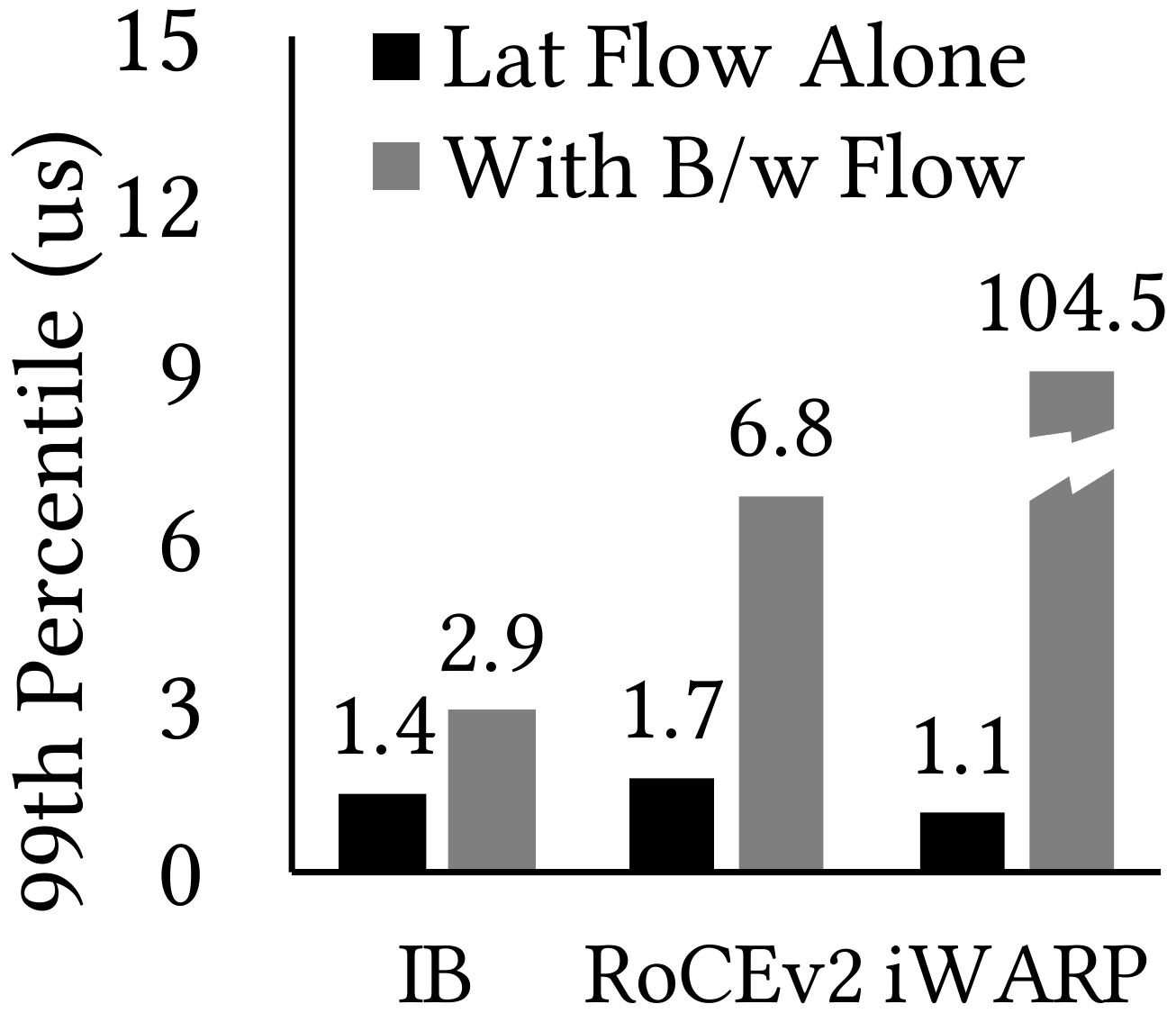}%
		}	
		\hfill
		\subfloat[][{Bandwidth Flow}]{%
			\label{fig:MOTI-EvL-bw}%
			\includegraphics[width=1.1in]{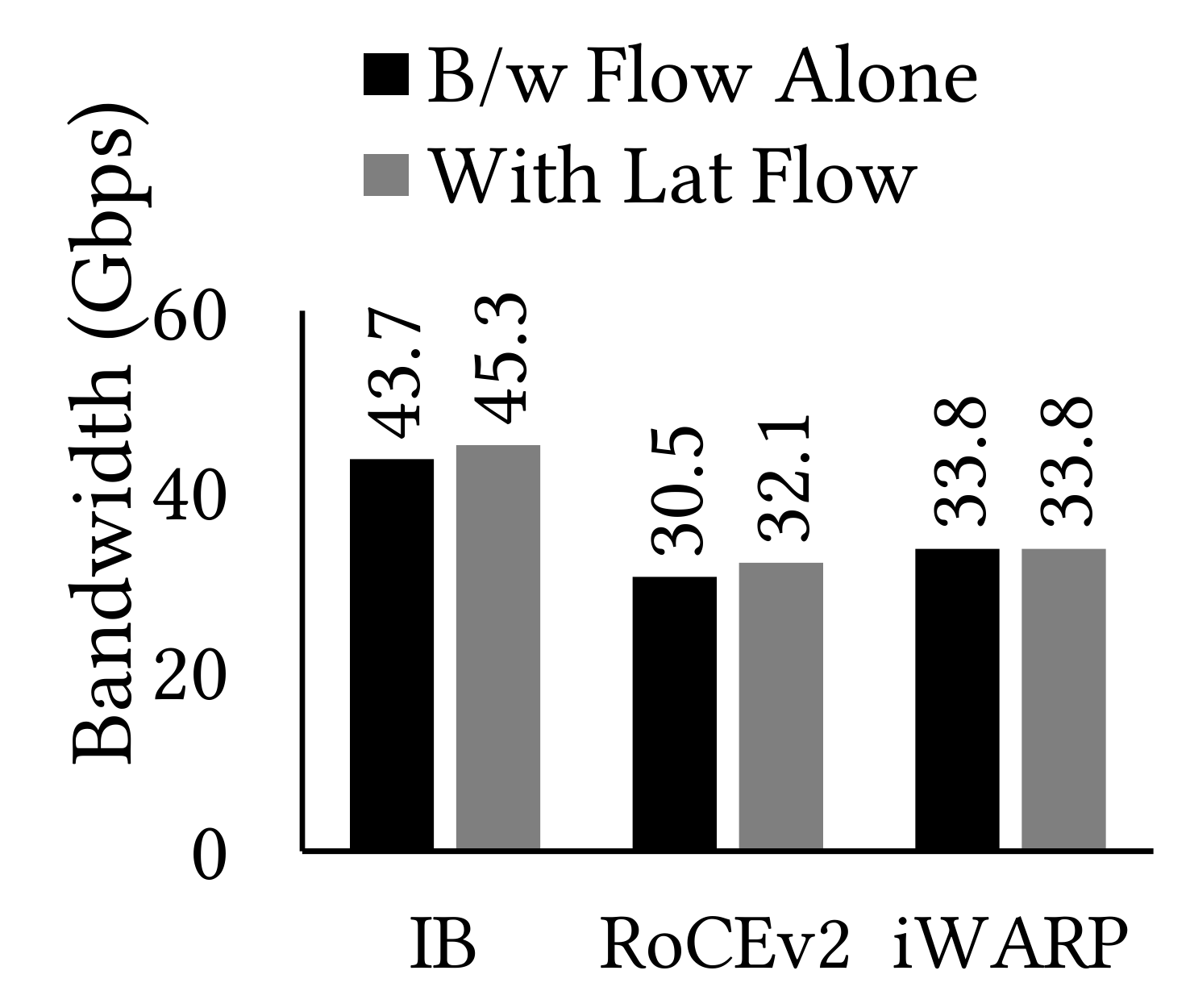}%
		}	
	\caption{Performance anomalies of a latency-sensitive flow running against a 1MB background bandwidth-sensitive flow.}
	\label{fig:sec3-E-vs-Lat}%
\end{figure}



\begin{figure*}[t]
    \mbox{
      \begin{minipage}{0.40\textwidth}
        \centering
		  \subfloat[][Throughput Flow]{%
			\label{fig:MOTI-EvT-ops}%
			\includegraphics[width=1.3in]{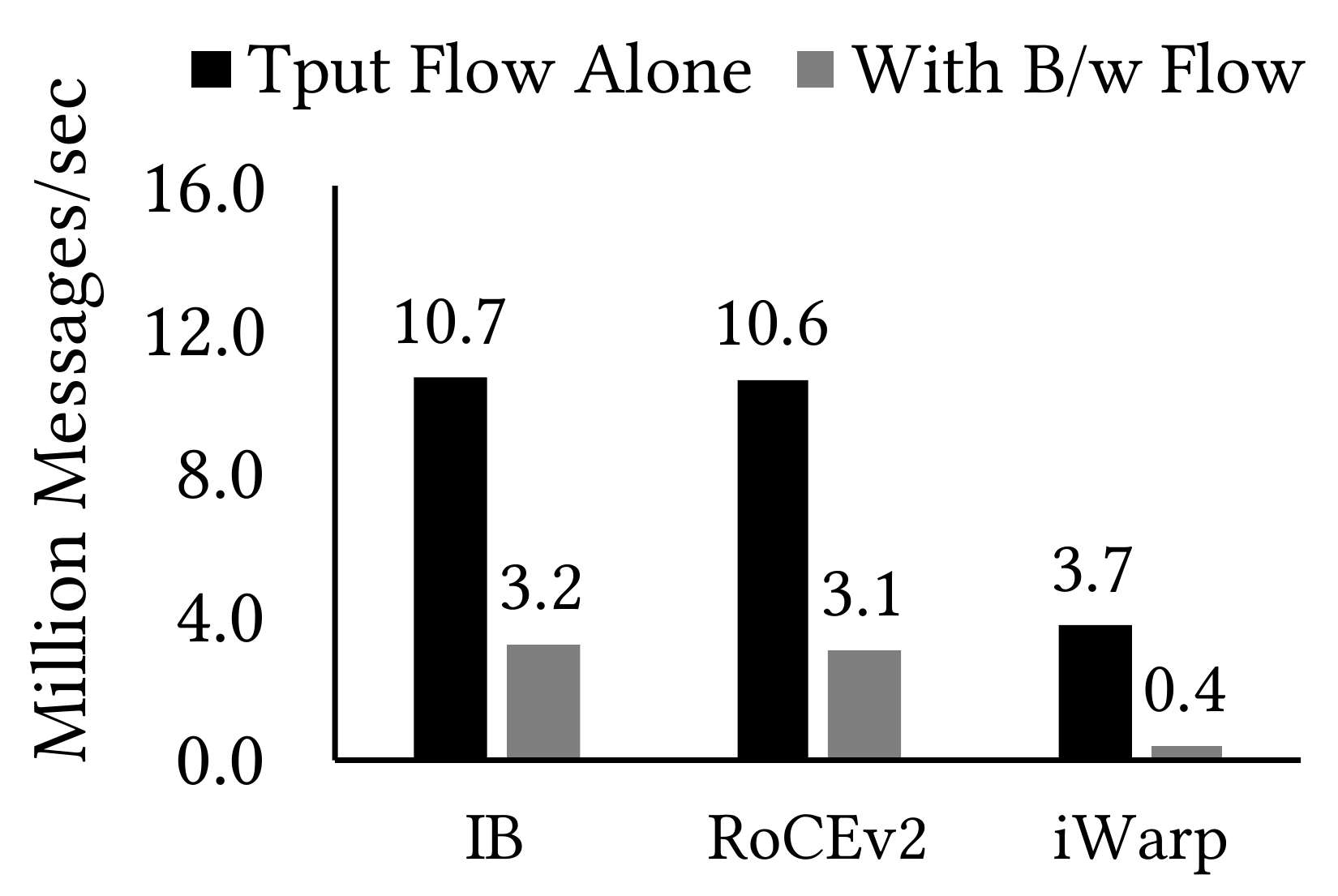}%
		  }
		\hfill
		  \subfloat[][Bandwidth Flow]{%
			\label{fig:MOTI-EvT-bw}%
			\includegraphics[width=1.3in]{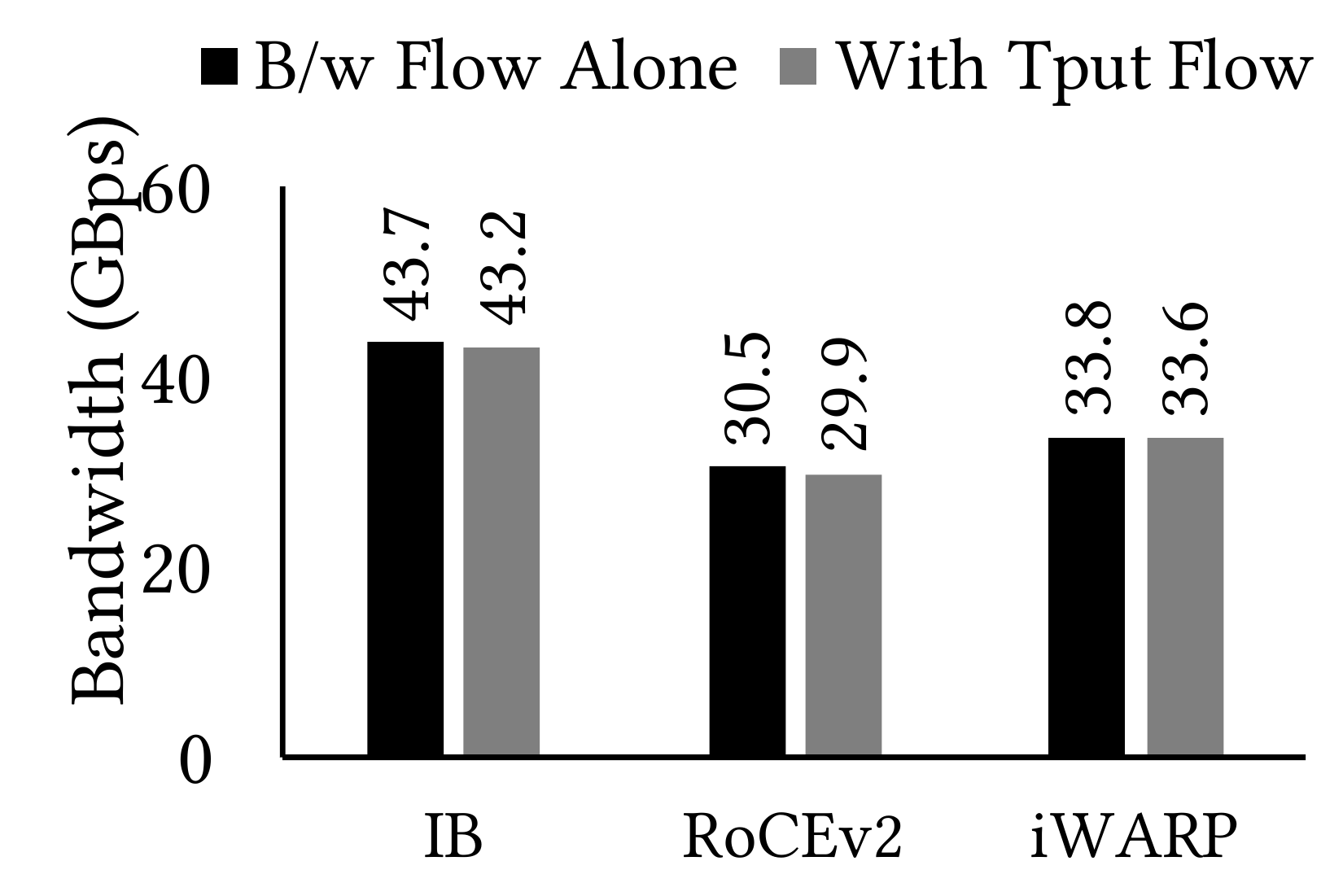}%
		  }	
	    \caption{Performance of a throughput-sensitive flow with and without a bandwidth-sensitive flow.}
	    \label{fig:sec3-E-vs-tput}%
      \end{minipage}
      \hfill
      \begin{minipage}{0.33\textwidth}
        \centering
        \includegraphics[width=2.0in]{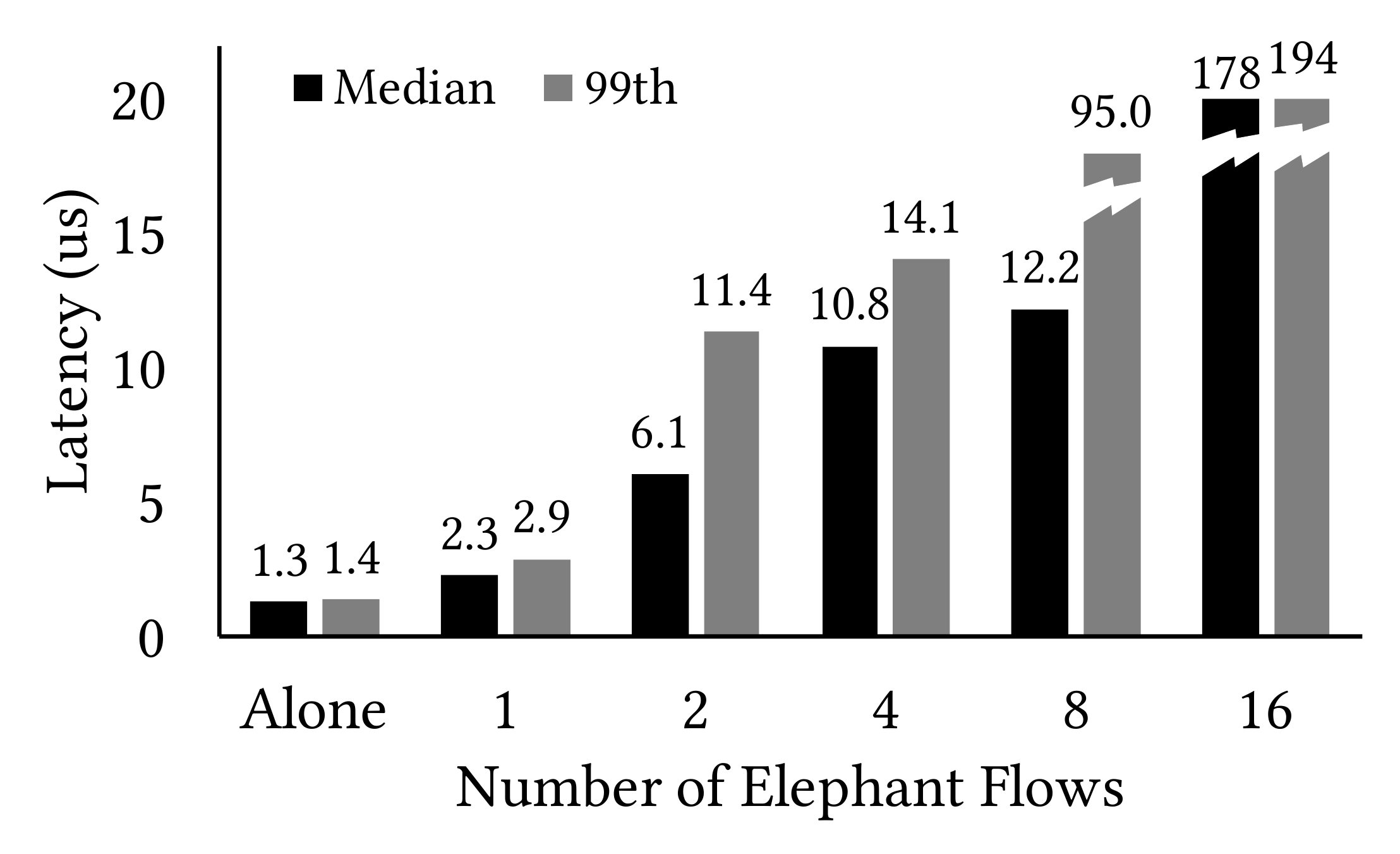}
        \caption{Impact of increasing 1MB background elephantson latency in InfiniBand.}%
        \label{fig:sec3-multiE-vs-Lat}%
      \end{minipage}
      \hfill
      \begin{minipage}{0.27\textwidth}
        \centering
        \includegraphics[width=1.8in]{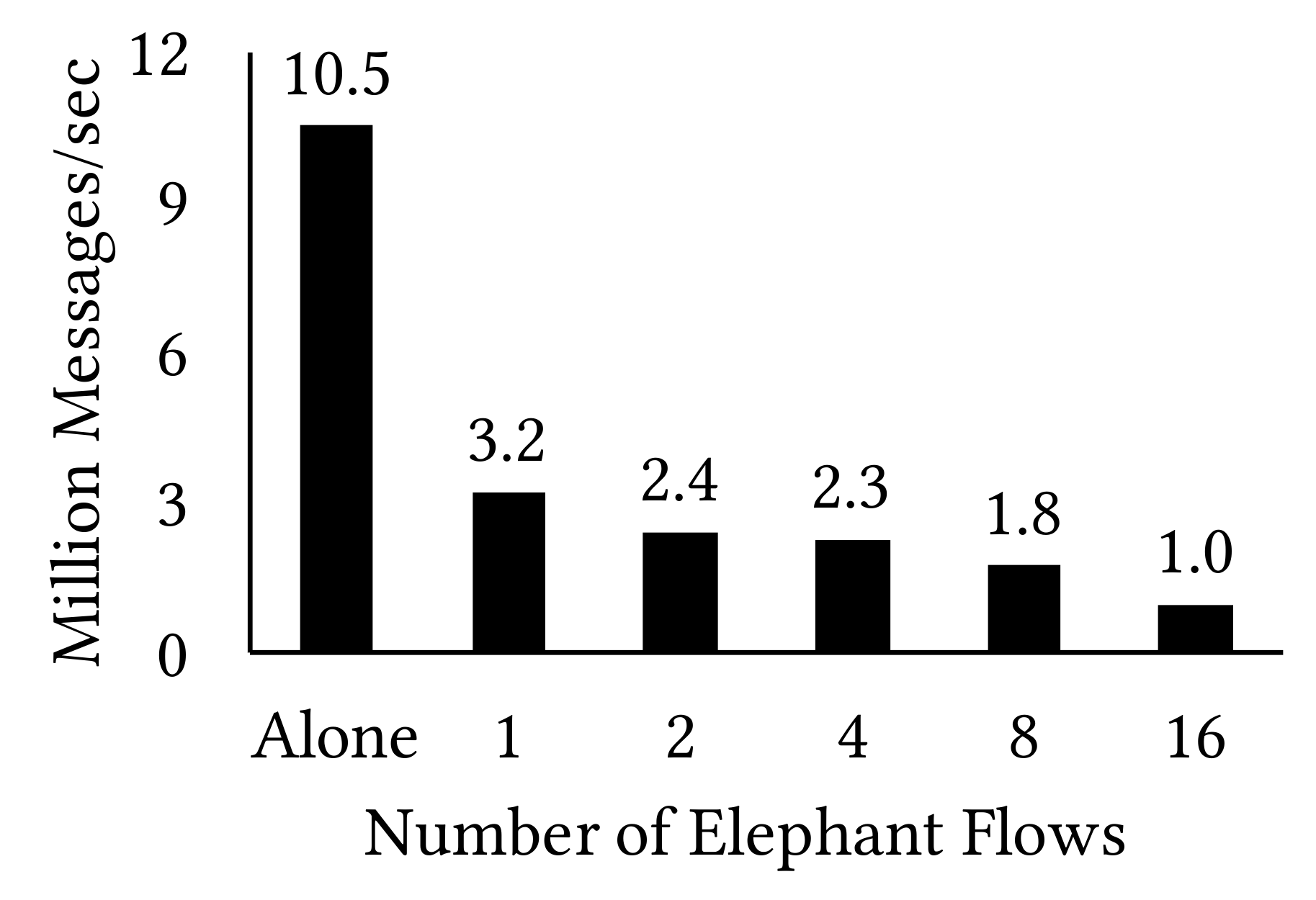}
        \caption{Impact of increasing background elephants on throughput.}%
        \label{fig:sec3-multiE-vs-Tput}%
      \end{minipage}
    }
\end{figure*}

\subsection{Flow-Level Analyses}
\label{sec3:flow-level-analyses}
We define a sequence of RDMA messages between the same pair of queue pairs (QPs) to be a \emph{flow}.
We focus on three primary types of flows and study how they affect each other. 
\begin{denseenum}
  \item \textbf{Latency-Sensitive:} Flows with small messages that care about individual message latencies. 

  \item \textbf{Throughput-Sensitive:} Flows with small messages trying to maximize the number of messages sent per second. 

  \item \textbf{Bandwidth-Sensitive:} Flows with large messages with high bandwidth requirements. 
\end{denseenum}

We performed microbenchmarks between two machines with the same type of RNIC, where both are connected to the same RDMA-enabled switch. 
For most of the experiments, we used 56 Gbps Mellanox ConnectX-3 Pro for InfiniBand, 40 Gbps Mellanox ConnectX-4 for RoCEv2, and 40 Gbps Chelsio T62100 for iWARP; 10 and 100 Gbps settings are described later.
All of the switches provide non-blocking forwarding at line-rate between ports, and we use a single switch in each experiment to avoid issues caused by path length asymmetry \cite{msr-rdma-15}. 
Further details of our hardware setups can be found in Table~\ref{tab:hw-spec} of Appendix~\ref{app:hw-summary}.

We used Mellanox perftest 4.2 \cite{perftool} as the benchmarking tool with minor modifications to enable latency and throughput logging and event-triggered polling in sending bandwidth-sensitive flows. 
Unless otherwise specified, latency-sensitive flows in our microbenchmarks send a continuous stream of 16B messages, throughput-sensitive ones send a continuous stream of batches with each batch having 64 16B messages, and bandwidth-sensitive flows send a continuous stream of either 1MB or 1GB messages. 
Latency- and throughput-sensitive flows use busy polling, whereas bandwidth-sensitive flows use event-triggered polling. 
Although all flows send data using RDMA WRITEs over reliable connection (RC) QPs in the observations below, other verbs show similar anomalies as well. 
Experiments on iWARP use RDMA Communication Manager to create and connect QPs.
We do not enable hardware virtual lanes in these experiments.


\subsubsection{Latency-Sensitive Flows are Unprotected}
\label{sec3:lat-flow-anomalies}
The biggest isolation issue appears to be the performance degradation of latency-sensitive flows in the presence of bandwidth-sensitive flows.
The performance of the former deteriorate for all RDMA implementations (Figure~\ref{fig:sec3-E-vs-Lat}).
Out of the three implementations we benchmarked, InfiniBand and RoCEv2 observes $1.85\times$ and $3.82\times$ degradations in median latency and $2.23\times$ and $4\times$ at the 99th percentile.
While iWARP performs well in terms of median latency, its tail latency degrades dramatically ($95\times$) in the presence of a bandwidth-sensitive flow.
The background bandwidth-sensitive flows were not affected across all three implementations.

\subsubsection{Throughput-Sensitive Flows Require Isolation}
\label{sec3:tput-flow-anomalies}
Throughput-sensitive flows also suffer.
When a background bandwidth-sensitive flow is running, the throughput-sensitive ones observe a throughput drop of $2.85\times$ or more across all RDMA implementations (Figure~\ref{fig:sec3-E-vs-tput}).

\subsubsection{Adding More Flows Exacerbates the Anomalies}
\label{sec3:add-more-flows}
The lack of protection for the latency-sensitive flows further exacerbates as more elephant flows (or equivalently more QPs) are created.
We increase the number of bandwidth-sensitive flows in our experiment to simulate more realistic datacenter applications.
Although InfiniBand performs relatively well in the presence of a single background bandwidth-sensitive flow (Figure~\ref{fig:sec3-E-vs-Lat}), adding one more flow incurs an additional drop of $2.65\times$ and $3.79\times$ in median and 99th percentile latencies (Figure~\ref{fig:sec3-multiE-vs-Lat}).
With 16 or more bandwidth-sensitive flows, the latency-sensitive flow can barely make any progress.
We observed a similar trend in other RDMA technologies.

Similarly, a throughput-sensitive flow experiences a continuous falloff in performance with the increasing number of background bandwidth-sensitive flows, losing 90\% of its original throughput with 16 elephant flows (Figure~\ref{fig:sec3-multiE-vs-Tput}).

Those anomalies illustrate RNIC's inability to handle multiple types of flows, which could stem from the limited number of queues inside the RNIC hardware, increasing head-of-line (HOL) blocking of small flows.

\subsubsection{Latency-Sensitive Flows Coexist Well; So Do \\ Throughput-Sensitive Flows}
\label{sec3:good-cases}
We observe no obvious anomalies among latency- or throughput-sensitive flows, or a mix of the two.
Detailed results can be found in Appendix~\ref{sec:lat-and-tput-anomalies}.

\subsubsection{Bandwidth-Sensitive Flows Hurt Each Other}
\label{sec3:bw-flow-anomalies}
\begin{figure}[!t]
	\centering
		\subfloat[][Different message sizes]{%
			\label{fig:sec3-E-vs-E}%
			\includegraphics[width=1.65in]{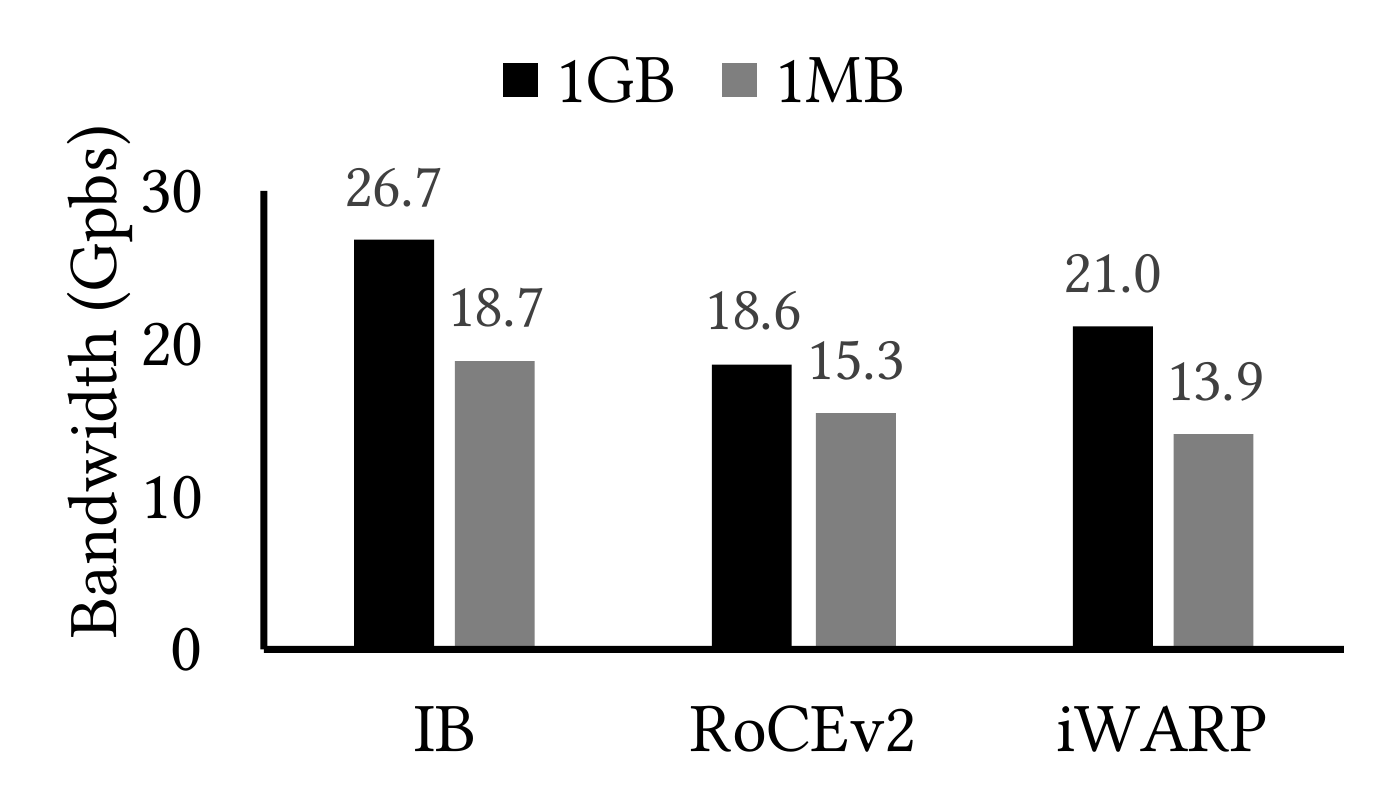}%
		}
		\hfill
		\subfloat[][Single-QP vs. multi-QP]{%
			\label{fig:sec3-multiE-vs-E}%
			\includegraphics[width=1.65in]{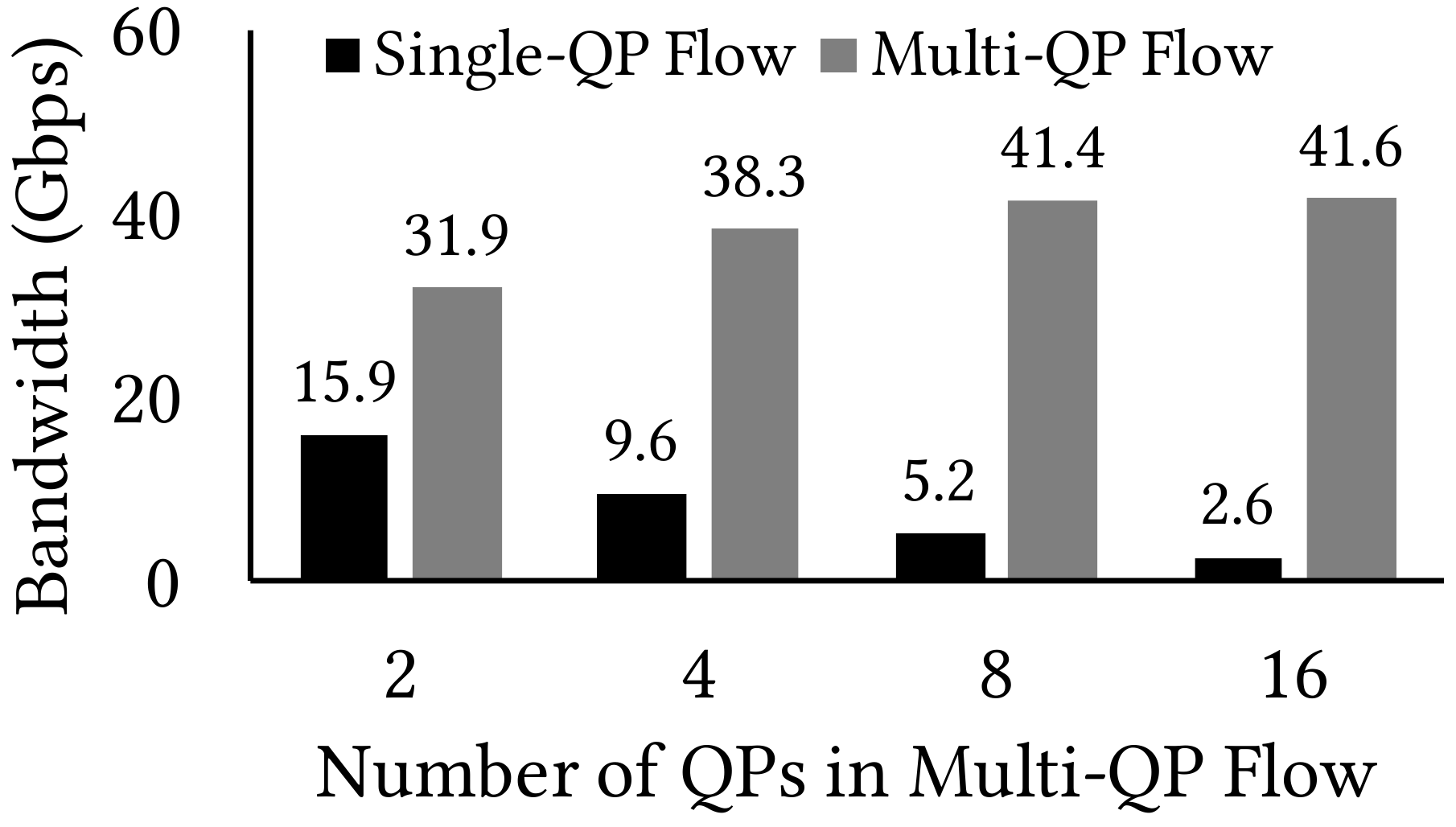}%
		}	
	\caption{Anomalies among Bandwidth-sensitive flows. (a) Two flows with different message sizes. (b) Single-QP vs. multi-QP bandwidth-sensitive flows in InfiniBand.}%
	\label{fig:sec3-bw-anomalies}%
\end{figure}

Unlike latency- and throughput-sensitive flows, bandwidth-sensitive flows with different message sizes do affect each other, especially when using event-triggered polling of completion events.
Although busy-polling can mitigate the unfairness in some cases \cite{iso-anomalies}, using busy-polling -- especially for bandwidth-sensitive flows where throughput is not the primary issue -- leads to unnecessary CPU waste.
Figure~\ref{fig:sec3-E-vs-E} shows that a bandwidth-sensitive flow using 1MB messages receive smaller share than one using 1GB messages.
The larger flow receives $1.42\times$, $1.22\times$ and $1.51\times$ more bandwidth in InfiniBand, RoCEv2, and iWARP, respectively.

Moreover, the current RNIC allocates bandwidth resources based on the unit of QPs without distinguish which application those QPs come from((Figure~\ref{fig:sec3-multiE-vs-E})).
In other words, users can use more QPs (similar to multiple connections in TCP/IP) to gain more bandwidth.
Althoguh we assume a cooperative datacenter, it is hard to restrain users from using a certain number of QPs in their applications, especially when an application indeed needs to establish connections to multiple receivers.
Without a proper control on the bandwidth share, multiple bandwidth-sensitive applications can result in unexpected bandwidth share.


%

\subsubsection{Anomalies are Present in Faster Networks Too}
We performed the same benchmarks on 100 Gbps InfiniBand, only to observe that most of the aforementioned anomalies are still present. 
Appendix~\ref{app:100gbps} has the details.

\begin{figure}[!t]
	\centering
		\subfloat[][FaSST throghput]{%
			\label{fig:MOTI-fasst-tput}%
			\includegraphics[width=1.0in]{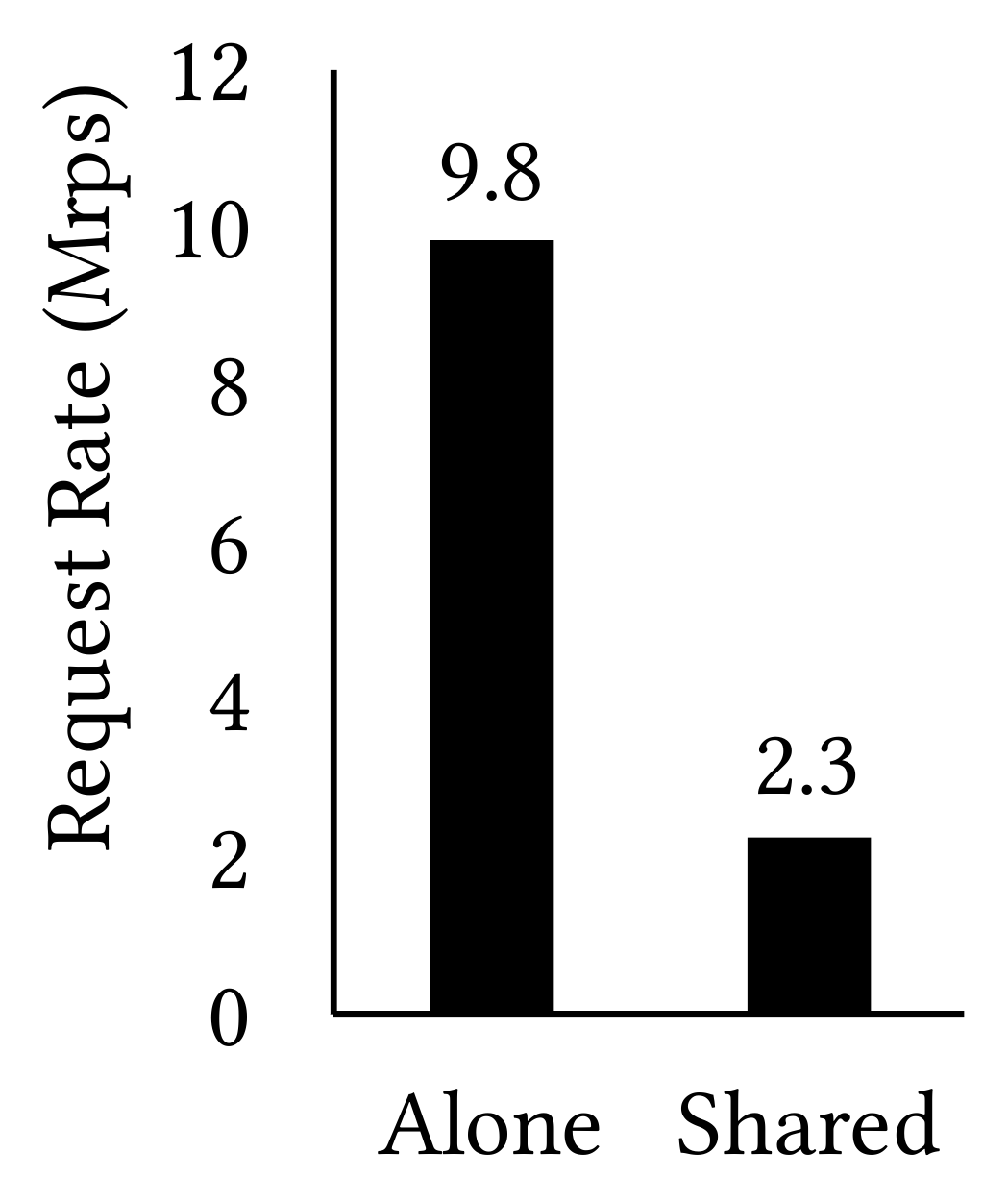}%
		}
		\hfill
		\subfloat[][eRPC throghput]{%
			\label{fig:MOTI-erpc-tput}%
			\includegraphics[width=1.0in]{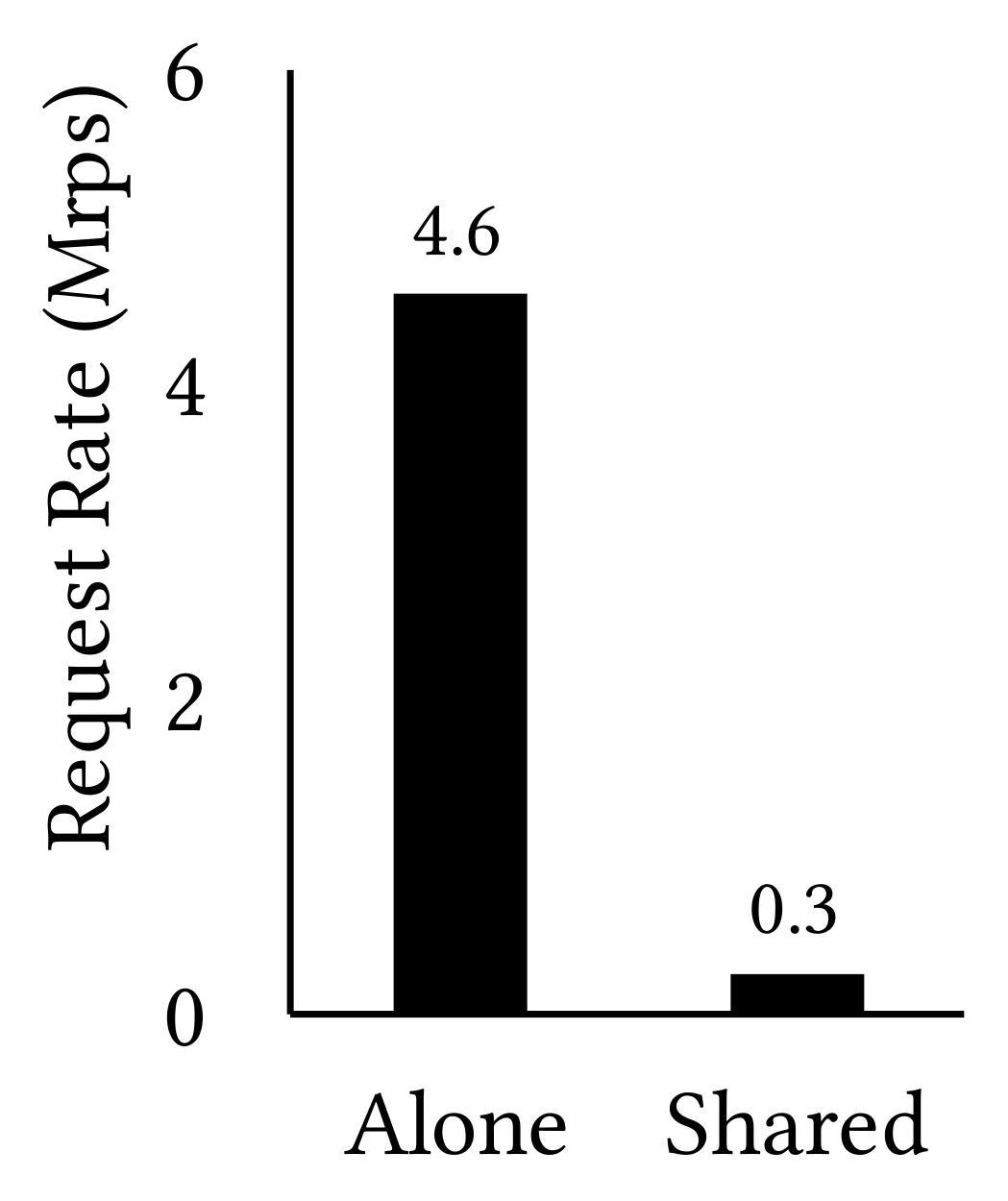}%
		}
		\hfill
		\subfloat[][eRPC latency]{%
			\label{fig:MOTI-eprc-lat}%
			\includegraphics[width=1.2in]{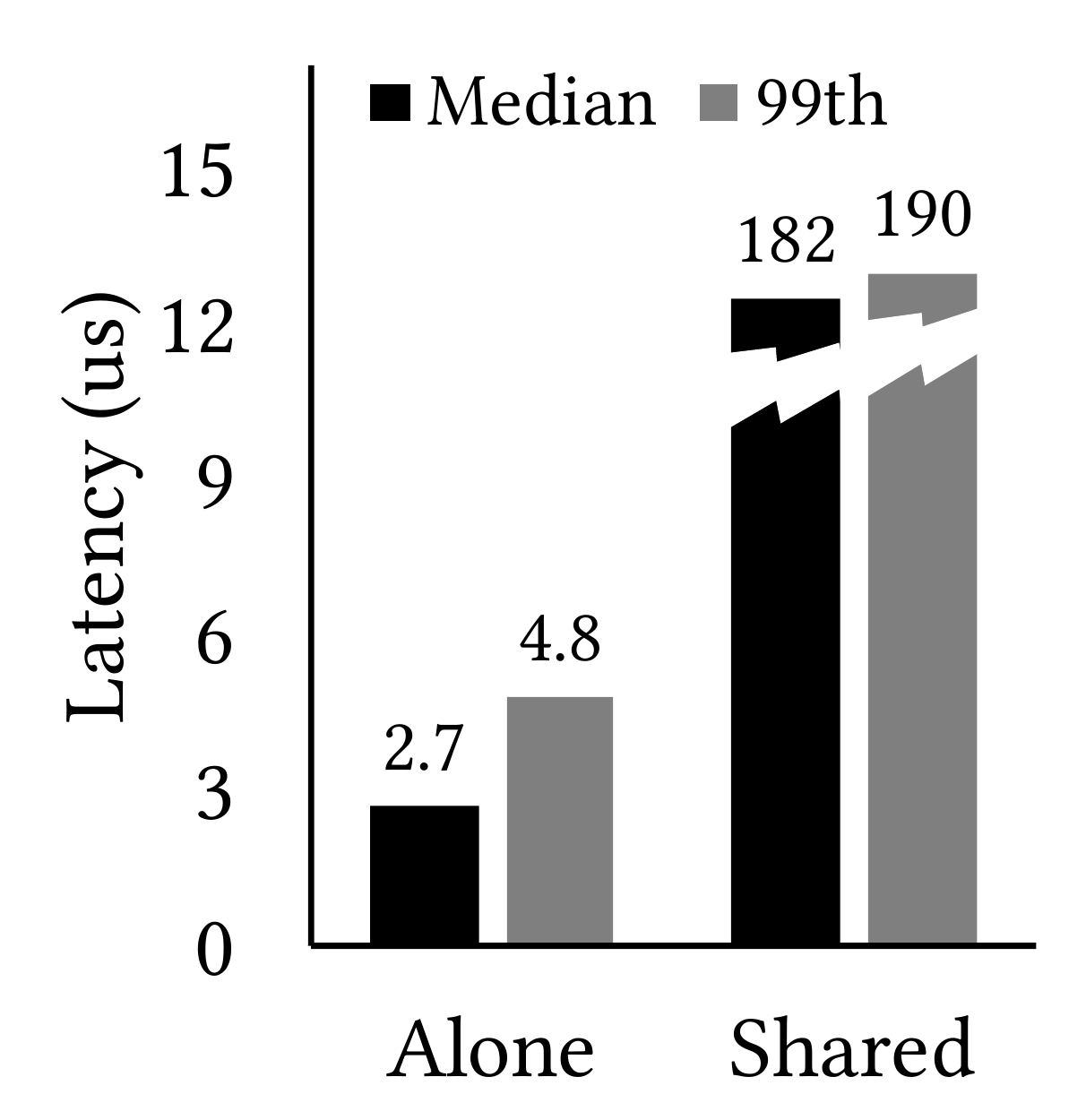}%
		}	
	\caption{FaSST's and eRPC's performance degradation at the presence of a bandwidth-sensitive storage application.}%
	\label{fig:sec3-eprc-micro}%
\end{figure}

\subsection{Application-Level Analyses}
\label{sec3:iso-in-apps}

In this section, we demonstrate how real RDMA-based systems fail to preserve their performance in the presence of the aforementioned anomalies.

\subsubsection{RDMA-Based Blob Storage}
To generate background traffic, we have implemented a simple RDMA-based blob storage backend across 16 machines.
Users read/write data to this storage using a PUT/GET interface via frontend servers. 
Objects larger than 1MB are divided into 1MB splits and distributed across the backend servers. 
This generates a stream of 1MB transfers, and the following RDMA-optimized systems have to compete with them in our experimental setup.

\subsubsection{FaSST}
FaSST \cite{fasst-osdi16} is an RDMA-based RPC system optimized for high message rate.
We deploy FaSST in 2 nodes with message size of 32 bytes and a batch size of 8.
We use 4 threads to saturate FaSST's message rate at 9.8 Mrps.
In the presence of the storage application, FaSST's throughput experiences a 74\% drop (Figure~\ref{fig:MOTI-fasst-tput}).

\subsubsection{eRPC}
eRPC \cite{erpc} is a brand-new RPC system built on top of RDMA.
We deploy eRPC in 2 nodes with message size of 32 bytes.
We evaluate eRPC's latency and throughput using the microbenchmark provided by its authors.
For the throughput experiment, we use 2 worker threads with a batch size of 8 on each node because 2 threads are enough to saturate the message rate in our 2-node setting.
In the presence of the storage application, eRPC's throughput drops by 93\% (Figure~\ref{fig:MOTI-erpc-tput}), and its median and tail latencies increase by $67\times$ and $40\times$, respectively (Figure~\ref{fig:MOTI-eprc-lat}).

\subsection{Congestion Control Does Not Fix It}
\label{sec3:cc-no-good}
To demonstrate that  DCQCN~\cite{msr-rdma-15} and PFC do not fix these
anomalies, we performed the benchmarks again with PFC enabled at both
the NICs and switch ports,  DCQCN~\cite{msr-rdma-15} enabled at the NICs,
and ECN markings enabled on a Dell 10 Gbps Ethernet switch (S4048-ON).
In these experiments, latency- and throughput-sensitive flows still suffer unpredictably
(\S\ref{sec:dcqcn-exp}).  

\subsection{Source of RDMA Performance Anomalies}
\label{sec3:anomalies-source}
We perform all our flow-level analyses in a simple 1-switch 2-node setting.
These anomalies occur even though the
switch is non-blocking and there are only two active ports on the switch.
This implies that the network is not the source of
anomalies in these experiments, and thus explains why DCQCN does not fix those anomalies. 
Rather, at the end hosts, RNICs' immediately processing all ready-to-consume messages to achieve work conservation 
is very likely to cause head-of-line (HOL) blocking of the smaller messages by the larger ones.
As a result, message latencies increase unpredictably, flows receive unfair bandwidth shares, and throughputs drop. 


%
%
%
%
%


\subsection{Summary}
We summarize our key observations as follows:
\begin{denseitemize}
    \item Both latency- and throughput-sensitive flows need isolation from the bandwidth-sensitive flows (\S\ref{sec3:lat-flow-anomalies}--\S\ref{sec3:add-more-flows}).
  
    \item If only latency- or throughput-sensitive flows (or a mix of the two) compete, they are isolated from each other (\S\ref{sec3:good-cases}). 
    
    \item Multiple bandwidth-sensitive flows can lead to unfair bandwidth allocations depending on their message sizes or number of QPs in use (\S\ref{sec3:bw-flow-anomalies}).

    \item Highly optimized, state-of-the-art RDMA-based systems also suffer from the anomalies we discovered (\S\ref{sec3:iso-in-apps}).

    \item The presence of a congestion control protocol is no panacea to isolate latency- or throughput-sensitive flows from the bandwidth-sensitive ones (\S\ref{sec3:cc-no-good}).
 
    \item The performance anomalies we discovered stem from end hosts and are very likely caused by HOL Blocking in RNICs (\S\ref{sec3:anomalies-source}).

\end{denseitemize}

%% file: properties.tex
\section{Requirements}
\label{sec:props}

\textbf{Goals.}
An ideal RDMA performance isolation solution should satisfy the following goals:


\begin{denseitemize}
  \item \textbf{Performance Isolation w/o Sacrificing Utilization:} Performance isolation and work conservation are known to be at odds in network-level scenarios \cite{hug, faircloud} even though max-min fairness \cite{jaffe-maxmin, wfq, wf2q} provides both on a single link.
    The latter, however, only holds when all flows are bandwidth-sensitive and have packets with bounded size differences \cite{drr}; for latency-sensitive flows, one must plan for the worst case \cite{hfsc}.
    Given that RDMA messages can range from bytes to gigabytes, relying on max-min fairness is not enough. 
    We should strive for increasing utilization without sacrificing isolation.
  
  \item \textbf{Traffic-Agnostic, Simple Service Interface:} Applications cannot be expected to change the nature of their traffic.
    Hence, we must accommodate all three types of flows.
    Applications should not have to specify traffic volume either.
    It is thus preferable to provide a narrow interface -- \eg, have applications choose one of the three classes of service when creating a flow.
  
  \item \textbf{No Changes to Applications or Hardware:} Although an ongoing body of work focuses on programmable NICs and switches \cite{p4}, large-scale deployments of these techniques are yet to happen.
    On a traditional life cycle, changes to the RNICs or switches are expensive, time-consuming, and are hard to deploy. 
    If possible, simple edge- and software-based solutions that are application- and hardware-independent are preferable. 
  
  \item \textbf{Scalability w/ Low Resource Usage:} The proposed software solution should scale to a large number of flows without large resource consumption to remain practical.
  
\end{denseitemize}

\textbf{Non-Goals.}
Users/tenants/applications gaming the public cloud network is a well-studied topic \cite{faircloud, mogul-popa, hug}, and RDMA will likely experience similar challenges in such an environment. 
Nonetheless, given the extent of RDMA performance isolation anomalies even in a controlled, non-adversarial environment (\S\ref{sec:anomalies}), we restrict our focus on a \emph{cooperative} datacenter environment in this paper. 
We consider the need for strategyproofness \cite{drf, drfq, faircloud} to mitigate adversarial/malicious behavior to be a non-goal.

%% file: design.tex
\section{{\name}}
\label{sec:design}
{\name} provides performance isolation between latency-, throughput-, and bandwidth-sensitive flows while maximizing RNIC resource utilization. 
In this section, we first present {\name}'s design principles (\S\ref{sec:design-principles}). 
Next, we present {\name}'s overall architecture in terms of its two core components: the {\name} daemon (\S\ref{sec:architecture}) and {\name} shapers (\S\ref{sec:shapers}). 
Finally, we extend {\name} to handle remote READs via inter-machine coordination (\S\ref{sec:remote-control}) and to further increase utilization when latency-sensitive flows cannot be helped (\S\ref{sec:new_algorithm}).

\subsection{Design Principles}
\label{sec:design-principles}
{\name}'s design principles follow from its requirements.
\begin{denseitemize}
  \item \textbf{Isolation via Sharing Incentive:} Given the isolation-vs-utilization tradeoff, instead of picking either one, we opt for guaranteeing each of the $n$ flows at least $\frac{1}{n}$th of one of the two resources and then maximize the total utilization until latencies may be affected. 
    This ensures that we are not unfairly penalizing one specific type of flows.
  
  \item \textbf{Soft Admission Control of Latency-Sensitive Flows:} An
      implication of enforcing a sharing incentive is that providing latency guarantees may become untenable (\eg, when the number of other flows are high). 
    In such cases, {\name} informs an application that a new latency-sensitive flow will not meet its latency target.

  \item \textbf{Sender-Side Multi-Resource Shaping in Software:} Instead of keeping separate queues for each bandwidth- or throughput-sensitive flow, {\name} relies on a host-wide daemon that arbitrates between all resource-hungry flows from the sender side.
    It splits large messages to roughly equal-sized chunks, which helps avoid HOL blocking.
    We do not use hardware rate limiters in the RNIC because they are limited in number and slow when setting new rates (2 milliseconds in our setup).

\end{denseitemize}

\begin{figure}[!t]
  \centering
  \includegraphics[width=\columnwidth]{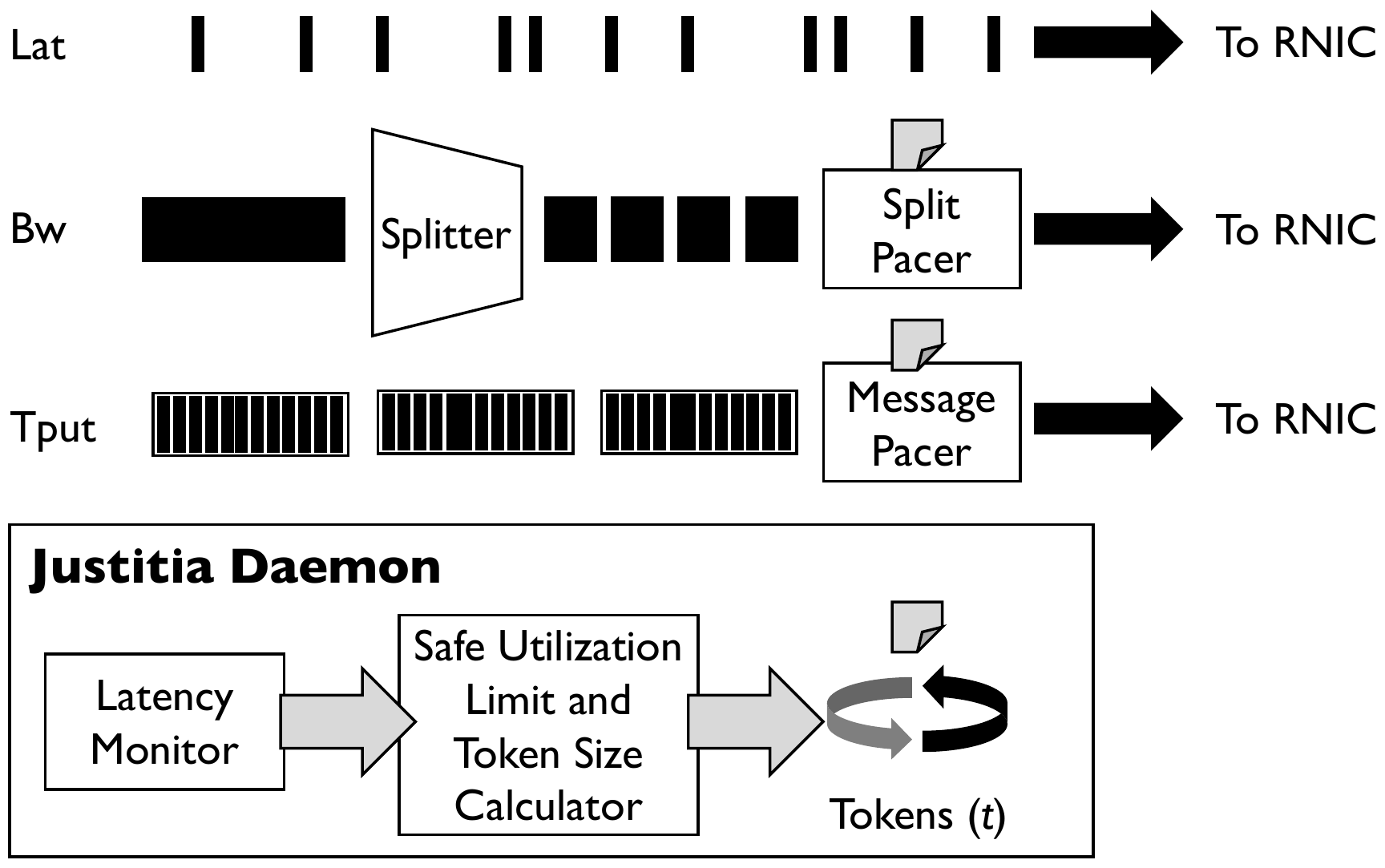}
  \caption{{\name} architecture.
    Bandwidth- and throughput-sensitive flows are shaped by tokens generated at a regular interval by {\name}. 
    Latency-sensitive flows are not paced at all.}%
  \label{fig:architecture}%
\end{figure}

\subsection{The {\name} Daemon}
\label{sec:architecture}

Figure~\ref{fig:architecture} presents a high-level overview of {\name}.
Each machine has a {\name} daemon that performs latency monitoring and proactive rate management, and applications create QPs using the existing API to perform RDMA communication.
{\name} relies on applications to optionally identify the type of a flow when creating the corresponding QP.\footnote{We implement this by passing an optional flag in \texttt{ibv\_qp\_init\_attr} structure in the \texttt{ibv\_create\_qp()} function (done in one line of code).}
By default, flows are treated as bandwidth-sensitive.
In the following, we first provide a high-level overview of how {\name} works and then elaborate on its different components. 

To monitor latency, {\name} does not interact with latency-sensitive flows at all. 
They can send messages/data whenever they want because they cannot saturate either of the two RNIC resources.
Instead, {\name} maintains a system-wide reference latency-sensitive flow to estimate the 99th percentile (\curtail) latencies for small messages (\S\ref{sec:reference-flow}). 
This works well in estimating the impact of resource-hungry (bandwidth- and throughput-sensitive) flows on the latency-sensitive ones because all latency-sensitive flows get affected when resource utilization is very high.
Moreover, by monitoring its own reference flow instead of the flows from applications, {\name} does not need to wait on latency-sensitive applications to send a large enough number of sample messages for accurate tail latency estimation.
It does not add additional delay into those applications by probing their flows either, which is significant in a microsecond-scale network.

{\name} performs proactive rate management of all bandwidth- and throughput-sensitive flows from the sender side. 
At its heart, the key idea is maximizing the safe total utilization (\safeutil) of all resource-hungry flows without violating system-wide latency target: \targettail, while guaranteeing sharing incentive.
Using \curtail as a signal, {\name} uses an Additive Increase Multiplicative Decrease (AIMD) algorithm to maximize \safeutil (\S\ref{sec:safeutil}).

{\name} enforces \safeutil among bandwidth- and throughput-sensitive flows using \emph{multi-resource} tokens.
The {\name} daemon generates multi-resource tokens every \tokengentime interval to limit the total utilization of all resource-hungry flows to \safeutil (\S\ref{sec:token-management}).
Each token corresponds to a \emph{fixed} amount of bytes (\tokenbytes) and a \emph{fixed} number of messages (\tokenops).
Because of the coupled nature of the two RNIC resources, a flow can completely spend only one resource of a token -- a bandwidth-sensitive flow will exhaust its associated bytes, while a throughput-sensitive flow will exhaust the number of operations that token allows to send. 
Tokens are distributed in a fair fashion among the active resource-hungry flows by the {\name} daemon up to \safeutil (\S\ref{sec:token-management}). 

Given the tokens, each flow shapes/paces itself (\S\ref{sec:shapers}).
A token is large enough for a flow not to bottleneck on token generation and distribution. 
Large messages from bandwidth-sensitive flows are divided into equal-sized chunks, which are then paced based on token availability. 
Splitting is necessary to avoid HOL blocking caused by bandwidth-sensitive flows.
Batches of small messages from throughput-sensitive flows are paced by per-flow pacers too. 

\subsubsection{Handling Latency-Sensitive Flows}
\label{sec:reference-flow}
{\name} does not interrupt or interact with latency-sensitive flows. 
Instead, in the presence of at least one latency-sensitive flow, it runs a reference flow that keeps sending 10B messages to another machine in the cluster in periodic intervals (by default, \refperiod=0.5 ms).
\refperiod is chosen to send the reference flow at a rate that adds no additional delay to other latency-sensitive flows, but still is frequent enough to monitor latency anomalies.
{\name} then measures the latency between posting a message and when its work completion is generated. 

Given the measurements, {\name} maintains a sliding window of the most recent \refcount (=10000) measurements, and it uses a count-min sketch \cite{countmin} on that window to estimate \curtail.
This is fed into the \safeutil computation algorithm described below.

If \curtail is higher than \targettail, {\name} can perform soft admission control (\eg, returning a warning code) when creating new latency-sensitive flows. 
If \targettail cannot be met at all, {\name} can opt for maximizing utilization (\S\ref{sec:new_algorithm}).

\floatname{algorithm}{Pseudocode}
\begin{algorithm}[t!]

\begin{small}
\begin{algorithmic}[1]

\Procedure{OnLatencyFlowUpdate}{$L$, \curtail}
  \If{$L=0$}\Comment{Reset if no latency-sensitive flows}
    \State{\safeutil = \maxrate}
  \Else
    \If{\curtail $>$ \targettail}
      \State{\safeutil = max($\frac{\mbox{\safeutil}}{2}$, $\frac{B+T}{L+B+T} \times \mbox{\maxrate}$)}
    \Else
      \State{\safeutil = \safeutil + 1}
    \EndIf
  \EndIf
  \State{\tokengentime = \tokenbytes / \safeutil}
\EndProcedure

\end{algorithmic}
\end{small}
\caption{Maximize \safeutil}
\label{alg:safeutil}
\end{algorithm}

\subsubsection{Maximizing \safeutil}
\label{sec:safeutil}
In the absence of latency-sensitive flows, \safeutil is set to total RNIC bandwidth (\maxrate), where \maxrate is pre-determined on a per-RNIC basis using the benchmark flows from Section~\ref{sec:anomalies}.
Because the ratio between \maxrate to \maxtput (\ie, the total ops/second) is fixed for a given RNIC, calculating \safeutil in terms of bandwidth is sufficient.

In the presence of latency-sensitive flows, the overarching goal of {\name} boils down to continuously maximizing \safeutil based on the current \curtail estimation (Pseudocode~\ref{alg:safeutil}). 
At the same time, it must ensure that each resource-hungry flow -- assume there are $L$ latency-, $B$ bandwidth-, $T$ throughput-sensitive flows -- receives at least $\frac{1}{L+B+T}$th of the RNIC resources.
Instead of attempting to achieve this on a per-flow basis, {\name} focuses on maximizing \safeutil, where \safeutil is $\frac{B+T}{L+B+T}$ or a higher fraction of \maxrate.

To continuously update \safeutil, {\name} uses a simple AIMD scheme that reacts to \curtail every \refperiod interval as follows.
If the estimation is above \targettail, {\name} decreases \safeutil by half; \safeutil is guaranteed to be at least $\frac{B+T}{L+B+T}$ of \maxrate.
If the estimation is below \targettail, {\name} slowly increases \safeutil.
Because \safeutil ranges between $\frac{B+T}{L+B+T}$ to the total RNIC resources and latency-sensitive flows are highly sensitive to too high a utilization level, our conservative AIMD scheme, which drops utilization quickly to meet \targettail, works well in practice.

\subsubsection{Token Generation And Distribution}
\label{sec:token-management}
{\name} uses multi-resource tokens to enforce \safeutil among the $B$ bandwidth- and $T$ throughput-sensitive flows in a fair manner.
Each token represents amount of a fixed amount of bytes (\tokenbytes) and a fixed number of messages (\tokenops).
In other words, the size of \tokenbytes determines the chunk size a bandwidth-sensitive flow is split into.
A token is generated every \tokengentime interval, but the value of \tokengentime depends on \safeutil as well as on the size of each token.
For example, given 48 Gbps application-level bandwidth and 30 Million operations/sec on a 56 Gbps RNIC, if \tokenbytes is set to 1MB, then we set \tokenops=5000 operations and \tokengentime=167 microseconds.

{\name} daemon continuously generates one token every \tokengentime interval and distributes it among the active resource-hungry flows in a round-robin fashion. 
Each flow independently enforces its rate using one of the shapers (\S\ref{sec:shapers}).
Note that introducing the notion of \emph{weighted} round-robin is straightforward.
If a flow's weight is $w_i$, {\name} can ensure it receives $w_i$-proportional tokens during each round. 

\subsection{{\name} Shapers}
\label{sec:shapers}
{\name} shapers -- implemented in the RDMA driver -- enforce utilization limits provided by the {\name} daemon-calculated tokens.
There are two shapers in {\name}: one for bandwidth- and another for throughput-sensitive flows. 

\textbf{Shaping Bandwidth-Sensitive Flows.}
This involves two steps: \emph{splitting} and \emph{pacing}.
For any bandwidth-sensitive flow, {\name} \emph{transparently} divides any message larger than \tokenbytes into \tokenbytes-sized chunks to ensure that the RNIC only sees roughly equal-sized messages.
Splitting messages for diverse RDMA communication verbs -- \eg, one-sided vs. two-sided -- requires careful designing (\S\ref{sec:splitting-impl}).

Given chunk(s) to send, the pacer requests for token(s) from the {\name} daemon by marking itself as an active flow.
Upon receiving a token, it transfers chunk(s) until that token is exhausted and repeats until there is nothing left to send. 

The application is notified of the completion of a message only after all of its chunks have been successfully transferred.

\textbf{Shaping Throughput-Sensitive Flows.}
These flows typically deal with (batches of) small messages.
Consequently, there is no need for message splitting. 
Instead, a pacer ensures that the flow can send at most \tokenops messages corresponding to each token.
Each token is large enough so as not to bottleneck on token generation and distribution.

\textbf{Mitigating Head-of-Line Blocking.}
One of the foremost goals of {\name} is to mitigate HOL blocking caused by the bandwidth-sensitive flows to provide good isolation.
To achieve this goal, we need to split messages into smaller chunks and pace them at a certain rate (enforcing \safeutil) with enough spacing between them to minimize the blocking.
However, this simple approach creates a dilemma.
On the one hand, too large a chunk may not resolve HOL Blocking.
On the other hand, too small a chunk may not be able to reach \safeutil.
It also leads to increased CPU overhead from using a spin loop to fetch tokens generated in a very short period in which context switches are not affordable.
Note that this is another manifestation of the performance isolation-work conservation tradeoff.
We discuss how to pick the chunk size in Section~\ref{sec:autotuner} and how to reduce CPU overhead in Section~\ref{sec:reduce-cpu}.

\subsection{Handling READs via Remote Control}
\label{sec:remote-control}
So far we have discussed {\name} from a sender-side perspective. 
However, RDMA allows remote machines to read from a local machine using the RDMA READ verb. 
RDMA READs operations from machines $A$ to read data from $B$ compete with all sending operations (\eg, RDMA WRITE) from machine $B$.
Consequently, {\name} must consider remote READs as well.

One possible design to achieve this would be sending tokens from $B$ to $A$ so that $A$'s {\name} daemon can pace the READs.
However, this requires tight coordination between many machines and susceptible to latency variations in sending/receiving tokens.
Instead, we opt for a simpler solution in {\name}, wherein $B$ sends the updated guaranteed utilization ($\frac{1}{L'+B'+T'}$, where $X'$ is the updated count of $X$ including remote READ flows) to each remote flow after each update, and $A$ locally enforces that rate.
Note that this can sometimes decrease utilization when remote READ flows do not completely use their assigned resources.

\subsection{What If \targettail Is Unattainable?}
\label{sec:new_algorithm}
A key consequence of the isolation-utilization tradeoff is that \targettail may sometimes be unattainable -- \eg, when it is set too low or in the presence of too many resource-hungry flows.
This can cause underutilization as {\name} continuously try to reduce \curtail without success while limiting resource-hungry flows to $\frac{1}{n}$th shares.

We address this issue by providing an option to the operator: if \curtail is higher than \targettail for $\delta$ period, {\name} assumes that \targettail is unattainable.
It can then ignore latency-sensitive flows altogether and focus on equally sharing all resources among resource-hungry flows.

It may need to come out of this state only when the $\frac{L}{B+T}$ ratio changes.
Specifically, when $\frac{L}{B+T}$ becomes even smaller -- \eg, $L$ decreasing or $B+T$ increasing -- it can stay in the same state.
Only when $\frac{L}{B+T}$ increases, {\name} can go back to the original algorithm and try to attain \targettail again.

The cluster operator can decide whether to use this option based on their experience and application expectations.

%% file: implementation.tex
\section{Implementation}
\label{sec:implementation}
We have implemented the {\name} daemon as a user-space process in 3,100 lines of C, and the shapers are implemented inside individual RDMA drivers with 5,200 lines of C code.


\subsection{Determining Token Size for Bandwidth Target}
\label{sec:autotuner}
One of the key steps in determining \safeutil is deciding the size of each token.
Because the RNIC can become throughput-bound for smaller messages instead of bandwidth-bound, we cannot use arbitrarily small messages to resolve HOL blocking.
At the same time, given a utilization target, we want to use the smallest \tokenbytes value to achieve that target to reduce HOL blocking while maximizing utilization.

Instead of dynamically determining it using another AIMD-like process, we observe that 
(i) this is an RNIC-specific characteristic and 
(ii) the number of RNICs is small.
With that in mind, we maintain a pre-populated dictionary; {\name} simply uses the mappings during runtime.
When latency-sensitive flows are not present, a large token size (1MB) is used.
Otherwise, {\name} switched to the smallest chunk with which bandwidth-sensitive flows can use to saturate most of line rate (to enforce \safeutil) when sending them in a batch (Figure~\ref{fig:pick-chunk-size} in the Appendix).
To avoid the variation caused by chunk sizes in different hardwares, we set the chunk size to be 5 KB by default.
 
\subsection{Transparently Splitting RDMA Messages}
\label{sec:splitting-impl}
{\name} splitter transparently divides large messages of bandwidth-sensitive flows into smaller chunks for pacing. 
It ensures that an application posts to a QP in a fully transparent manner and does not notice any difference when posting a Work Queue Element (WQE) or polling for Completion Queue Element (CQE) of that request from the Completion Queue (CQ) associated with that QP. 

%

Our splitter uses a custom QP called a \emph{Split QP} to handle message splitting, which is created when the original QP of a bandwidth-sensitive flow is created.
A corresponding \emph{Split CQ} is used to handle CQEs for the WQEs posted to a Split QP. 
A custom completion channel is used to poll those CQEs in an event-triggered fashion to preserve low CPU overhead of native RDMA. 

\begin{figure}[!t]
  \centering
  \includegraphics[width=\columnwidth]{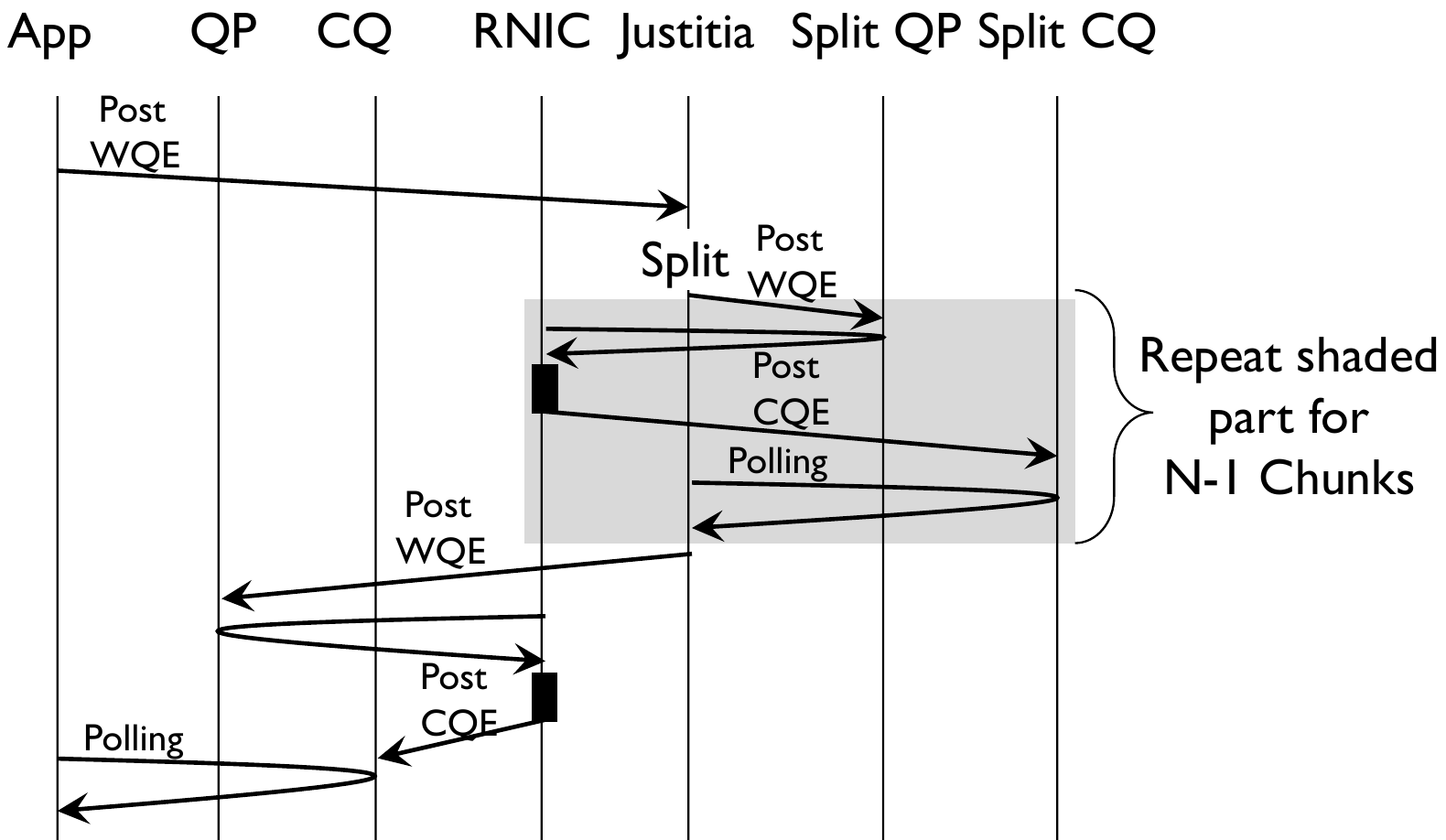}
  \caption{High-level overview of transparent message splitting in {\name} for one-sided verbs.
    Times are not drawn to scale.
    Two-sided verbs involve extra bookkeeping.}%
  \label{fig:split-sequence}%
\end{figure}

To handle one-sided RDMA operations, when detecting a message larger than \tokenbytes, we divide the original message into chunks and only post the last chunk to the application's QP (Figure~\ref{fig:split-sequence}). 
The rest of the chunks are posted to the Split QP.
Split QP ensures all chunks have been successfully transfered before the last chunk handled by the application's QP.
This makes sure the user cannot poll the CQE until the entire message has done transferring.
The two-sided RDMA operations such as SEND are handled in a similar way, with additional flow control messages for the chunk size change and receive requests to be pre-posted at the receiver side.
The WRITE\_WITH\_IMM verb can be further simplified by using WRITE in the WQE handled by the Split QP.

\subsection{Reduce CPU Overhead From Using Small Tokens}
\label{sec:reduce-cpu}
As mentioned earlier, using small tokens lead to CPU overhead mainly from busy spinning to fetch tokens generated at a short period (around 1us) which precludes any context switches.
We solve this challenge by decoupling token generation (TG) with token enforcement (TE).
We move the discussion to Appendix~\ref{sec:reduce-cpu-details} due to limited space.

%% file: evaluation.tex
\begin{figure}[!t]
	\centering
		\subfloat[][Latency-sensitive flow]{%
			\label{fig:EVAL-E-vs-lat-latency}%
			\includegraphics[width=2.2in]{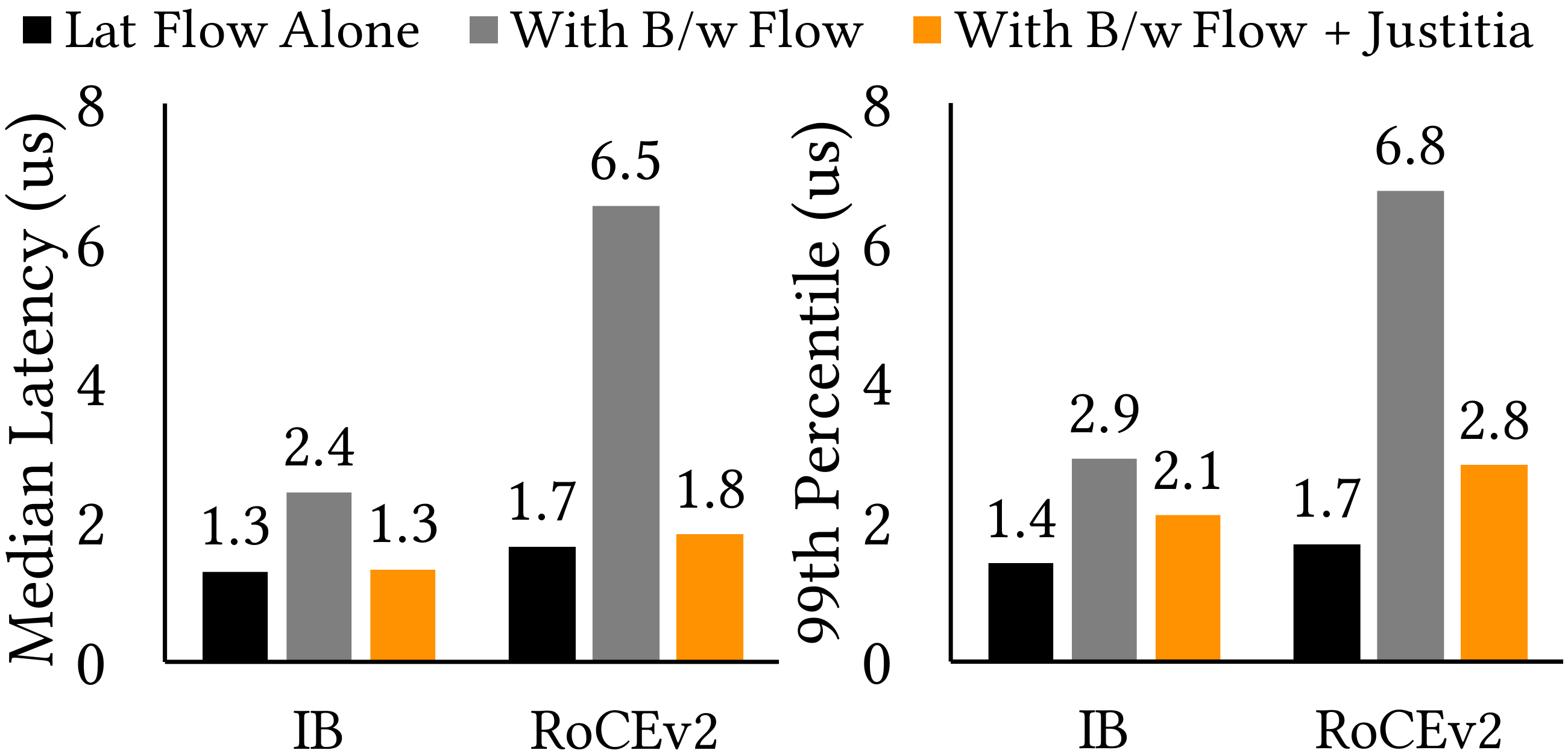}%
		}
		\hfill
		\subfloat[][{Bandwidth flow}]{%
			\label{fig:EVAL-E-vs-lat-bw}%
			\includegraphics[width=1.1in]{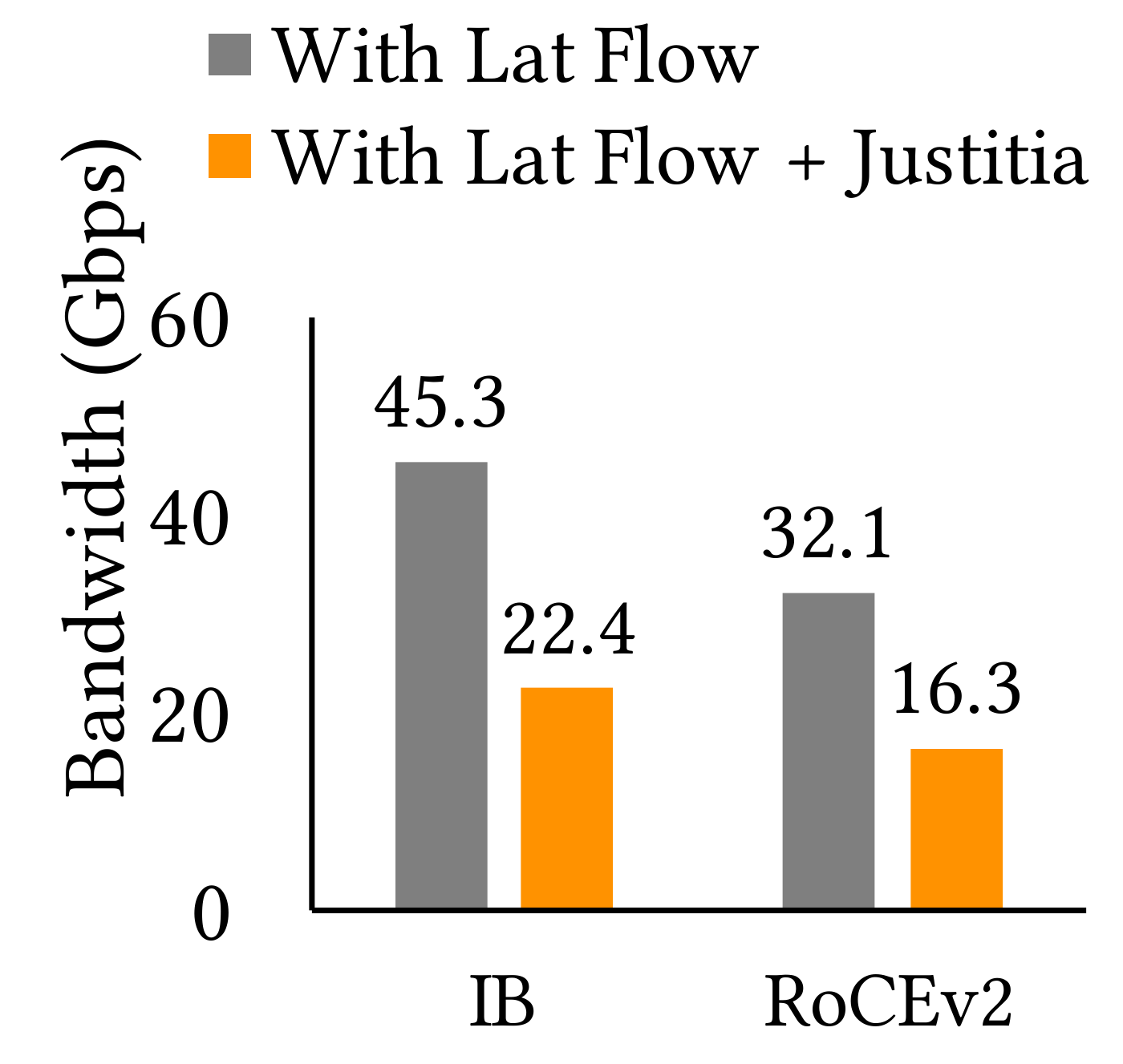}%
		}	
	\caption{Performance isolation of a latency-sensitive flow running against a 1MB background bandwidth-sensitive flow.}
	\label{fig:sec7-E-vs-lat}%
\end{figure}

\begin{figure}[!t]
	\centering
		\subfloat[][Latency-sensitive flow]{%
			\label{fig:EVAL-E-vs-lat-latency-large-target}%
			\includegraphics[width=2.2in]{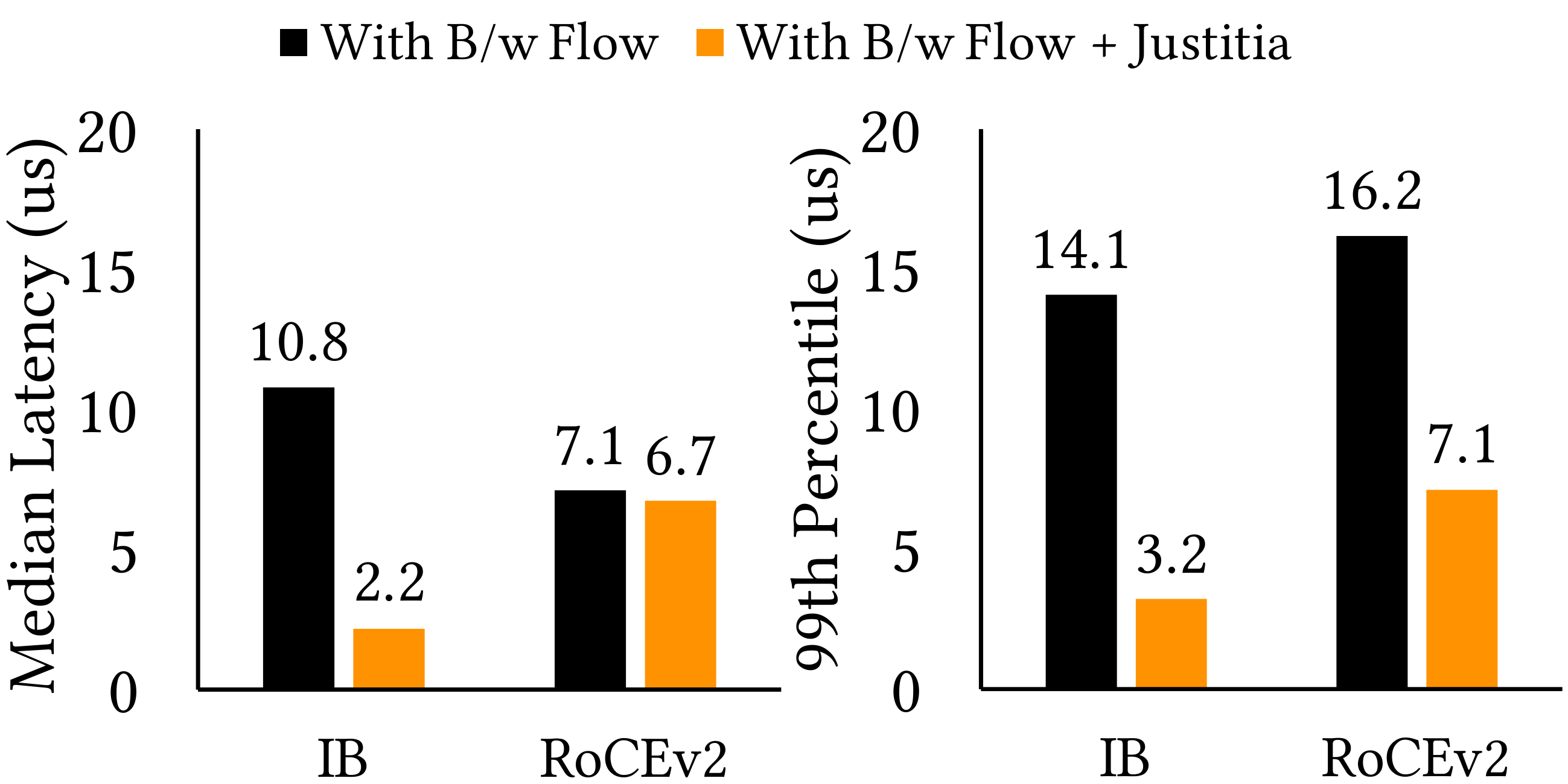}%
		}
		\hfill
		\subfloat[][{Bandwidth flow}]{%
			\label{fig:EVAL-E-vs-lat-bw-large-target}%
			\includegraphics[width=1.1in]{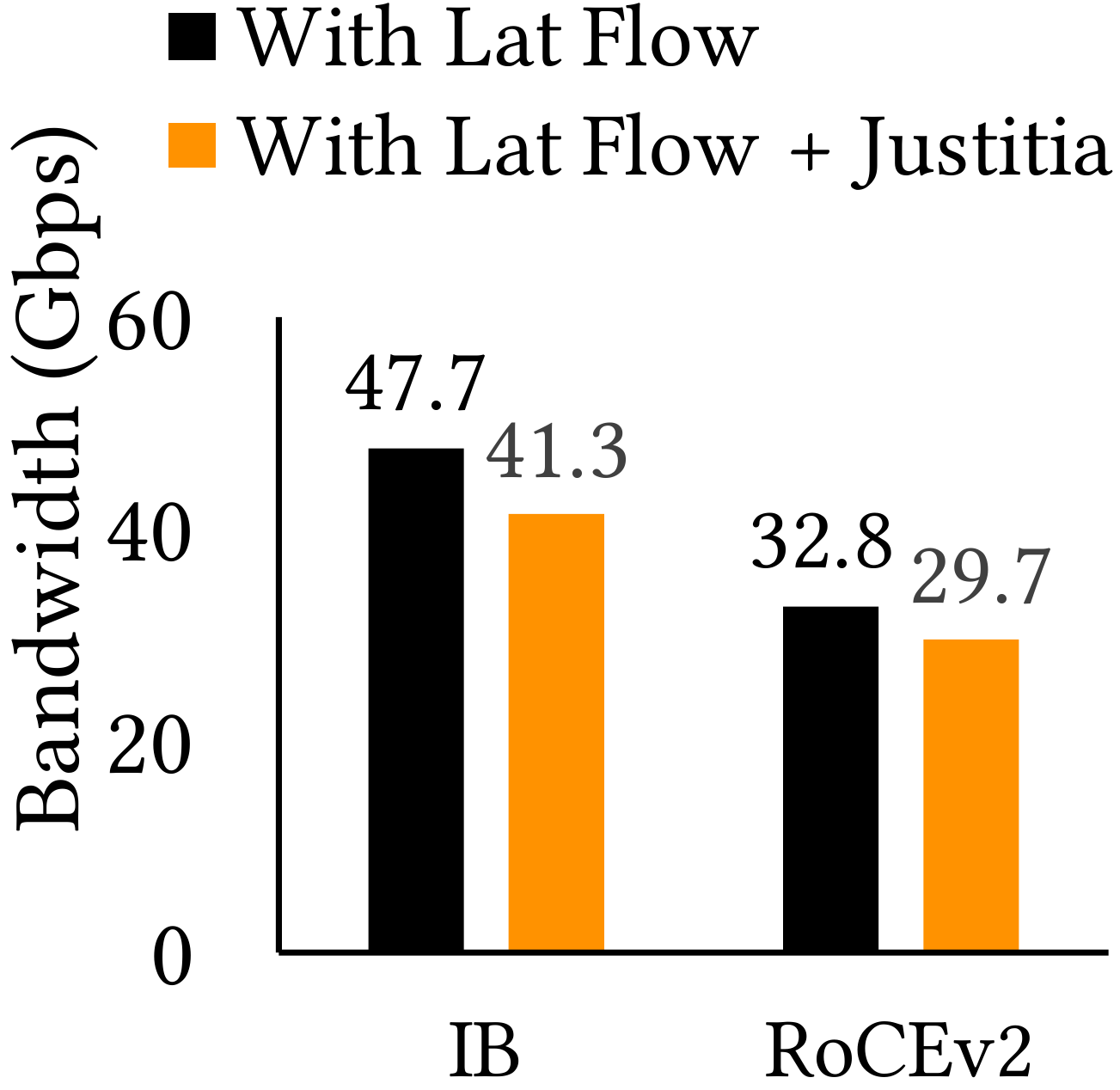}%
		}	
	\caption{Latency bound of a latency-sensitive flow running against a 1MB background bandwidth-sensitive flow using 4 QPs. Latency target is removed to maximize bandwidth allocation.}
	\label{fig:sec7-E-vs-lat-large-target}%
\end{figure}

\section{Evaluation}
\label{sec:eval}
In this section, we evaluate {\name}'s effectiveness in providing performance isolation between latency-, throughput-, and bandwidth-sensitive flows on InfiniBand and RoCEv2.

Our key findings can be summarized as follows:
\begin{denseitemize}
  \item {\name} can effectively mitigate RDMA performance isolation anomalies highlighted in Section~\ref{sec:anomalies} at both flow (\S\ref{sec:micro-eval}) and application levels (\S\ref{sec:apps-eval}).
  
  \item {\name} scales well to a large number of flows and works for a variety of settings (\S\ref{sec:deepdive-eval}); it complements DCQCN and hardware virtual lanes (\S\ref{sec:alternatives-eval}).
  
  \item {\name}'s benefits hold in long-running, dynamic scenarios with many latency- and bandwidth-sensitive flows (\S\ref{sec:long-eval}), in the presence of remote READs (\S\ref{sec:read-eval}), and in incast scenarios (\S\ref{sec:incast-eval}).

\end{denseitemize}

Unless specified, we do not use hardware virtual lanes.

To measure latency, we perform 5 consecutive runs and present their median.
Most of our results are very stable; we do not show error bars when they are too close to the median.

\textbf{Ethics.} This work does not raise any ethical issues.


\subsection{Preventing Isolation Anomalies}
\label{sec:micro-eval}
We start by revisiting the scenarios from Section~\ref{sec:anomalies} to understand how {\name} isolates different types of RDMA flows. 

\textbf{Experimental Setup.}
We use the same setups as those described in Section~\ref{sec:anomalies}, and unless otherwise specified, we set \targettail=2 microseconds on both InfiniBand and RoCEv2 for the latency-sensitive flows.
{\name} works well in 100 Gbps networks too (Appendix~\ref{app:100gbps}).
Unless otherwise specified, sharing incentive is strictly enforced.


\begin{figure}[!t]
	\centering
		\subfloat[][Different message sizes]{%
			\label{fig:sec7-E-vs-E}%
			\includegraphics[width=1.65in]{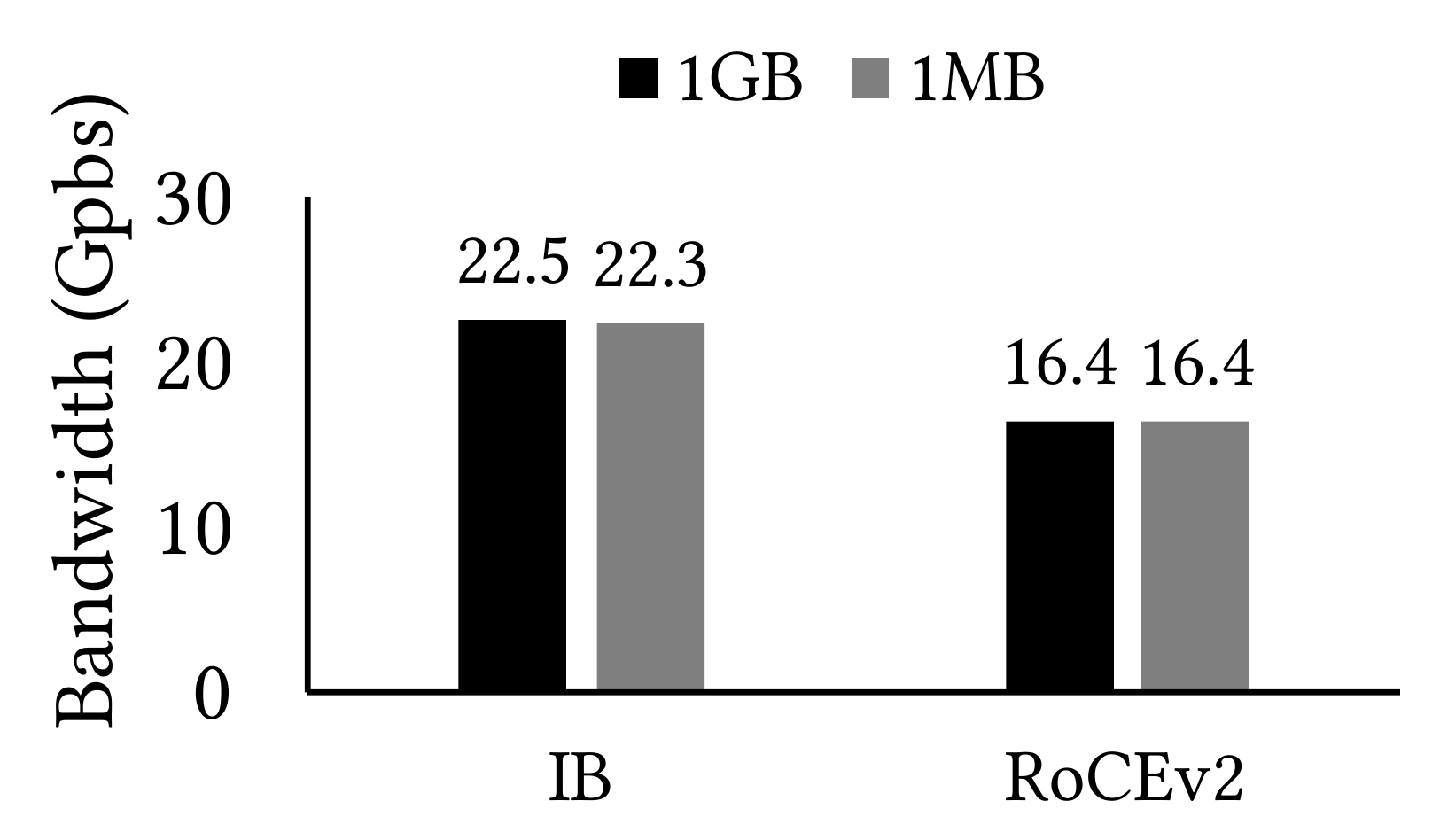}%
		}
		\hfill
		\subfloat[][Single-QP vs. 16-QP]{%
			\label{fig:sec7-multiE-vs-E}%
			\includegraphics[width=1.65in]{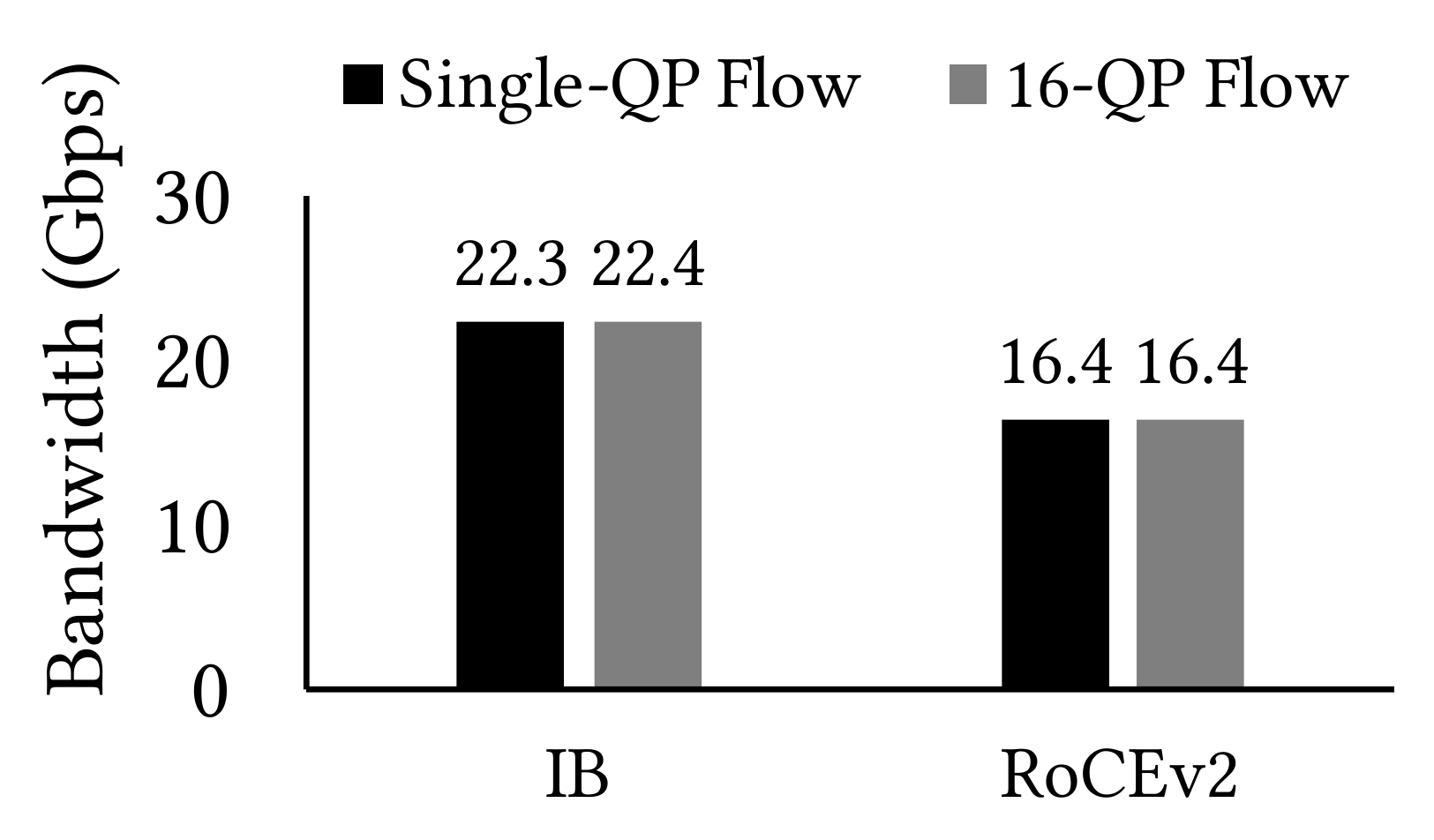}%
		}	
	\caption{Fair bandwidth allocations of bandwidth-sensitive flows. (a) Two flows with different message sizes. (b) Single-QP vs. 16-QP 1MB flows in InfiniBand.}%
	\label{fig:sec7-bw-anomalies}%
\end{figure}

\begin{figure}[!t]
	\centering
		\subfloat[][Throughput flow]{%
			\label{fig:EVAL-tput-vs-E-tput}%
			\includegraphics[width=1.65in]{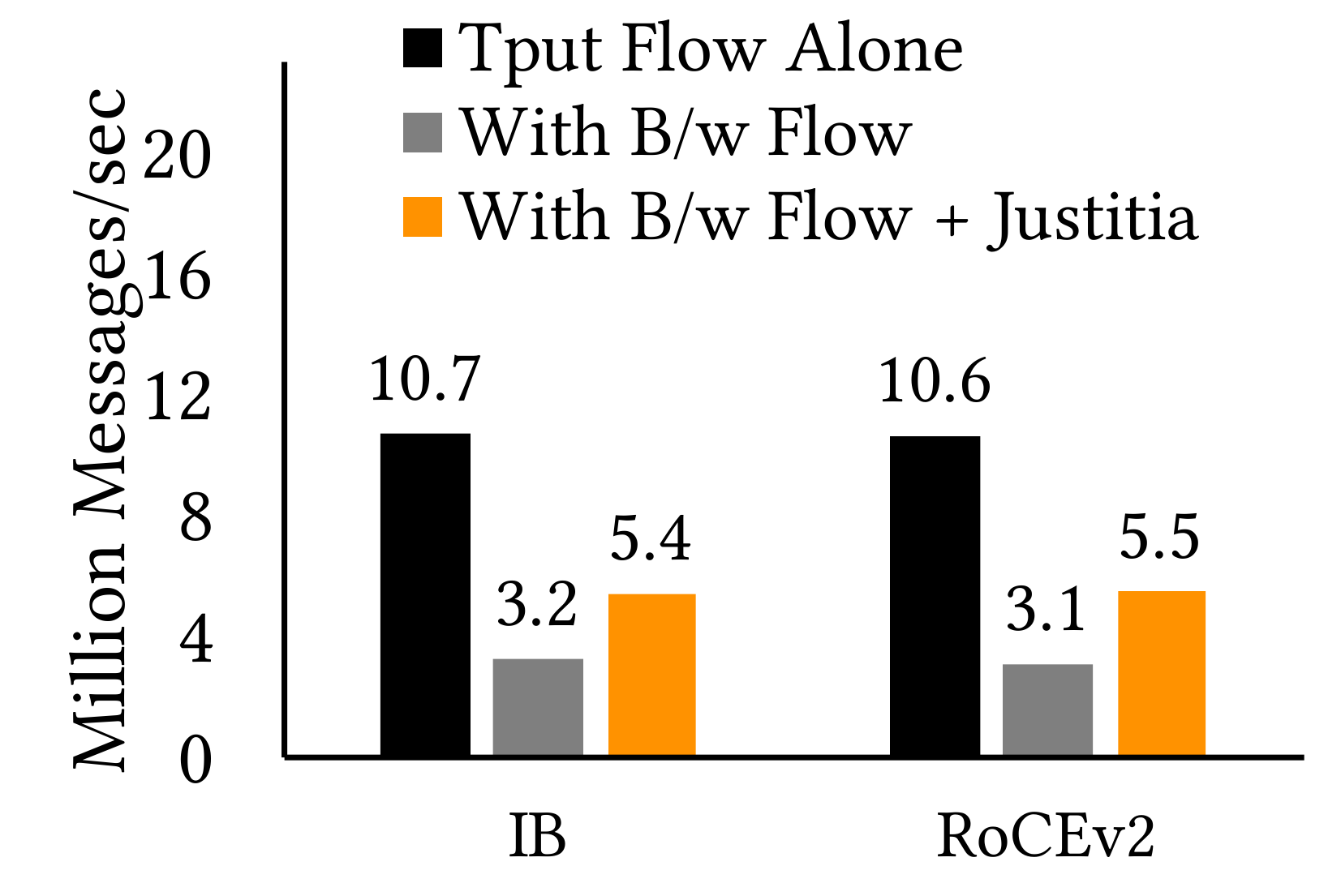}%
		}
		\hfill
		\subfloat[][{Bandwidth flow}]{%
			\label{fig:EVAL-tput-vs-E-bw}%
			\includegraphics[width=1.65in]{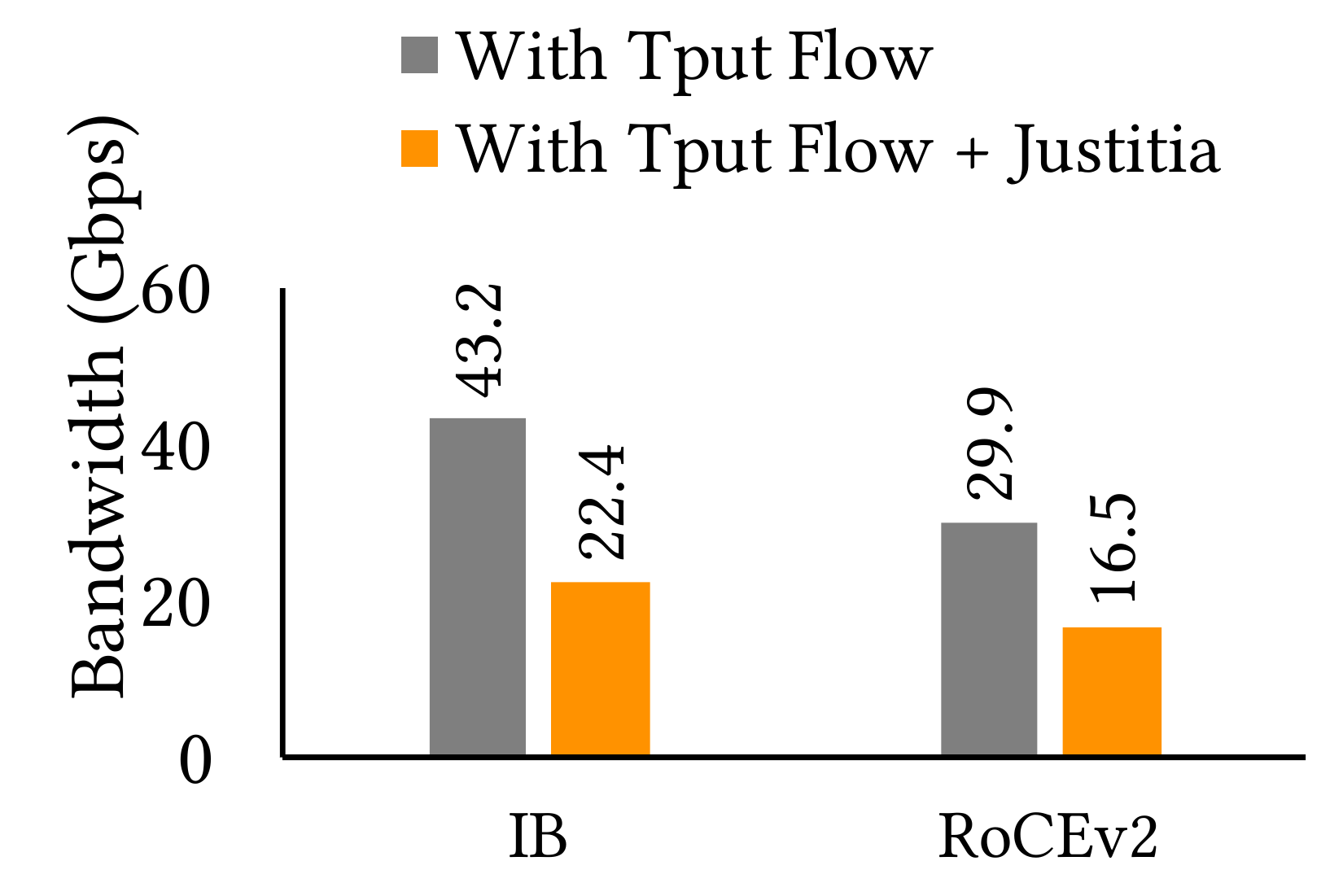}%
		}	
	\caption{Performance isolation of a throughput-sensitive flow running against a 1MB background bandwidth-sensitive flow.}
	\label{fig:sec7-E-tput-vs-E}%
\end{figure}

\subsubsection{Predictable Latency}

\paragraph{Maximizing Isolation.}
Recall that latency-sensitive flows are affected the most when they compete with a bandwidth-sensitive flow.
In the presence of {\name}, both median and tail latencies improve significantly in both InfiniBand and RoCEv2 (Figure~\ref{fig:EVAL-E-vs-lat-latency}).
In this experiment, we set the latency target to the value when the latency-sensitive is running alone.
By sharing incentive requirement, the bandwidth-sensitive flow is limited to half of its original bandwidth (Figure~\ref{fig:EVAL-E-vs-lat-bw}).
In other words, Figure~\ref{fig:EVAL-E-vs-lat-latency} shows the best latency isolation while maintaining sharing incentive.
Because {\name} treats all bandwidth-sensitive flows from the same application as one and distribute tokens among them in a round-robin fashion, introducing more flows will not affect isolation.

\textbf{Maximizing Work Conservation.}
Next we evaluate how {\name} performs when the latency target is set to a large value (\targettail=10 microseconds) that can always be met.
{\name} keeps increasing \safeutil toward the line rate until the target is violated.
Figure~\ref{fig:sec7-E-vs-lat-large-target} illustrates the latency bound that can be achieved in such case.

For a slightly high \targettail, {\name} can provide bounded latency for applications sharing the same RNIC without compromising high bandwidth allocation.
Note that as long as all applications go through {\name}, bandwidth-sensitive applications are all paced by {\name} with aggregate bandwidth set to line rate.
Thus latency numbers in Figure~\ref{fig:sec7-E-vs-lat-large-target} will not change regardless of the number of bandwidth-sensitive applications.

\subsubsection{Fair Bandwidth and Throughput Sharing}
{\name} ensures that bandwidth-sensitive flows receive equal shares regardless of their message sizes (Figure~\ref{fig:sec7-bw-anomalies}).
To achieve fair sharing, {\name} introduces small bandwidth overhead (less than 6\% on InfiniBand and 2\% on RoCEv2). 

{\name}'s benefits extends to the bandwidth- vs through-sensitive flow scenario as well.
In this case, it ensures that both receive roughly half of their resources. 
Figure~\ref{fig:sec7-E-tput-vs-E} illustrates this behavior.
In both InfiniBand and RoCEv2, the throughput-sensitive flow is able to achieve half of its original message rate of itself running alone (Figure~\ref{fig:EVAL-tput-vs-E-tput}).
The bandwidth-sensitive flow, on the other hand, is limited to half its original bandwidth as expected (Figure~\ref{fig:EVAL-tput-vs-E-bw}).

\begin{figure}[!t]
	\centering
		\subfloat[][Latency-sensitive Flow]{%
			\label{fig:EVAL-TvL-median}%
			\includegraphics[width=2.2in]{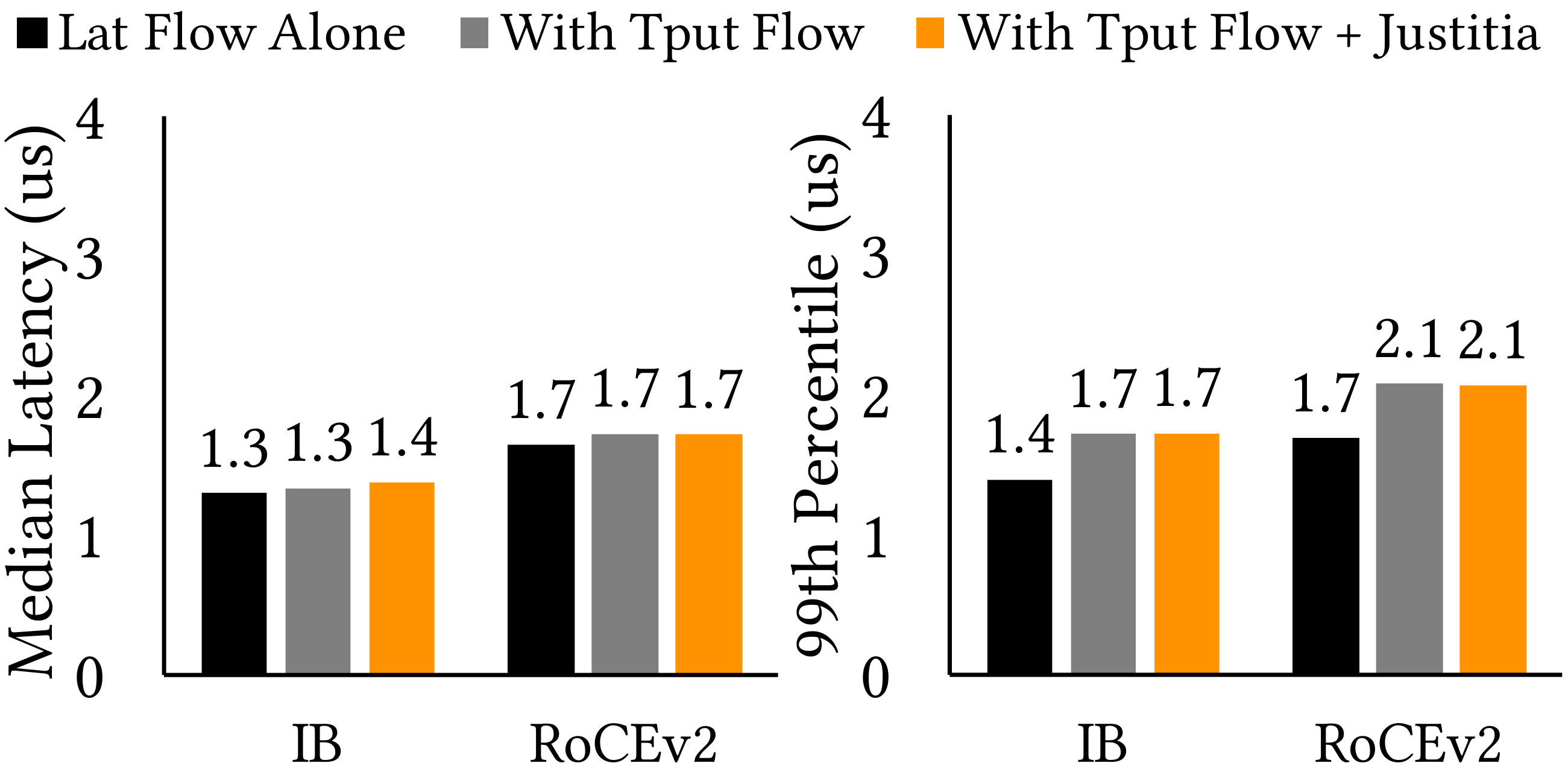}
		}
		\hfill
		\subfloat[][{Throughput Flow}]{%
			\label{fig:EVAL-TvL-ops}%
			\includegraphics[width=1.1in]{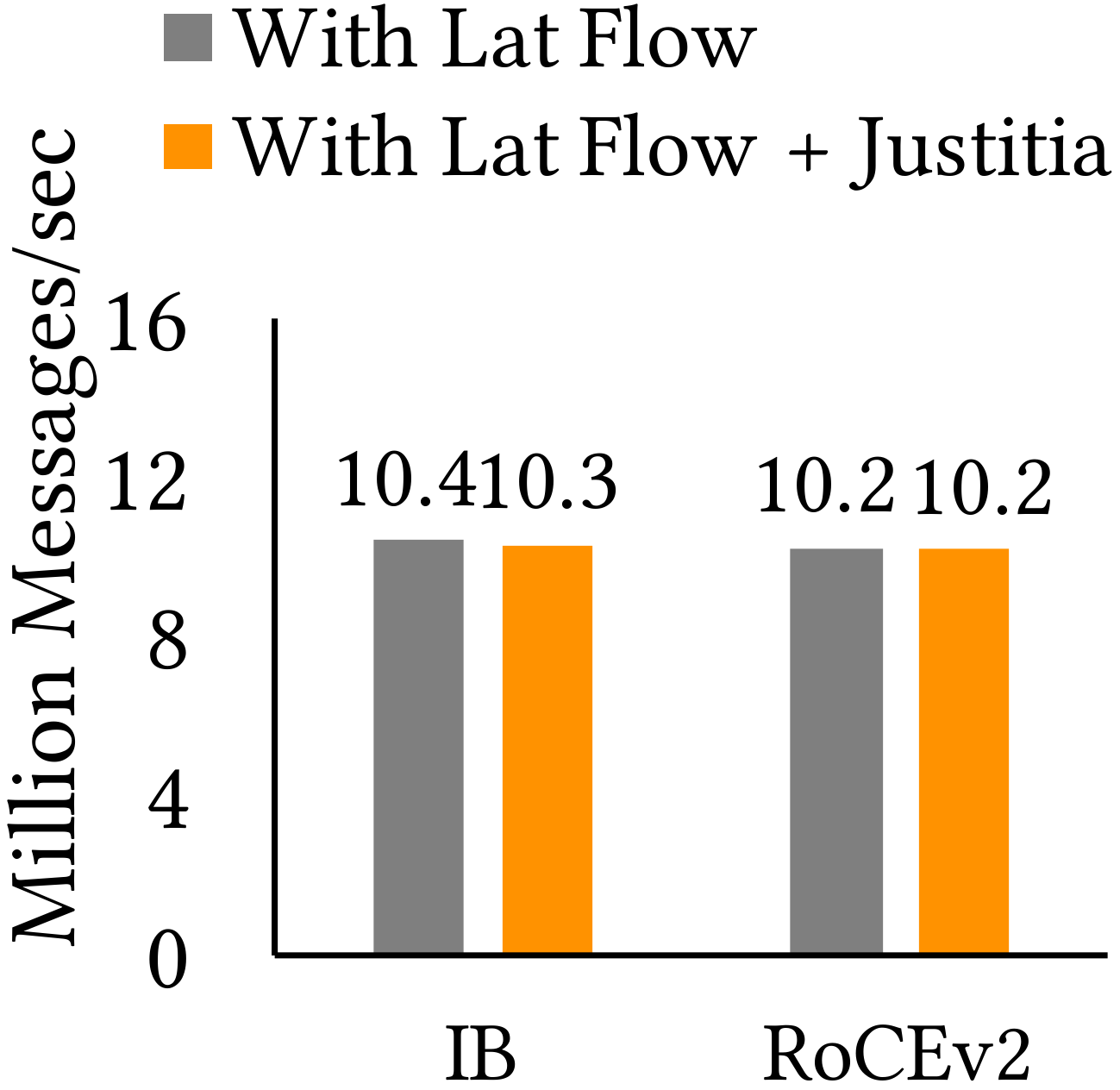}%
		}	
	\caption{Performance isolation of a latency-sensitive flow running against a background throughput-sensitive flow.}
	\label{fig:sec7-tput-vs-lat}%
\end{figure}

\subsubsection{Throughput- vs. Latency-Sensitive Flow}
We observed in Section~\ref{sec:anomalies} that latency- and throughput-sensitive flows do not significantly affect each other.
Adding {\name} into the mix does not change anything (Figure~\ref{fig:sec7-tput-vs-lat}).

\begin{figure}[!t]
  \centering
  \includegraphics[width=2.2in]{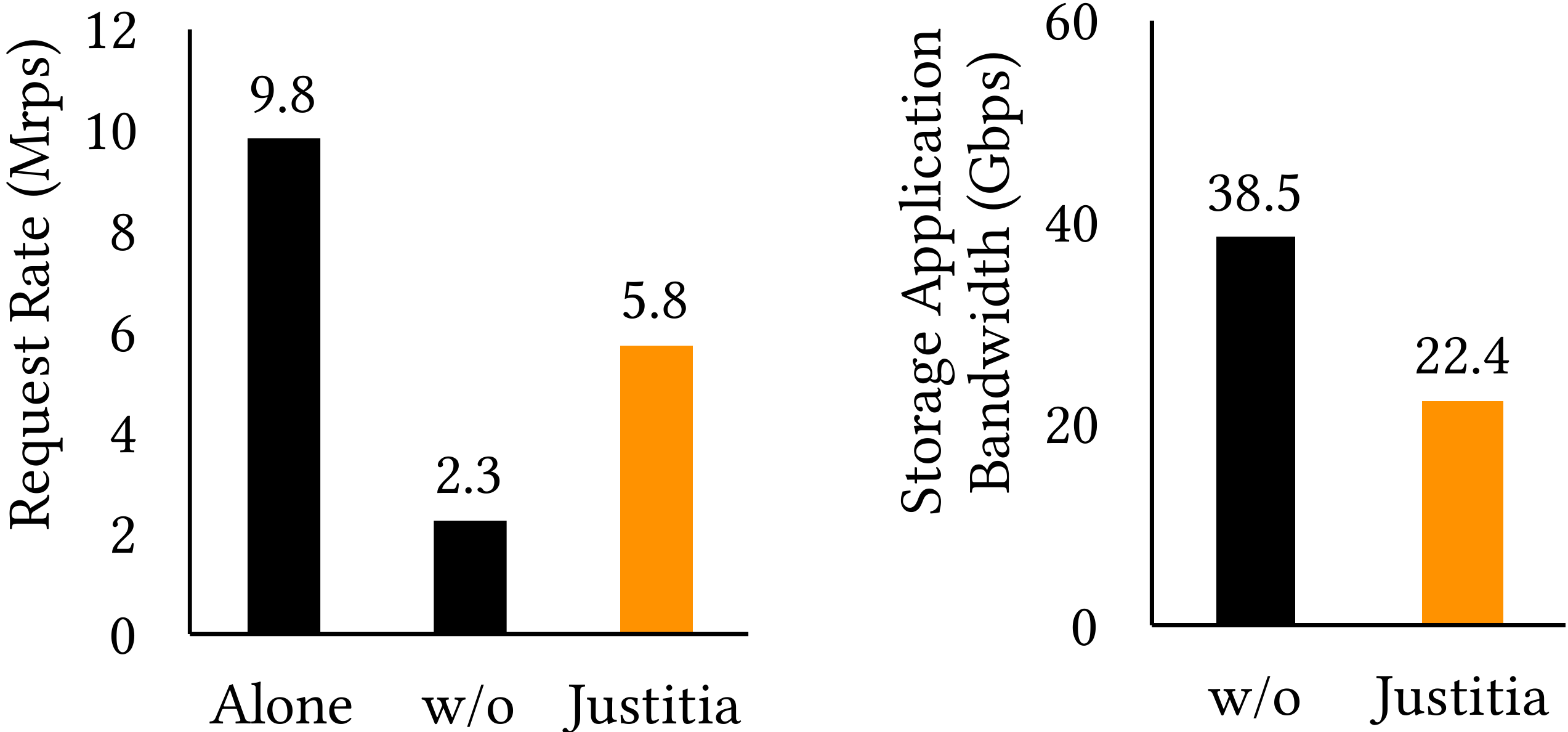}
  \caption{Performance isolation of FaSST running against a bandwidth-sensitive storage application.}
  \label{fig:EVAL-fasst}%
\end{figure}

\begin{figure}[!t]
	\centering
		\subfloat[][Latency]{%
			\label{fig:EVAL-erpc-lat}%
			\includegraphics[width=1.65in]{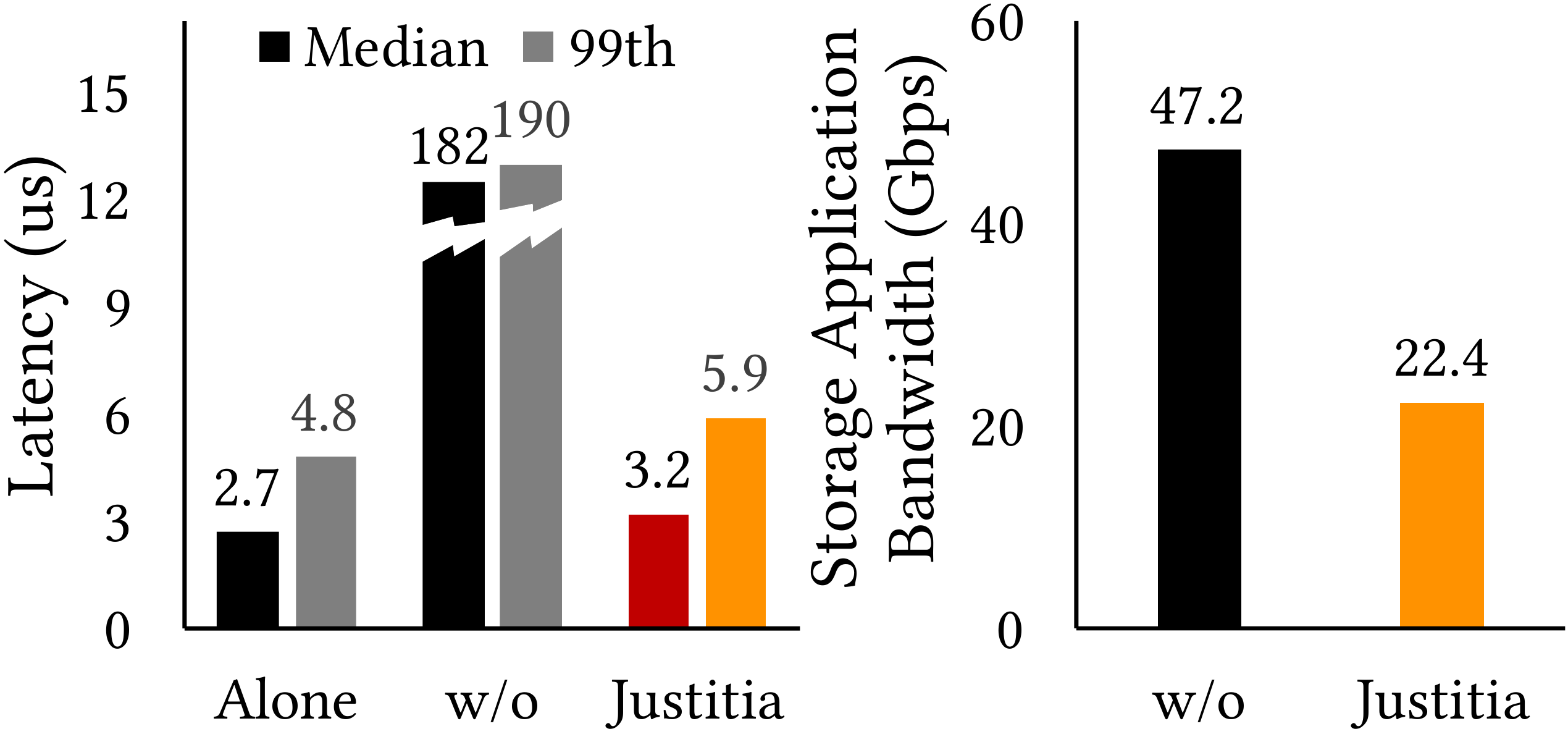}%
		}
		\hfill
		\subfloat[][Throughput]{%
			\label{fig:EVAL-erpc-tput}%
			\includegraphics[width=1.65in]{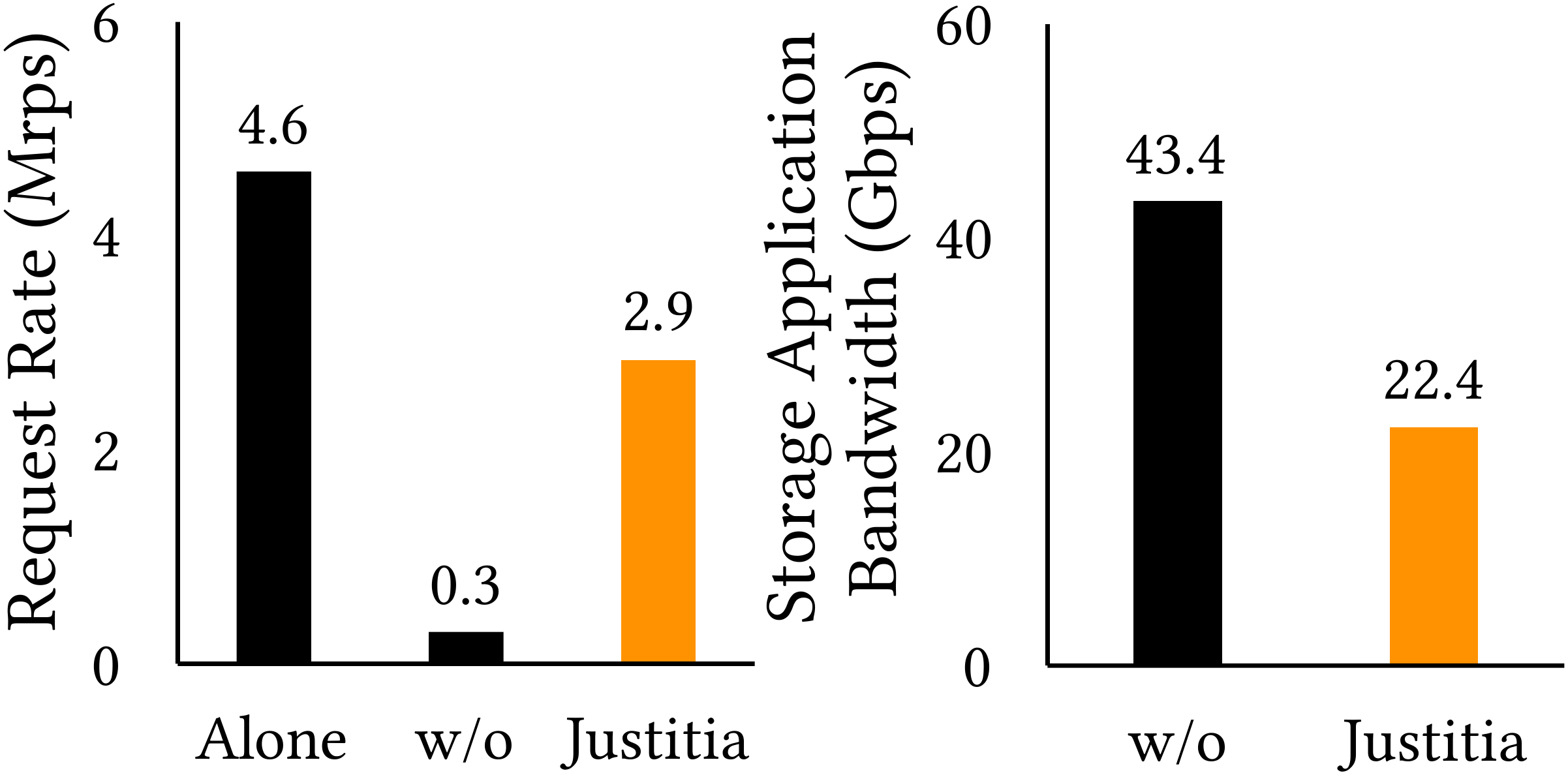}%
		}	
  \caption{Performance isolation of eRPC running against a bandwidth-sensitive storage application.}
	\label{fig:sec7-erpc-evel}%
\end{figure}

\subsection{{\name} and RDMA Applications}
\label{sec:apps-eval}
We now shift our attention to real applications (\S\ref{sec3:iso-in-apps}) and evaluate {\name}'s effectiveness at the  application level.
%
We observe that {\name} achieves better RNIC resource sharing when FaSST and the bandwidth-sensitive storage application coexist -- FaSST's throughput improves by $2.5\times$ with a $1.7\times$ decrease in storage application's bandwidth (Figure~\ref{fig:EVAL-fasst}).
{\name} also improves eRPC's median (tail) latency by $56.9\times$ ($32.2\times$) and its throughput by $9.7\times$ while still maintaining sharing incentive.

\begin{figure}[!t]
  \centering
    \subfloat[][Bandwidth-sensitive]{%
      \label{fig:bw-scale}%
      \includegraphics[scale=0.5]{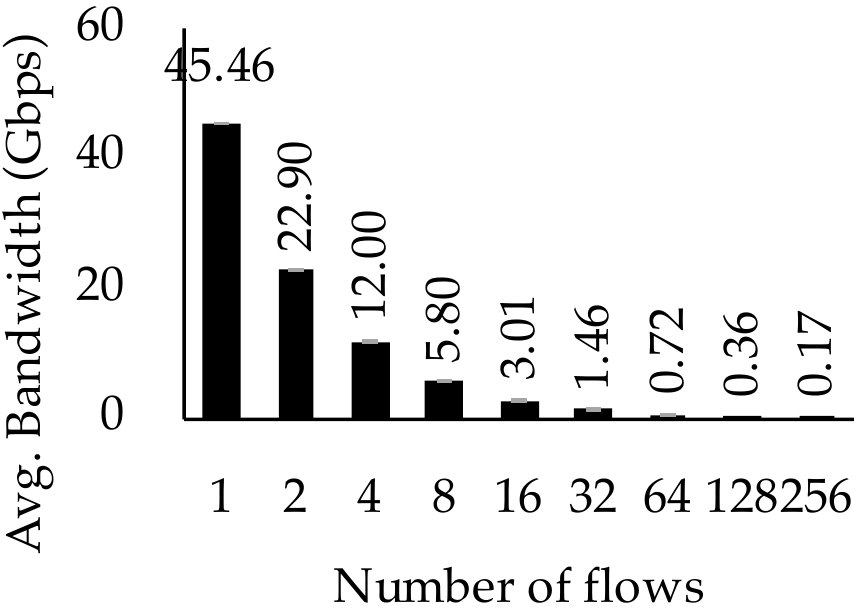}%
    }
    \hfill
    \subfloat[][Throughput-sensitive]{%
      \label{fig:tput-scale}%
      \includegraphics[scale=0.5]{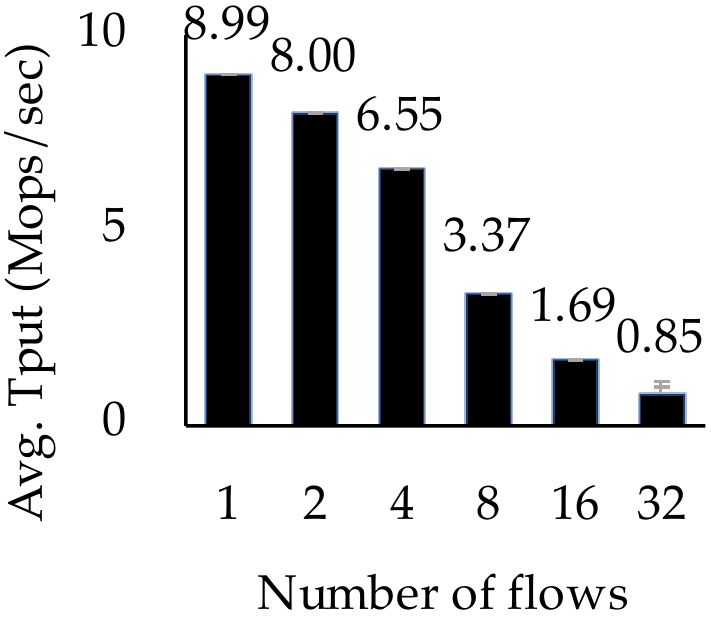}%
    }
  \caption{[InfiniBand] {\name} scales to a large number of flows and still provides equal share to all of them. 
    The error bars represent the minimum and the maximum values across all the flows.}%
  \label{fig:scalability}
\end{figure}

\subsection{{\name} Deep Dive}
\label{sec:deepdive-eval}

\subsubsection{Scalability and Rate Conformance}
Figure~\ref{fig:bw-scale} shows that as the number of bandwidth-sensitive flows increases, all flows receive the same amount of bandwidth using {\name}. 
The overall RNIC bandwidth utilization remains close to that of its maximum capacity. 

The same holds for throughput-sensitive flows (Figure~\ref{fig:tput-scale}), but with two caveats. 
First, a single throughput-sensitive flow cannot saturate the RNIC -- it takes four or more (refer to Figure~\ref{fig:sec3-multi-tput} in the Appendix). 
Hence, {\name} ensures that all throughput-sensitive flows send roughly equal number of messages. 
Second, throughput-sensitive flows are CPU-hungry because they drive a large number of messages. 

\subsubsection{CPU and Memory Consumption}
{\name} uses two dedicated CPU cores per machine:
one to generate and distribute tokens and the other for the reference latency-sensitive flow.
A detailed analysis on CPU overhead can be found in Appendix D.
Its memory footprint is not significant. 



%
%

\begin{figure}[!t]
  \centering
  \includegraphics[width=\columnwidth]{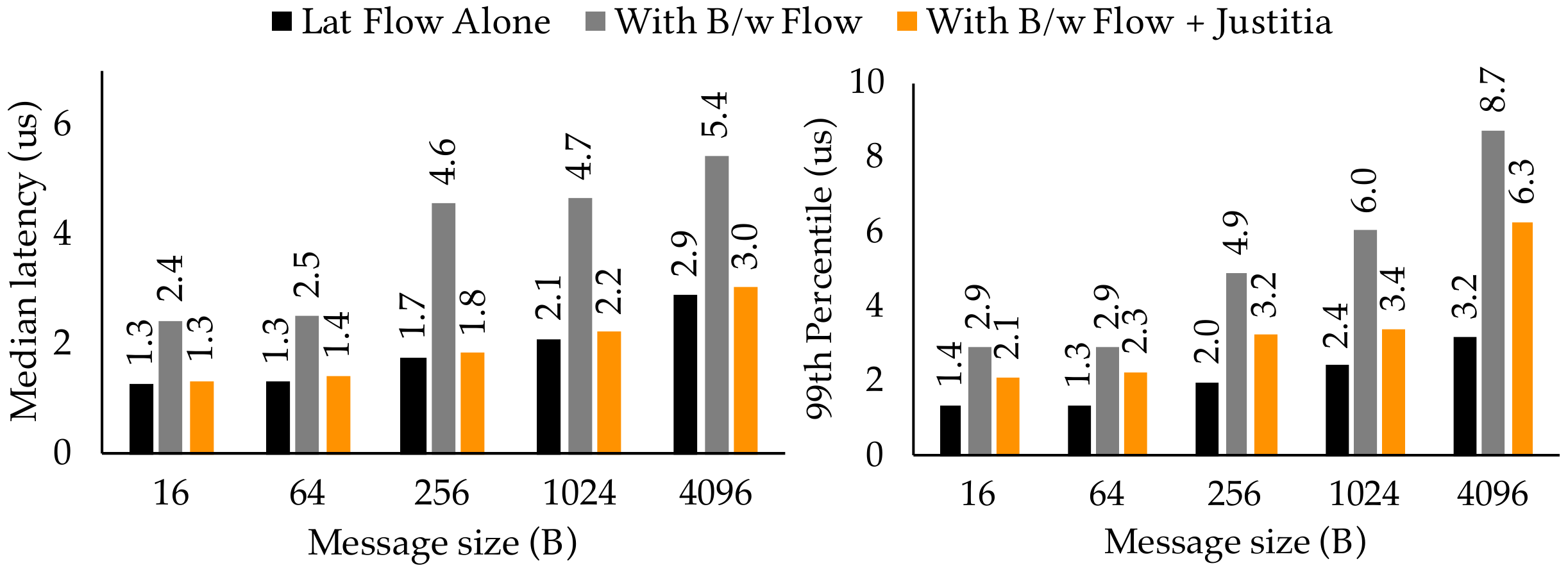}
  \caption{[InfiniBand] Latency-sensitive flows with different message sizes competing against a bandwidth-sensitive flow.}%
  \label{fig:diff-size-lat}%
\end{figure}

\subsubsection{Impact of Latency-Sensitive Flow's Message Size}
All our latency-sensitive experiments use small, 16B messages.
Here, we vary the message size and observe that {\name} can still meet the median latency of the flow running alone, and its tail performance is still limited due to the isolation-utilization tradeoff (Figure~\ref{fig:diff-size-lat}).
The bandwidth-sensitive flow receives half the bandwidth in all cases.

\begin{figure*}[t]
  \mbox{
    \begin{minipage}{0.33\textwidth}
        \centering
    		\subfloat[][Latency-sensitive Flow]{%
    			\label{fig:EVAL-dcqcn-EvL-lat}%
    			\includegraphics[width=1.3in]{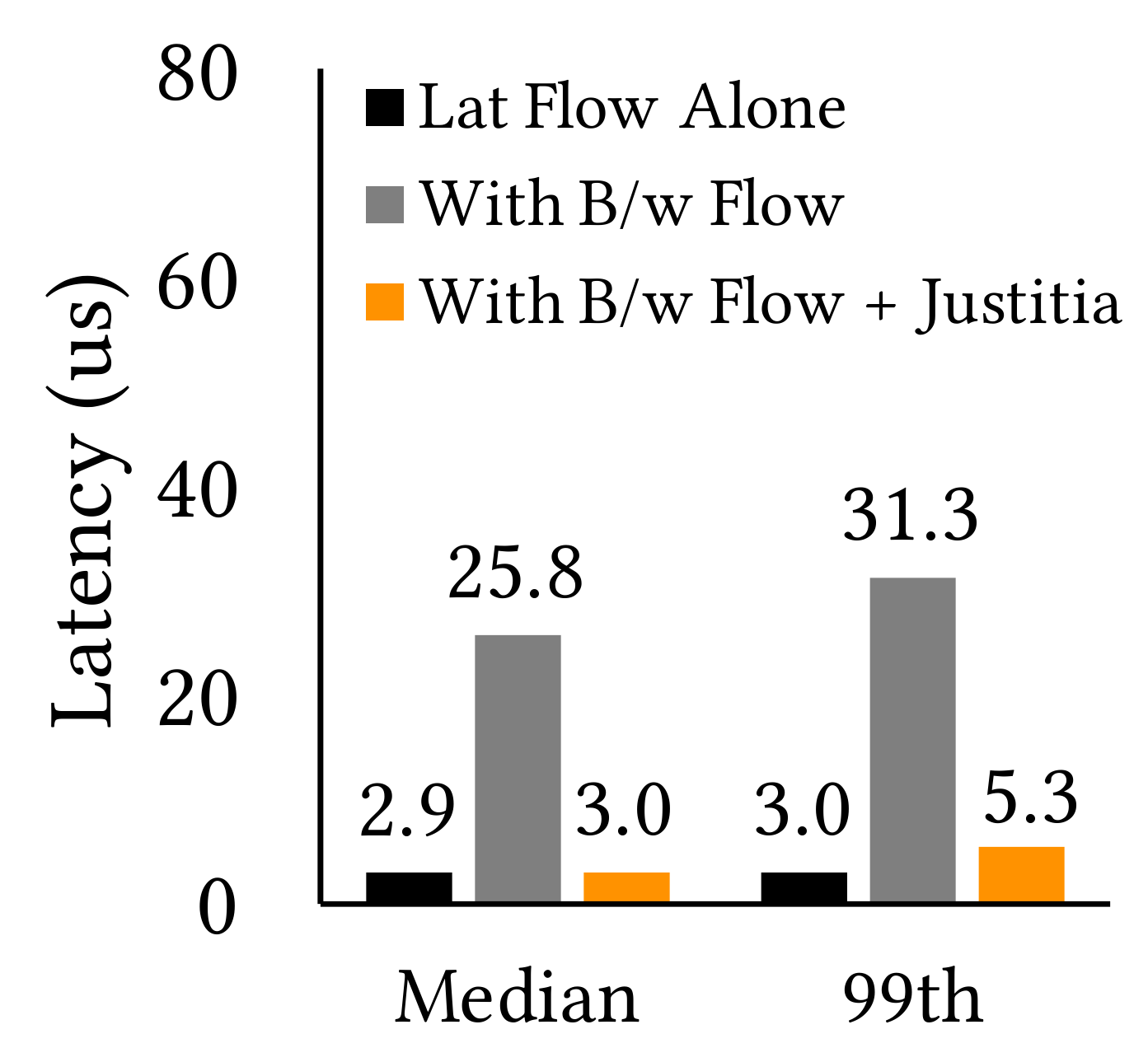}%
    		}
    		\hfill
    		\subfloat[][{Bandwidth Flow}]{%
    			\label{fig:EVAL-dcqcn-EvL-bw}%
    			\includegraphics[width=0.9in]{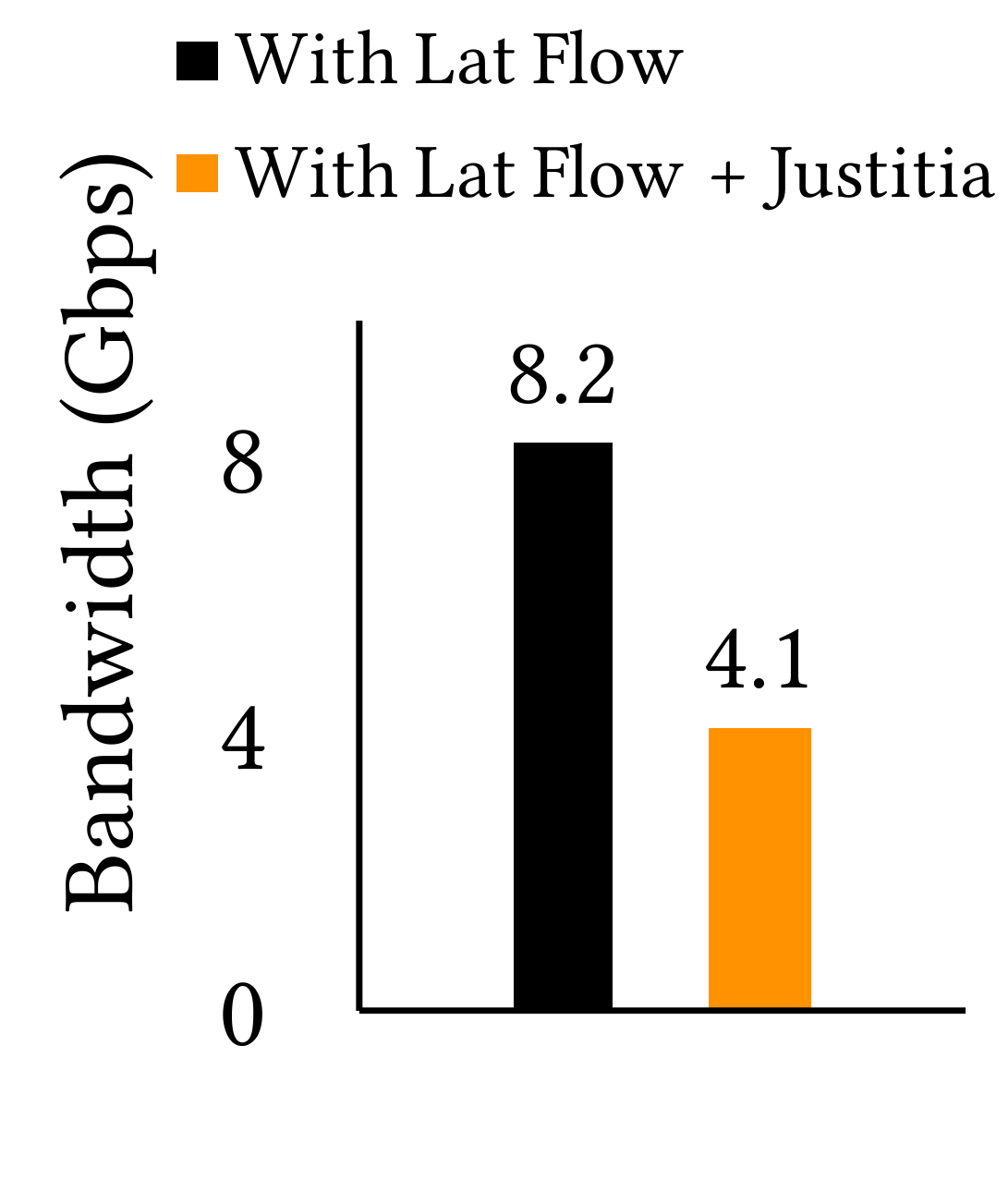}%
    		}	
    	\caption{[DCQCN] Latency-sensitive flow against a bandwidth-sensitive flow.}
    	\label{fig:dcqcn-bw-vs-lat}%
    \end{minipage}
    \hfill
    \begin{minipage}{0.33\textwidth}
          \centering
      		\subfloat[][Tput Flow]{%
      			\label{fig:EVAL-dcqcn-EvT-lat}%
      			\includegraphics[width=1.3in]{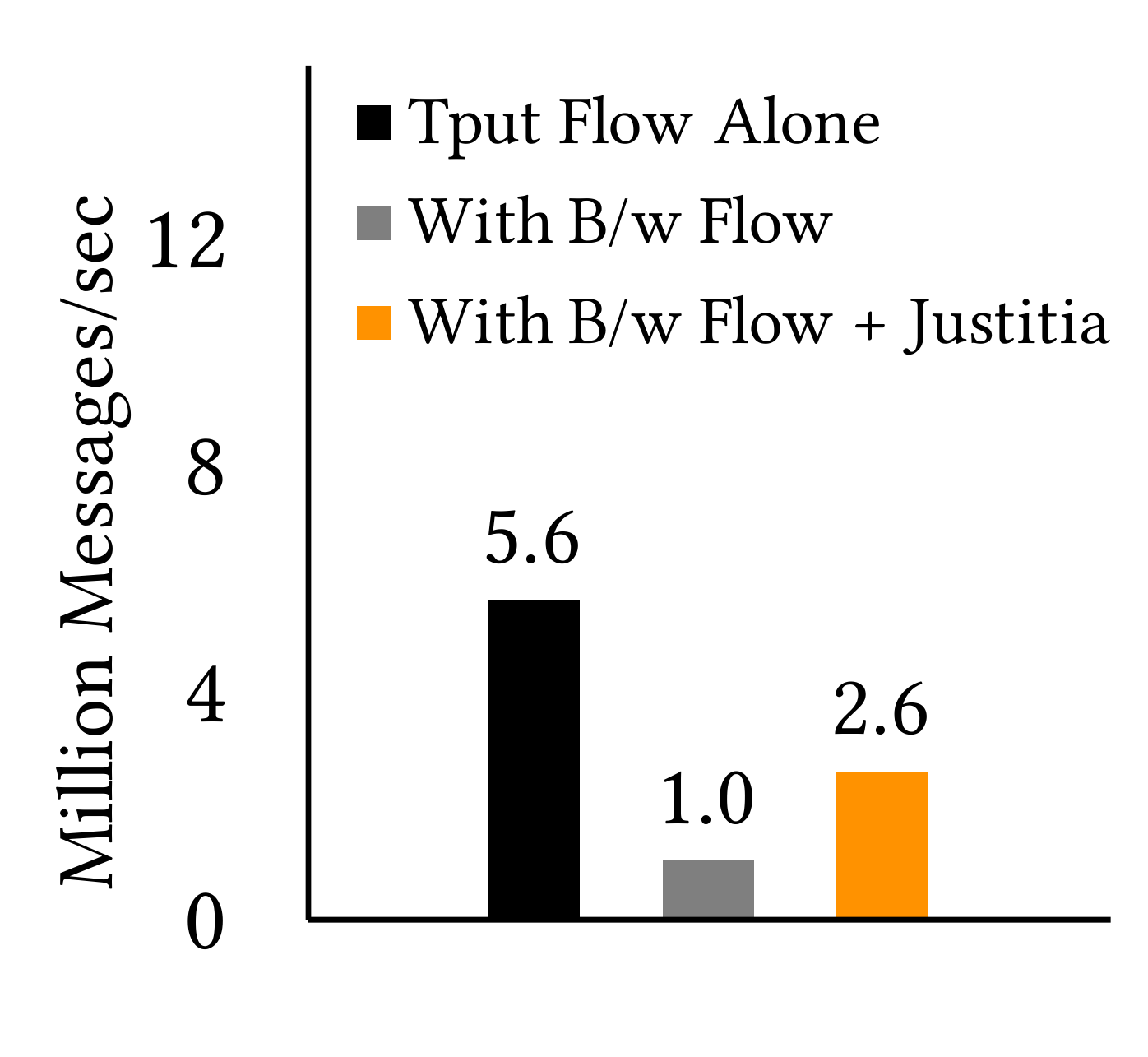}%
      		}
          \hfill
      		\subfloat[][{Bandwidth Flow}]{%
      			\label{fig:EVAL-dcqcn-EvT-bw}%
      			\includegraphics[width=0.9in]{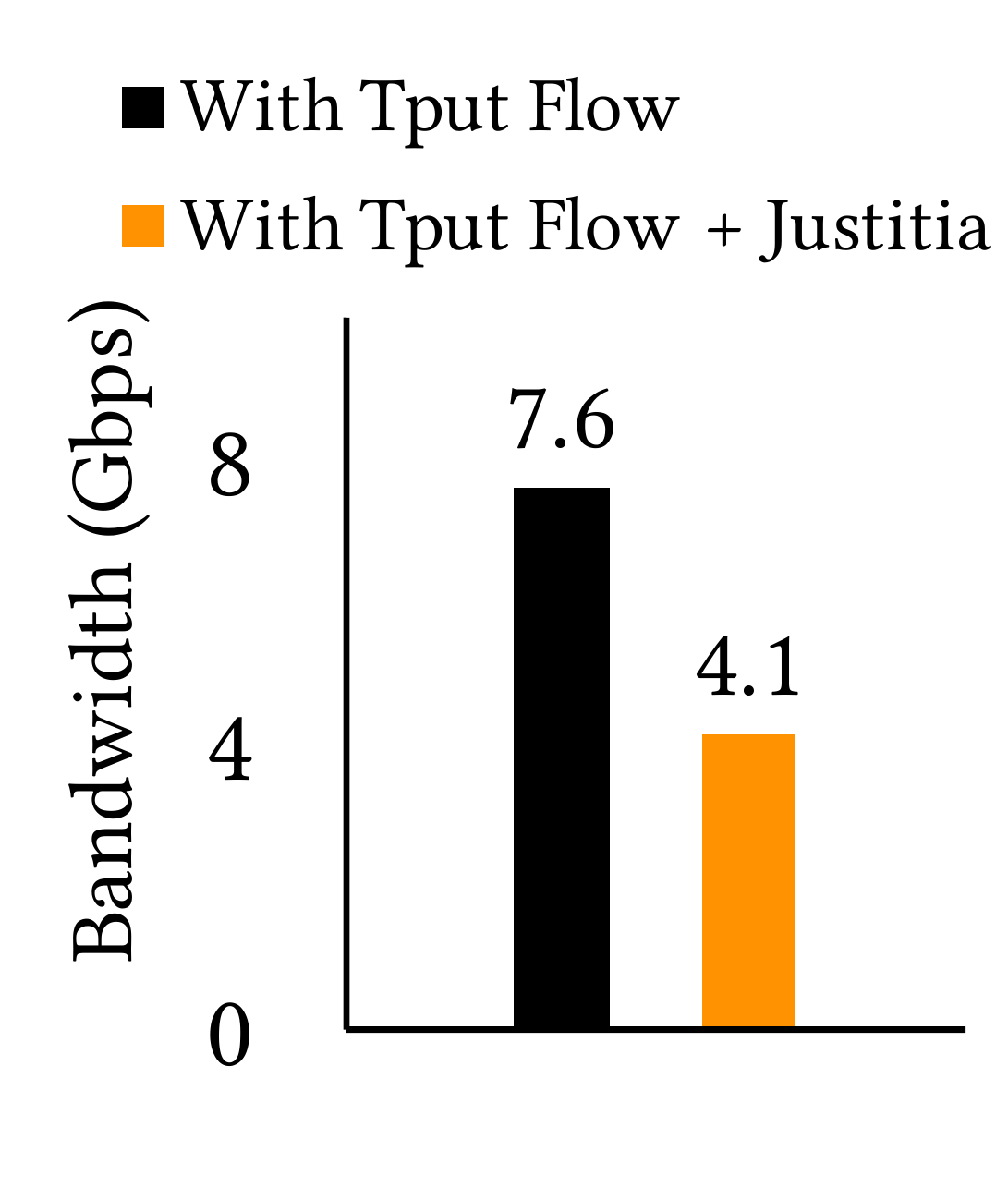}%
      		}	
      	\caption{[DCQCN] Throughput-sensitive flow against a bandwidth-sensitive flow.}
      	\label{fig:dcqcn-bw-vs-tput}%
    \end{minipage}
    \hfill
    \begin{minipage}{0.33\textwidth}
        \centering
    		\subfloat[][Latency-sensitive Flow]{%
    			\label{fig:EVAL-dcqcn-TvL-lat}%
    			\includegraphics[width=1.3in]{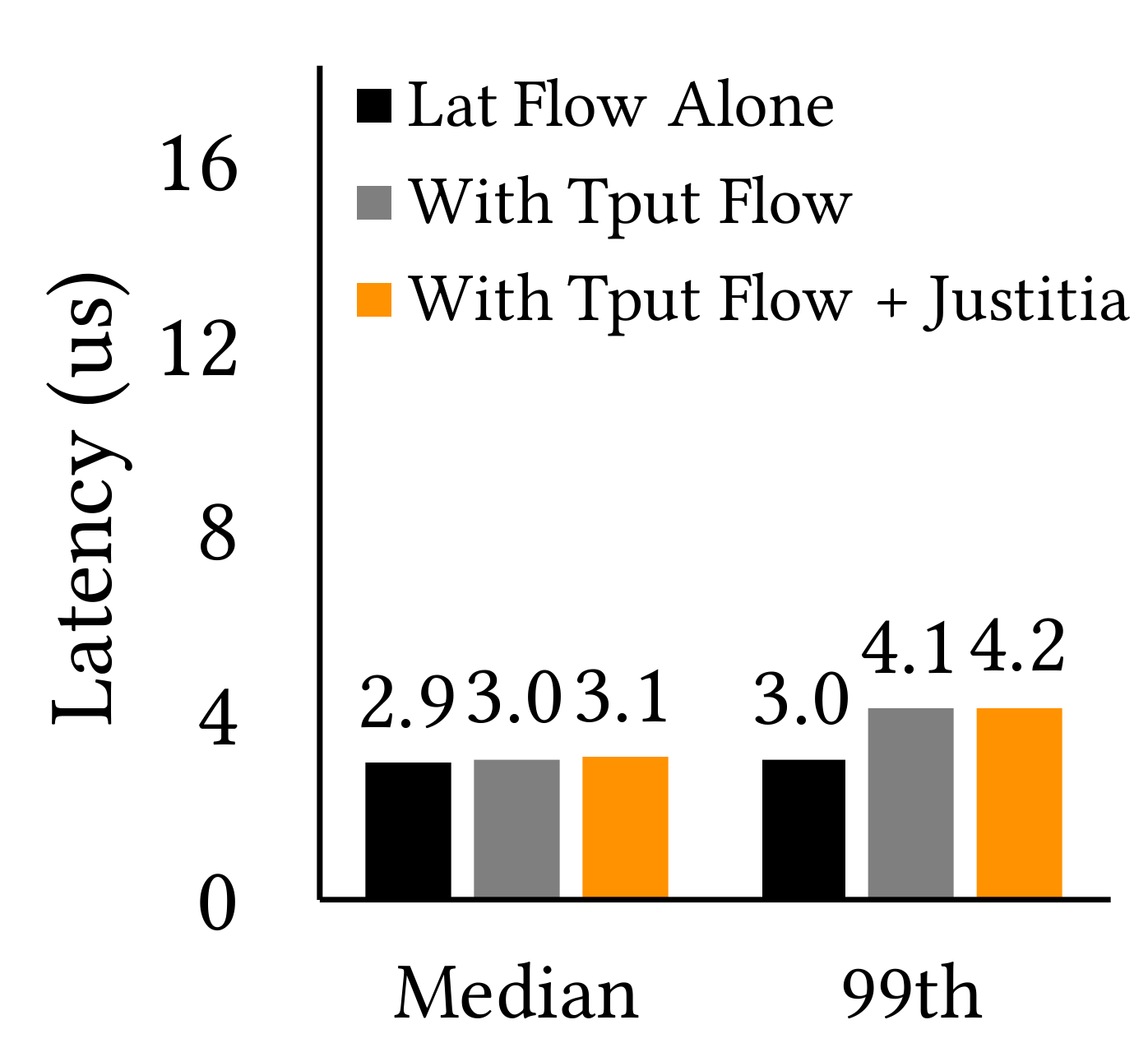}%
    		}
    		\hfill
    		\subfloat[][{Tput Flow}]{%
    			\label{fig:EVAL-dcqcn-TvL-tputu}%
    			\includegraphics[width=0.9in]{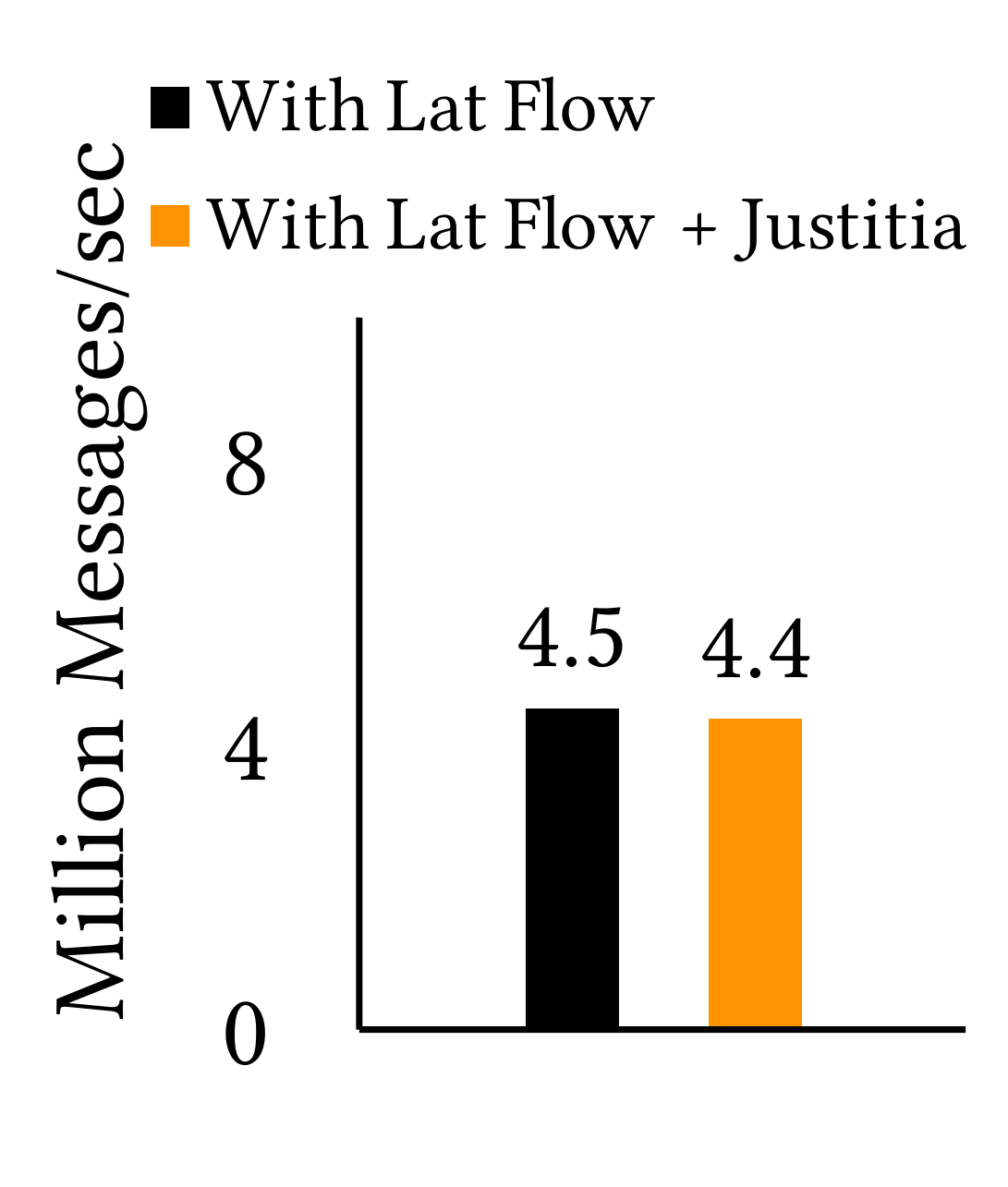}%
    		}	
    	\caption{[DCQCN] Latency-sensitive flow against a throughput-sensitive flow.}
      \label{fig:dcqcn-tput-vs-lat}%
    \end{minipage}
  }
\end{figure*}

\subsection{{\name} and Alternatives}
\label{sec:alternatives-eval}

\subsubsection{{\name} + DCQCN}
\label{sec:dcqcn-exp}
As discussed earlier (\S\ref{sec3:anomalies-source}), the anomalies we discover in this paper does not stem from the network congestion, but rather happens at the end hosts.
To further confirm our hypothesis, we deployed DCQCN (\S\ref{sec3:cc-no-good}) and found that it indeed falls short for latency- and throughput-sensitive flows (Figures~\ref{fig:dcqcn-bw-vs-lat},~\ref{fig:dcqcn-bw-vs-tput},~\ref{fig:dcqcn-tput-vs-lat}). 
{\name} mitigates them and complements DCQCN by improving latencies by up to $8.6\times$ and throughput by $2.6\times$.

\begin{figure}[!t]
  \centering
  \includegraphics[width=3.3in]{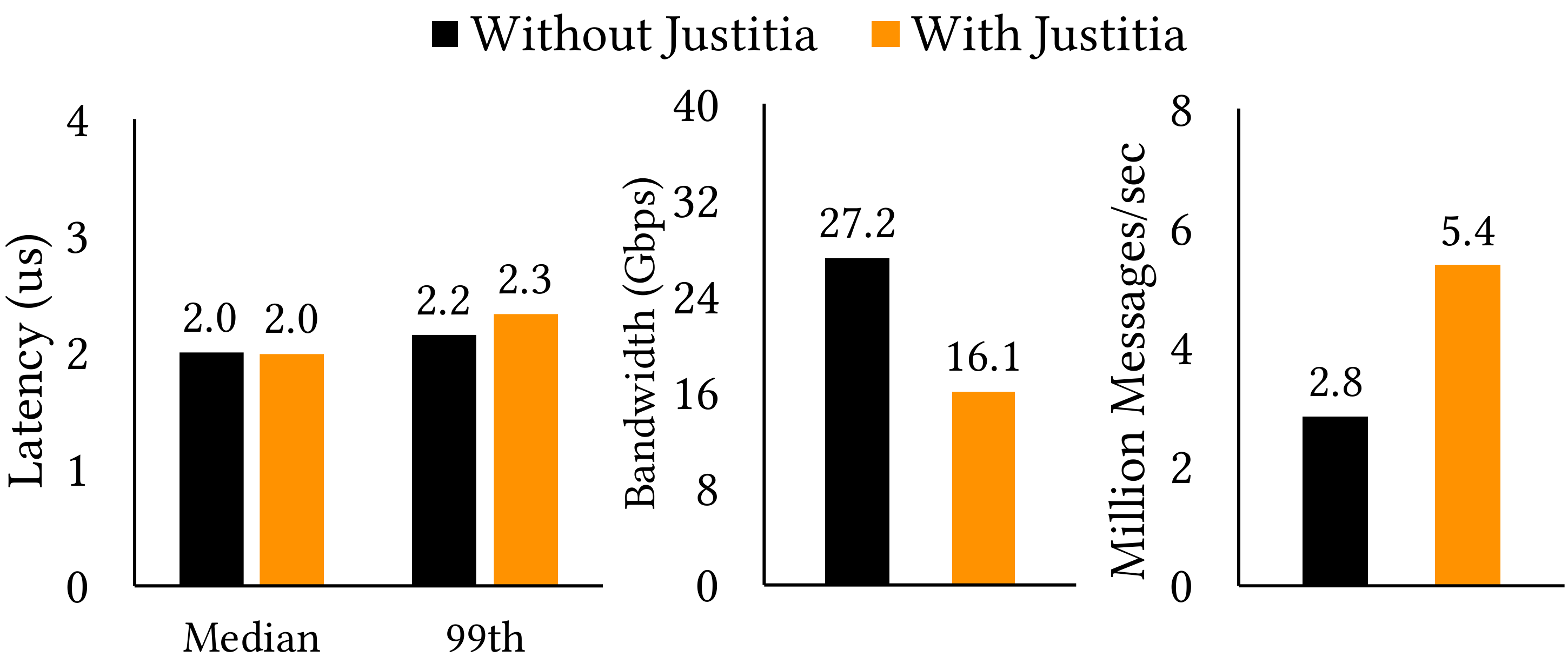}
  \caption{[RoCEv2] A bandwidth-, throughput-, and latency-sensitive flow running on two hardware priority queues at the switch.
  The latency-sensitive flow uses one queue, while the other two share the other queue.}%
  \label{fig:hw-prio}%
\end{figure}

\subsubsection{{\name} + Hardware Virtual Lanes}
\label{sec:hw-exp}
Hardware virtual lanes are limited in number \cite{virtualized-shapers-hotcloud, pfabric, mellanox-ib-switches}; \eg, our Ethernet switches support only two lossless traffic classes. 
In this experiment, we run three flows, one each for each of the three types (Figure~\ref{fig:hw-prio}).
Although the latency-sensitive flow remains isolated in its own class, the bandwidth- and throughput-sensitive flows compete in the same class.
As a result, the latter observes throughput loss (similar to Figure~\ref{fig:sec7-E-tput-vs-E}).
{\name} can effectively provide performance isolation between bandwidth- and throughput-sensitive flows in the shared queue.

\subsubsection{{\name} vs. LITE}
{\name} significantly outperforms LITE \cite{lite}, a software-based RDMA implementation.
See Appendix~\ref{sec:lite-eval} for details.


\begin{figure}[!t]
    \centering
    \includegraphics[width=1.5in]{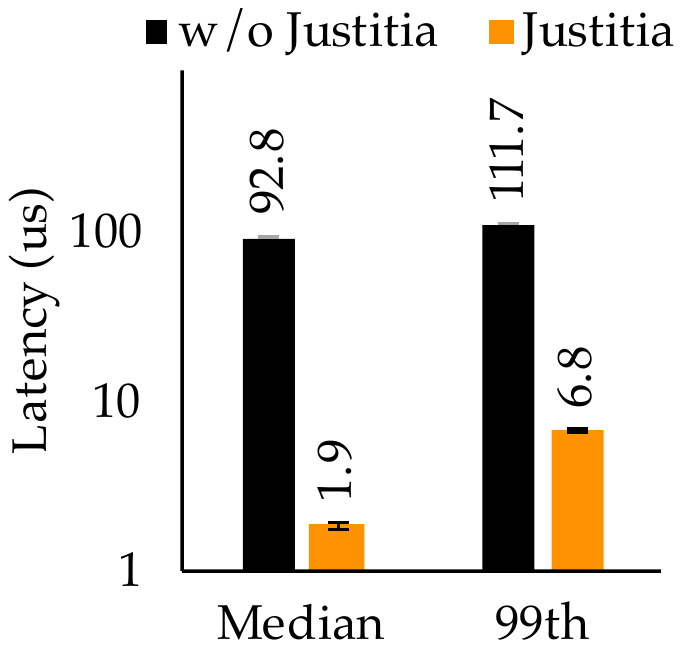}
    \caption{[InfiniBand] {\name} isolating 8 latency-sensitive flows from 8 bandwidth-sensitive ones.
    The error bars (almost invisible due to close proximity) represent the minimum and the maximum values across the 8 latency-sensitive flows.}%
    \label{fig:8e8mlats}%
  \end{figure}

\subsection{Dynamic, Long-Running Scenarios}
\label{sec:long-eval}
Here we extend our evaluation from microbenchmarks to two dynamic scenarios.
Both use \targettail = 2 microseconds.

\subsubsection{Sharing Incentive Enforcement}
First, we focus on {\name}'s effectiveness in isolating many flows with different requirements and performance characteristics. 
Specifically, we consider 8 long-running bandwidth-sensitive flows -- 2 each with message sizes: 1MB, 10MB, 100MB, and 1GB -- that arrive over time in pairs. 
When all of the bandwidth-sensitive flows are active, we start 8 latency-sensitive flows that run for a relatively short period of time (20 million samples) and finish.
Figure~\ref{fig:8e8mlats} shows the latency measurements.

In the absence of {\name}, latency-sensitive flows suffer large performance hits: individually each flow had median and 99th percentile latencies of 1.3 and 1.4 microseconds (Figures~\ref{fig:MOTI-EvL-median} and \ref{fig:MOTI-EvL-tail}).
With bandwidth-sensitive flows, they worsen by $71.4\times$ and $79.8\times$. 
{\name} improves median and tail latencies of latency-sensitive flows by $48.8\times$ and $16.4\times$ while guaranteeing sharing incentive among all the flows.




\subsubsection{When \targettail Is Unattainable}
\label{sec:new_algorithm-exp}
\begin{figure}[!t]
  \centering
  \includegraphics[scale=0.55]{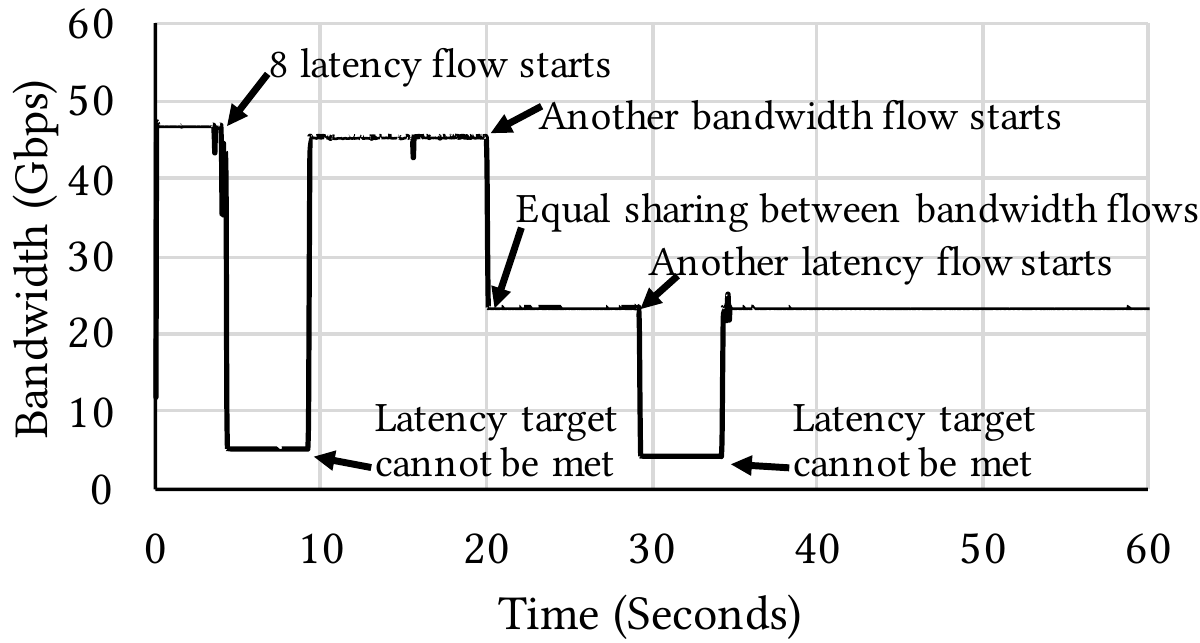}
  \caption{[InfiniBand] Bandwidth of a bandwidth-sensitive flow over time in dynamic setting. 
    {\name} can use up resources when the latency target is unattainable.}%
  \label{fig:new_algorithm-exp}
\end{figure}

In this experiment, we focus on {\name}'s dynamic adjustments to use up resources when \targettail cannot be achieved (Figure~\ref{fig:new_algorithm-exp}). 
{\name} first tries to ensure sharing incentive when the ratio of active latency-sensitive flows increases.
However, when it cannot meet the target for a long duration (in this case $\delta$=5 seconds), {\name} provides an option to opt for increasing utilization and equally shares bandwidth between the bandwidth-sensitive flows.
Note that the operator can choose the opposite as well.

\begin{figure}[!t]
  \centering
    \includegraphics[scale=0.5]{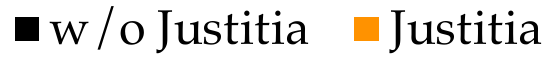}%
    \\
    \subfloat[][]{%
      \label{fig:lread-lat}%
      \includegraphics[scale=0.5]{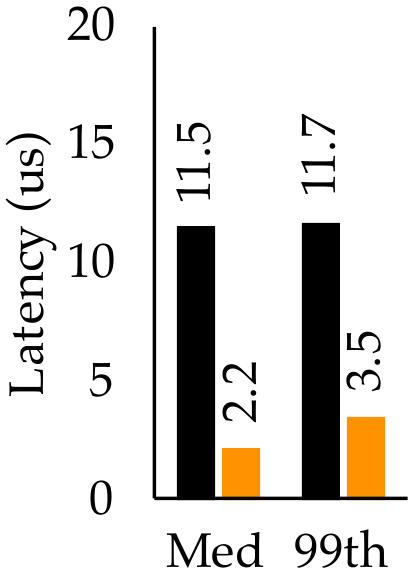}%
    }
    \hfill
    \subfloat[][]{%
      \label{fig:lread-bw}%
      \includegraphics[scale=0.5]{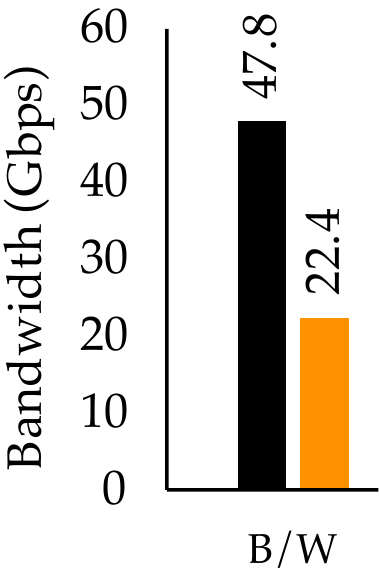}%
    }
    \hfill
    \subfloat[][]{%
      \label{fig:lwrite-lat}%
      \includegraphics[scale=0.5]{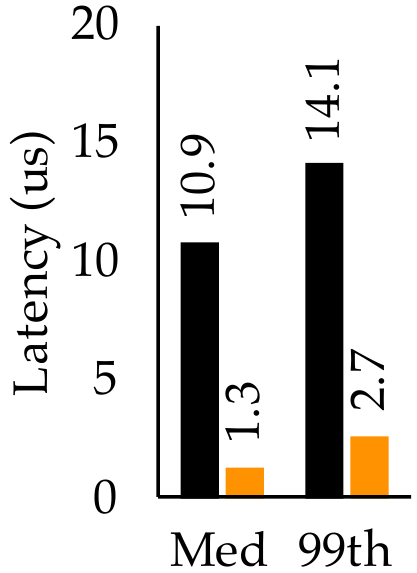}%
    }
    \hfill
    \subfloat[][]{%
      \label{fig:lwrite-bw}%
      \includegraphics[scale=0.5]{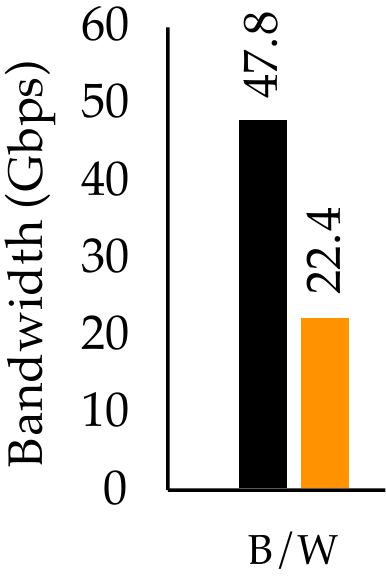}%
    }
  \caption{[InfiniBand] \protect\subref{fig:lread-lat}--\protect\subref{fig:lread-bw} {\name} isolating remote latency-sensitive READs from local bandwidth-sensitive WRITEs. \protect\subref{fig:lwrite-lat}--\protect\subref{fig:lwrite-bw} {\name} isolating local latency-sensitive WRITEs from remote bandwidth-sensitive READs.}%
  \label{fig:read-eval}
\end{figure}

\subsection{Handling Remote READs}
\label{sec:read-eval}
Unlike TCP/IP, RDMA provides READ verbs that allows a remote machine $B$ to read from a local machine $A$, where data flows in the $A\rightarrow B$ direction.
Consequently, they compete with WRITEs and SENDs from machine $A$ to $B$.
Figure~\ref{fig:read-eval} shows that, as expected (\S\ref{sec:remote-control}), {\name} can isolate latency-sensitive remote READs from local bandwidth-sensitive WRITEs and vice versa. 

\begin{figure}[!t]
  \centering
    \subfloat[][Experiment setup]{%
      \label{fig:incast-setup}%
      \includegraphics[width=1.1in]{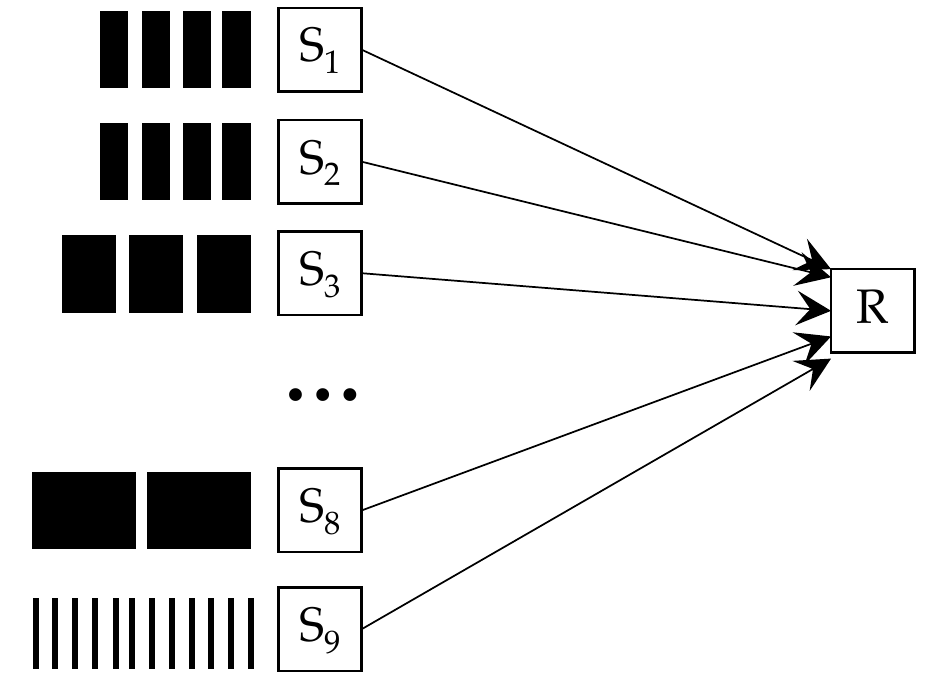}%
    }
    \hfill
    \subfloat[][Bandwidth]{%
      \label{fig:incast-numbers}%
      \includegraphics[width=0.8in]{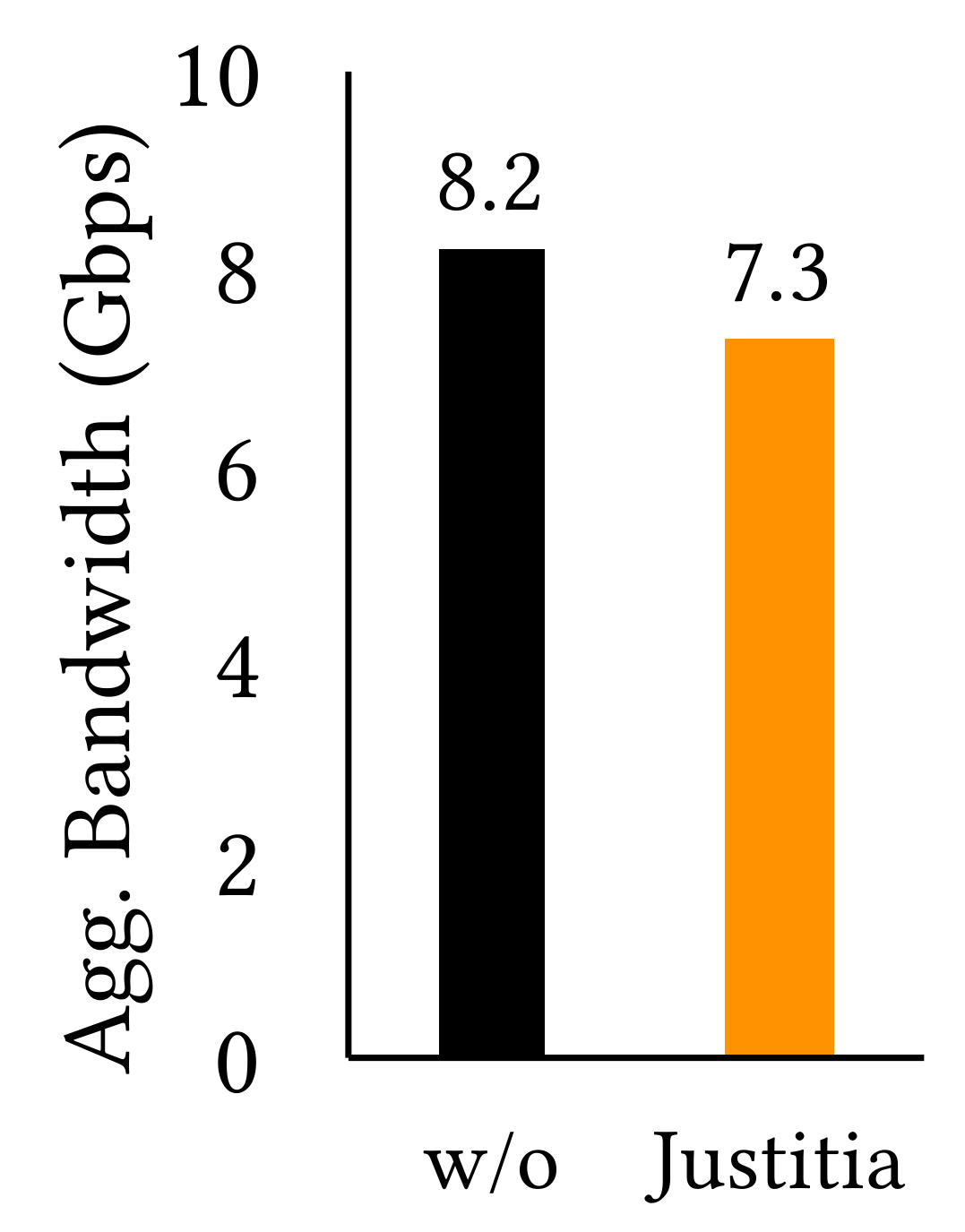}%
    }
    \hfill
    \subfloat[][Latency]{%
      \label{fig:incast-lat}%
      \includegraphics[width=1.4in]{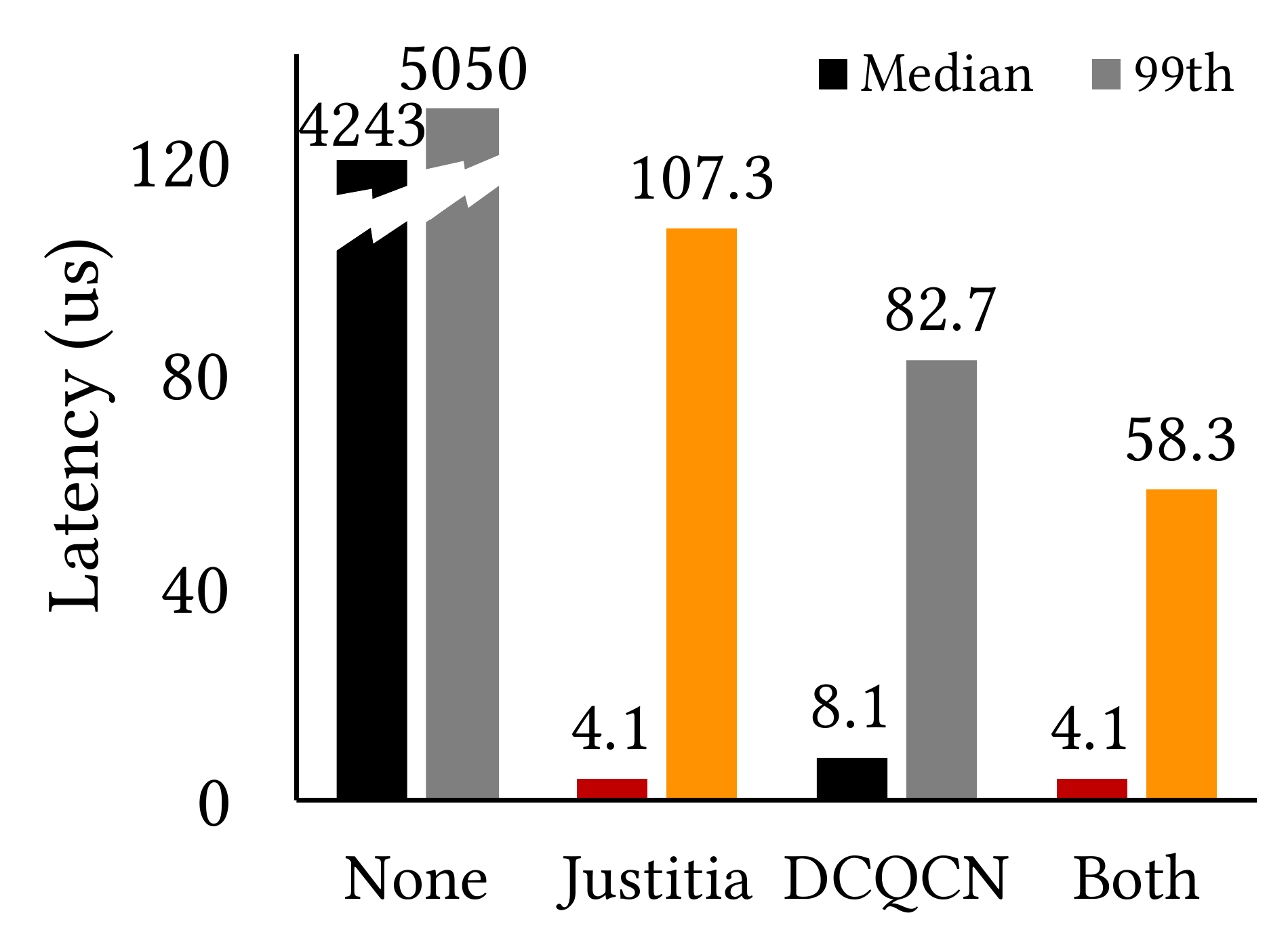}%
    }
  \caption{[DCQCN] Each pair of bandwidth-sensitive senders ($S_i$ and $S_{i+1}$ for $i=1,3,5,7$) send messages of 1MB, 10MB, 100MB, and 1GB sizes to a single receiver $R$. 
    $S_9$ sends a latency-sensitive flow.}%
  \label{fig:incast}
\end{figure}

\subsection{{\name}'s Impact on Incast Scenarios}
\label{sec:incast-eval}
So far, we have always focused on sender-side RNIC contentions.
In this experiment, we focus on {\name}'s impact on simple incast scenarios, where multiple senders $S_1$--$S_8$ continuously send messages of 1MB, 10MB, 100MB, and 1GB (two senders each) to a single receiver $R$ (Figure~\ref{fig:incast}).
Simultaneously, $S_9$ sends a latency-sensitive flow to $R$.
We extended {\name} daemons to continuously exchange receiver side views with the senders (similar to the RDMA READ case).
We compare four cases where (1) neither DCQCN or {\name} is applied, (2) only {\name} is applied, (3) only DCQCN is applied, (4) both DCQCN and {\name} is applied. 

We make two observations.
First, DCQCN indeed greatly improves incast. 
However, using {\name} alone can achieve similar performance as DCQCN.
Second, {\name} can complement with DCQCN to further improve the incast scenario.

%% file: related.tex
\section{Related Work}

\paragraph{RDMA Sharing}
Recently, large-scale RDMA deployment over RoCEv2 have received wide attention \cite{msr-rdma-15, msr-rdma-16, timely, irn}.
However, the resulting RDMA congestion control algorithms~\cite{timely, msr-rdma-15, rogue} primarily deal with Priority-based Flow Control (PFC) to provide fair sharing between bandwidth-sensitive flows inside the network. 
In contrast, {\name} focuses on RNIC isolation and is complementary to them (\S\ref{sec:dcqcn-exp}).

Similarly, {\name} is also complementary to FreeFlow~\cite{freeflow}, which solves a different problem: enabling \emph{untrusted} containers to securely gain some of the performance benefits of RDMA.
Because FreeFlow does not change how verbs are sent to queue pairs and only validates that it is secure to do so, it can still suffer from the performance isolation problems that {\name} addresses.
It can also potentially benefit from employing an approach similar to {\name}.  
Further, in scenarios where applications are \emph{trusted}, {\name} has the potential to achieve better performance than FreeFlow.

LITE \cite{lite} also addresses resource sharing and isolation issues in RNICs.
However, through experiments (\S\ref{sec:lite-eval}), we have found that LITE does not perform very well in the absence of hardware virtual lanes.
In contrast, {\name} is a software-only solution that appreciates the isolation-vs-utilization tradeoff to mitigate RDMA performance isolation anomalies.

%
{\name}'s goal is to enable such diverse workloads to coexist.
Although {\name} currently works at the level of flows, it can potentially be extended to handle application- and tenant-level isolation issues (\S\ref{sec:app-tenant}).

\paragraph{Link Sharing}
Max-min fairness \cite{jaffe-maxmin, wfq, wf2q, drr} is the well-established solution for link sharing that achieves both sharing incentive and work conservation, but it only considers bandwidth-sensitive flows. 
Latency-sensitive flows can rely on some form of prioritization for isolation \cite{pfabric, pdq, d3}.

Although DRFQ \cite{drfq} dealt with multiple resources, it considered cases where a packet sequentially accessed each resource, both link capacity and latency were significantly different than RDMA, and the end goal was equalizing utilization instead of performance isolation. 
Furthermore, implementing DRFQ required hardware changes.

Both Titan~\cite{titan} and Loom~\cite{loom} improve performance isolation on conventional NICs by programming on-NIC packet schedulers.
However, this is not sufficient to address all RDMA performance
isolation problems because it only schedules a single resource: the
outgoing link.
Further, {\name} works on existing RDMA NICs that do not have programmable packet schedulers.

\paragraph{Datacenter Network Sharing}
With the advent of cloud computing, the focus on link sharing has expanded to network sharing between multiple tenants \cite{mogul-popa, faircloud, hug, oktopus, seawall}.
Almost all of them -- except for static allocation -- deal with bandwidth isolation and ignore latency-sensitive flows.

Silo \cite{silo} dealt with datacenter-scale challenges in providing latency and bandwidth guarantees with burst allowances on Ethernet networks. 
In contrast, we focus on isolation anomalies in multi-resource RNICs between latency-, bandwidth-, and throughput-sensitive flows.

%% file: outro.tex
\section{Concluding Remarks}
\label{sec:outro}

We have demonstrated that performance isolation issues between bandwidth-, throughput-, and latency-sensitive RDMA flows are pervasive across InfiniBand, RoCEv2, and iWARP and in 10, 40, 56, and 100 Gbps RDMA networks. 
The root causes include the work-conserving nature of RDMA NICs (RNICs) and their multi-resource design. 
The overall impact is head-of-line (HOL) blocking when flows with diverse message sizes and performance requirements compete.
 
{\name} addresses these anomalies both at flow and application levels in two steps.
First, it guarantees each flow at least $\frac{1}{n}$th of the two RNIC resources (bandwidth and execution unit throughput). 
Second, it maximizes RNIC utilization across both dimensions without violating that guarantee.
{\name} is easily deployable, scales well, can handle remote READs, and performs well in simple incast scenarios. 
{\name} works well in isolating the performance of real-world RDMA applications such as FaSST and eRPC.
Furthermore, it complements RDMA congestion control protocols such as DCQCN and hardware virtual lanes (when present) well. 

{\name} is only a first step toward RNIC performance isolation and raises interesting research questions (Appendix~\ref{sec:cc-ext}).
 

%% file: appendix.tex
\appendix

\section{Hardware Testbed Summary}
\label{app:hw-summary}

Table~\ref{tab:hw-spec} summarizes the hardware we use for different RDMA protocols in our experiments.

\begin{table*}[!t]
    \begin{center}
	  \begin{tabular}{l|l|l|r}
		\textbf{Protocol} & \textbf{NIC} & \textbf{Switch} & \textbf{NIC Capacity}\\
		\hline
		InfiniBand & ConnectX-3 Pro & Mellanox SX6036G & 56 Gbps\\
		InfiniBand & ConnectX-4 & Mellanox SB7770 & 100 Gbps\\
		RoCEv2 & ConnectX-4 & Mellanox SX6018F & 40 Gbps\\
		RoCEv2 (DCQCN) & ConnectX-4 Lx & Dell S4048-ON & 10 Gbps\\
		iWARP & T62100-LP-CR & Mellanox SX6018F & 40 Gbps\\
	  \end{tabular}
	  \caption{Testbed hardware specification.}
	\label{tab:hw-spec}
    \end{center}
\end{table*}

\begin{figure}[!t]
	\centering
		\subfloat[][Latency]{%
			\label{fig:multi-mouse-lat}%
			\includegraphics[width=1.65in]{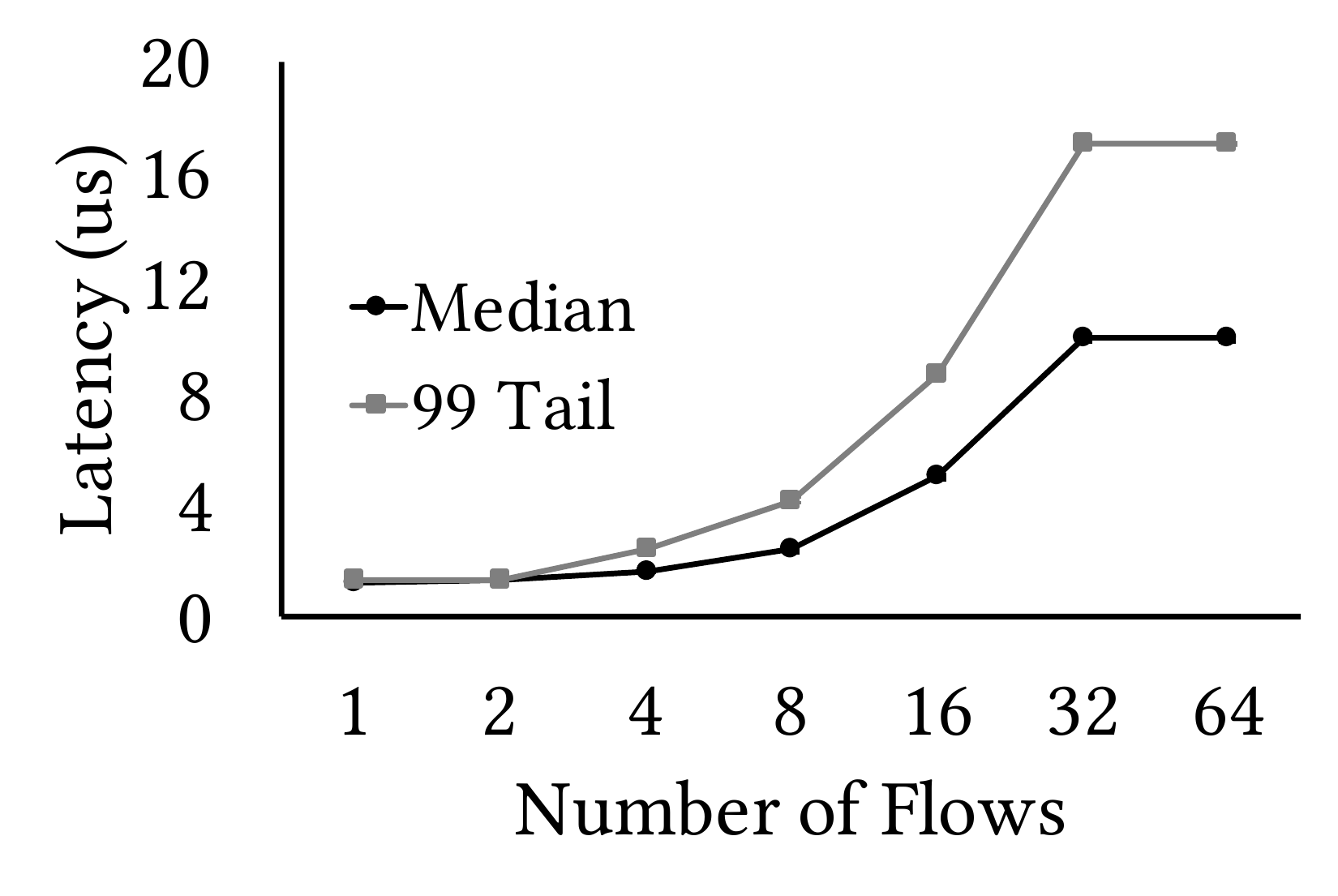}%
		}
		\hfill
		\subfloat[][Throughput]{%
			\label{fig:multi-mouse-tput}%
			\includegraphics[width=1.65in]{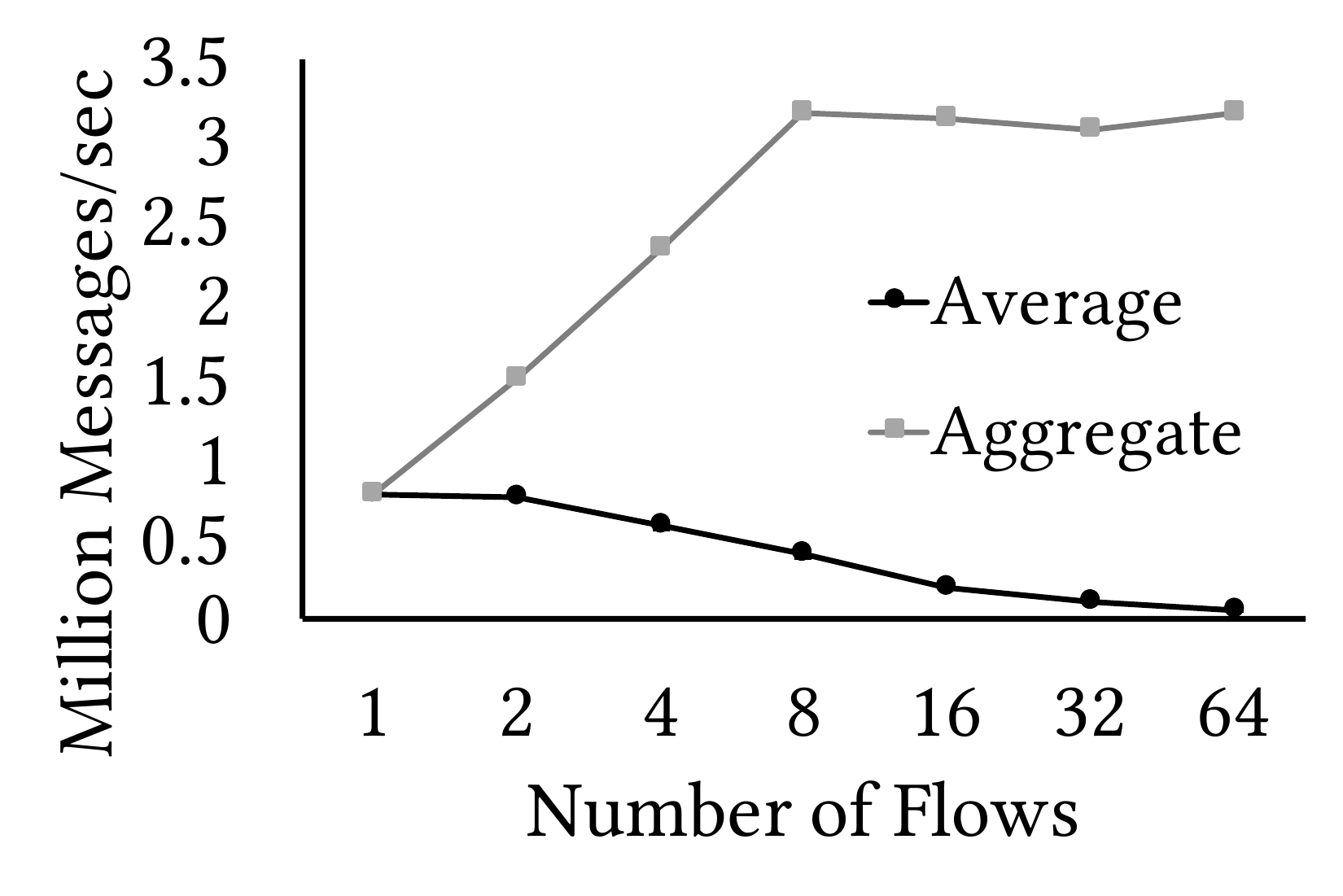}%
		}	
	\caption{Latencies and throughputs of multiple latency-sensitive flows in InfiniBand. 
    Error bars (almost invisible due to close proximity) represent the flows with the lowest and highest values.}
	\label{fig:sec3-multi-M}%
\end{figure}

\begin{figure}[!t]
  \centering
  \includegraphics[width=1.9in]{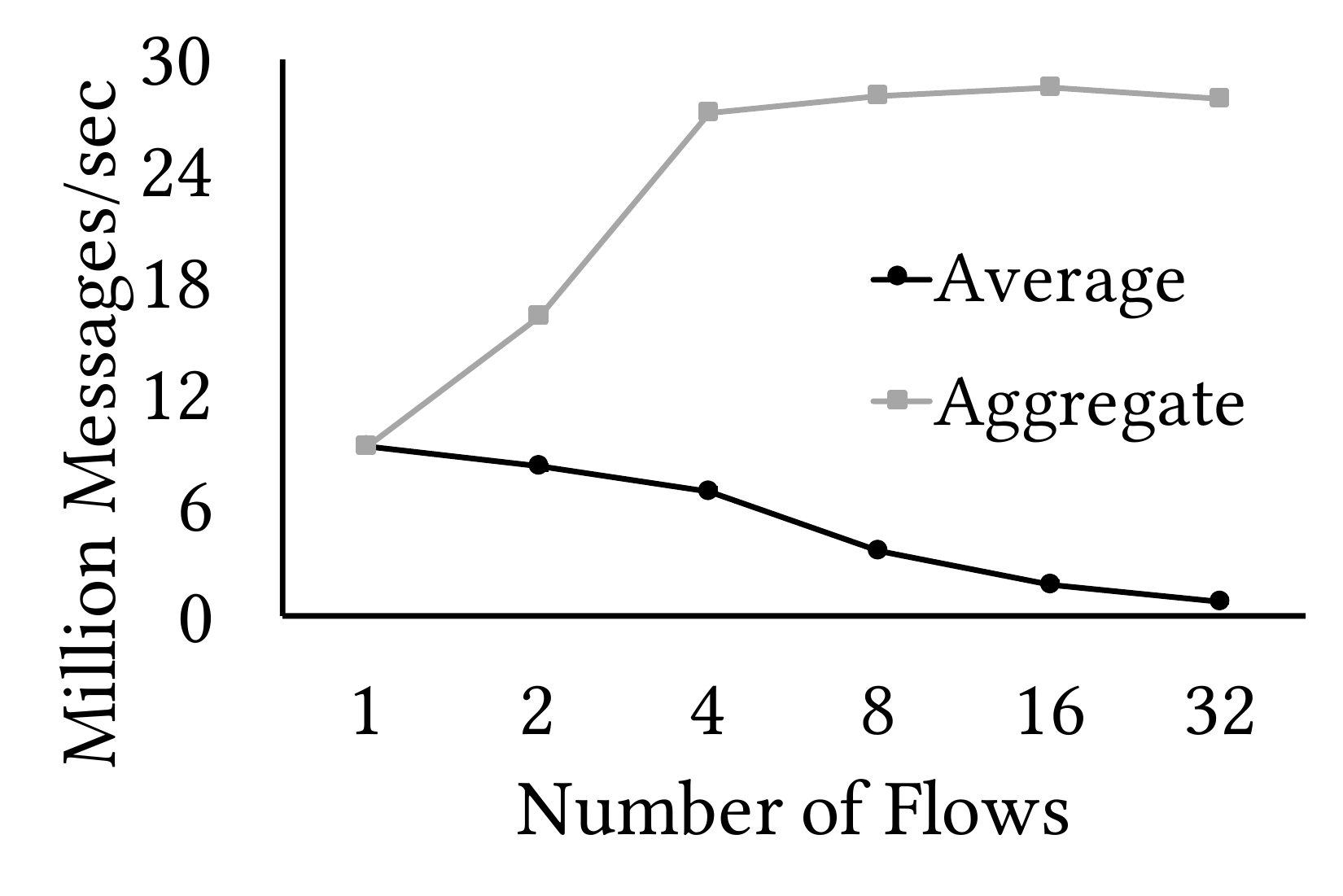}
  \caption{Throughputs of multiple throughput-sensitive flows in InfiniBand.
  Error bars (almost invisible due to close proximity) represent the flows with the lowest and highest values.}%
  \label{fig:sec3-multi-tput}%
\end{figure}

\begin{figure}[!t]
	\centering
		\subfloat[][Latency Flow (Med)]{%
			\label{fig:MOTI-TvL-median}%
			\includegraphics[width=1.1in]{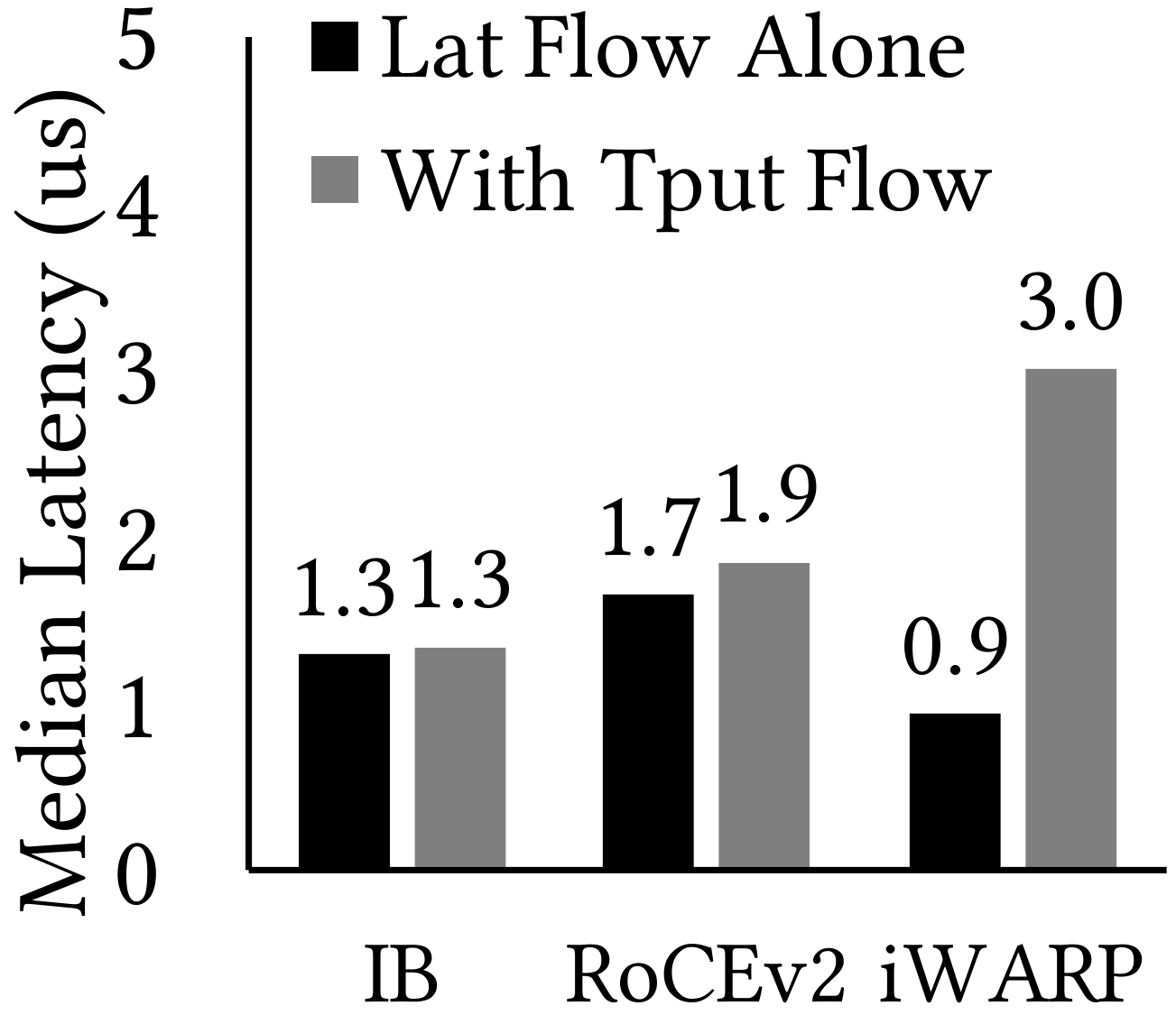}%
		}
		\hfill
		\subfloat[][{Latency Flow (99th)}]{%
			\label{fig:MOTI-TvL-tail}%
			\includegraphics[width=1.1in]{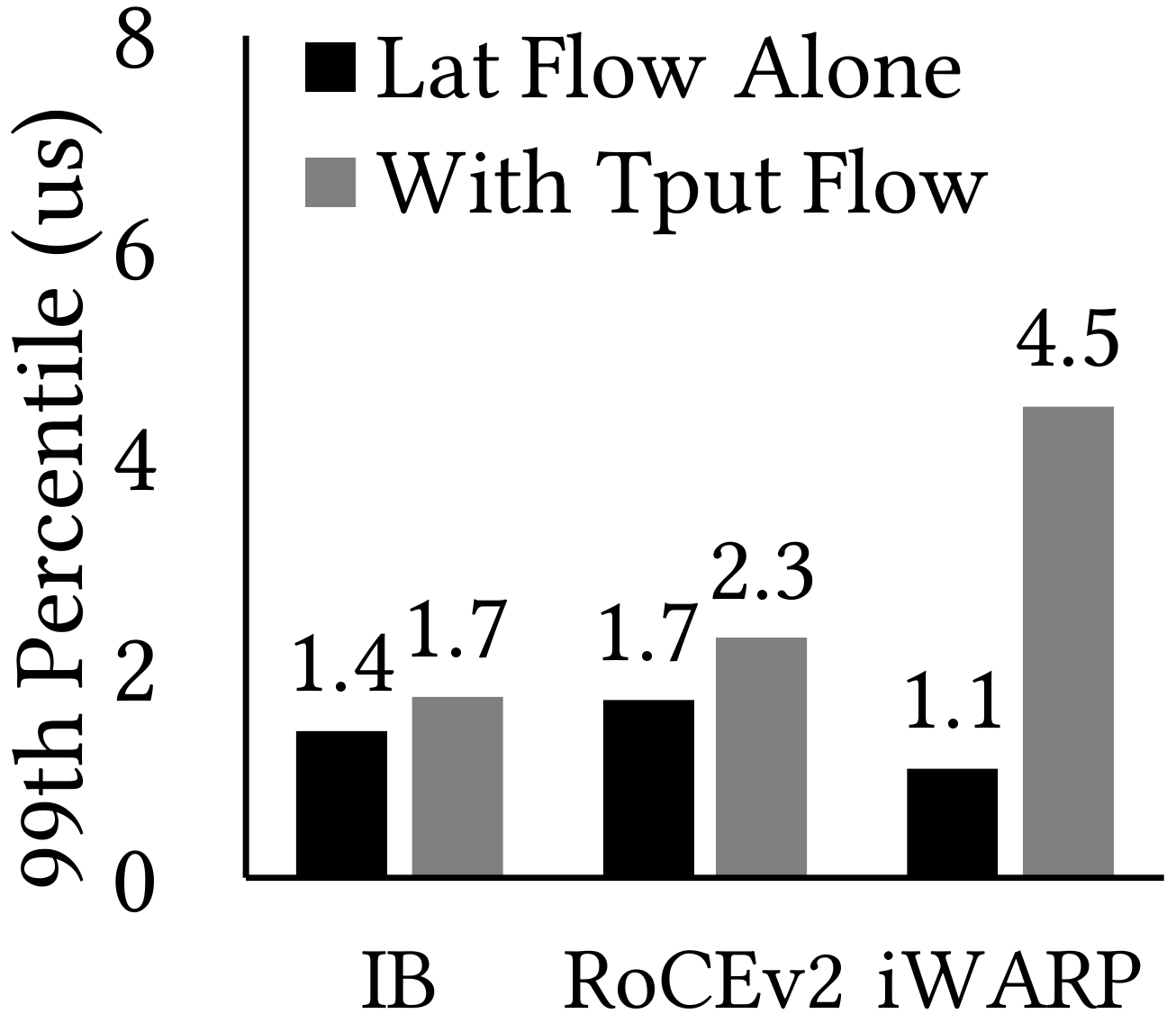}%
		}	
		\hfill
		\subfloat[][{Throughput Flow}]{%
			\label{fig:MOTI-TvL-ops}%
			\includegraphics[width=1.1in]{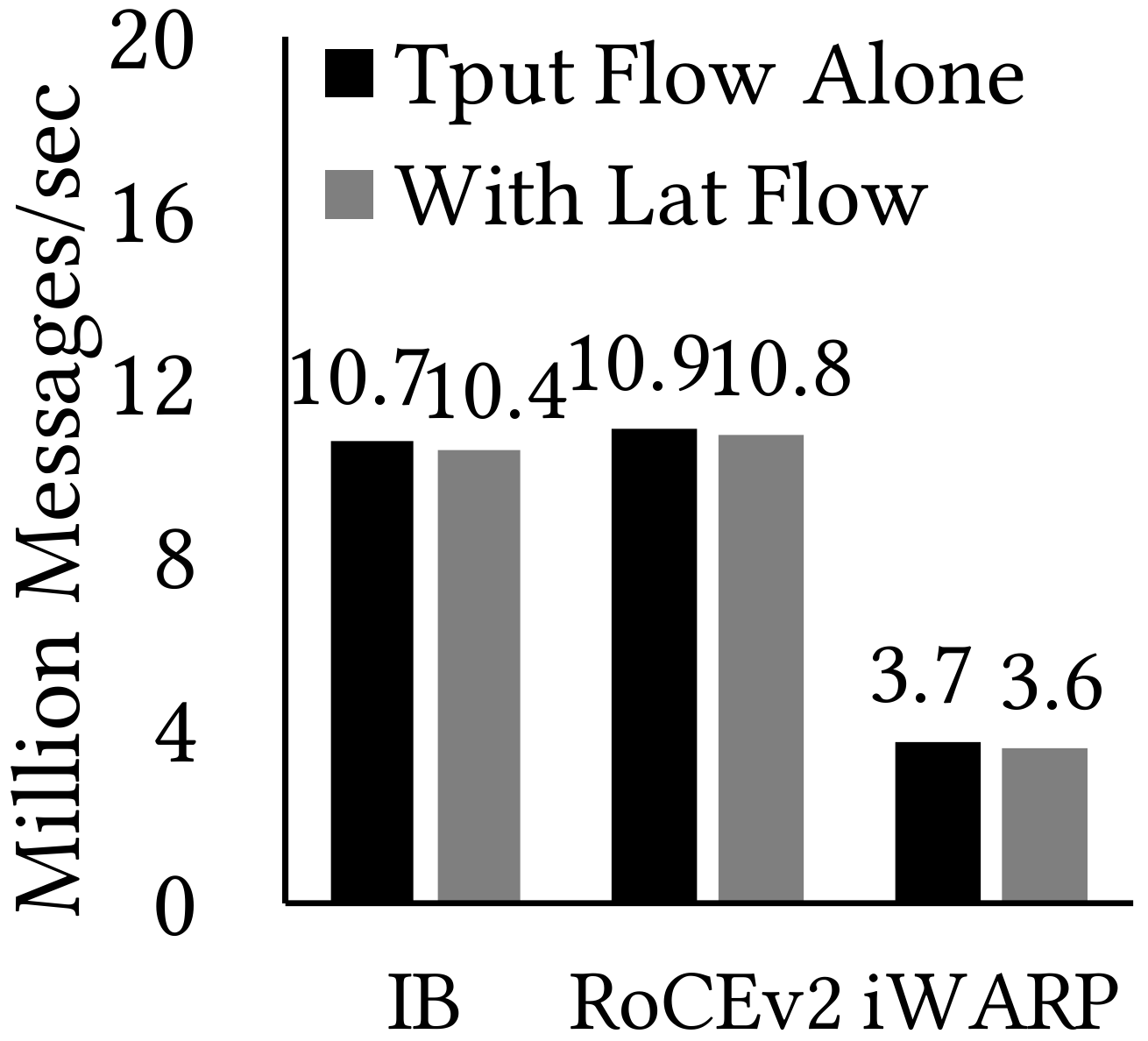}%
		}	
	\caption{Performance anomalies of a latency-sensitive flow running against a background throughput-sensitive flow.}
	\label{fig:sec3-tput-vs-lat}%
\end{figure}

\begin{figure}[!t]
    \centering
    \includegraphics[width=2.4in]{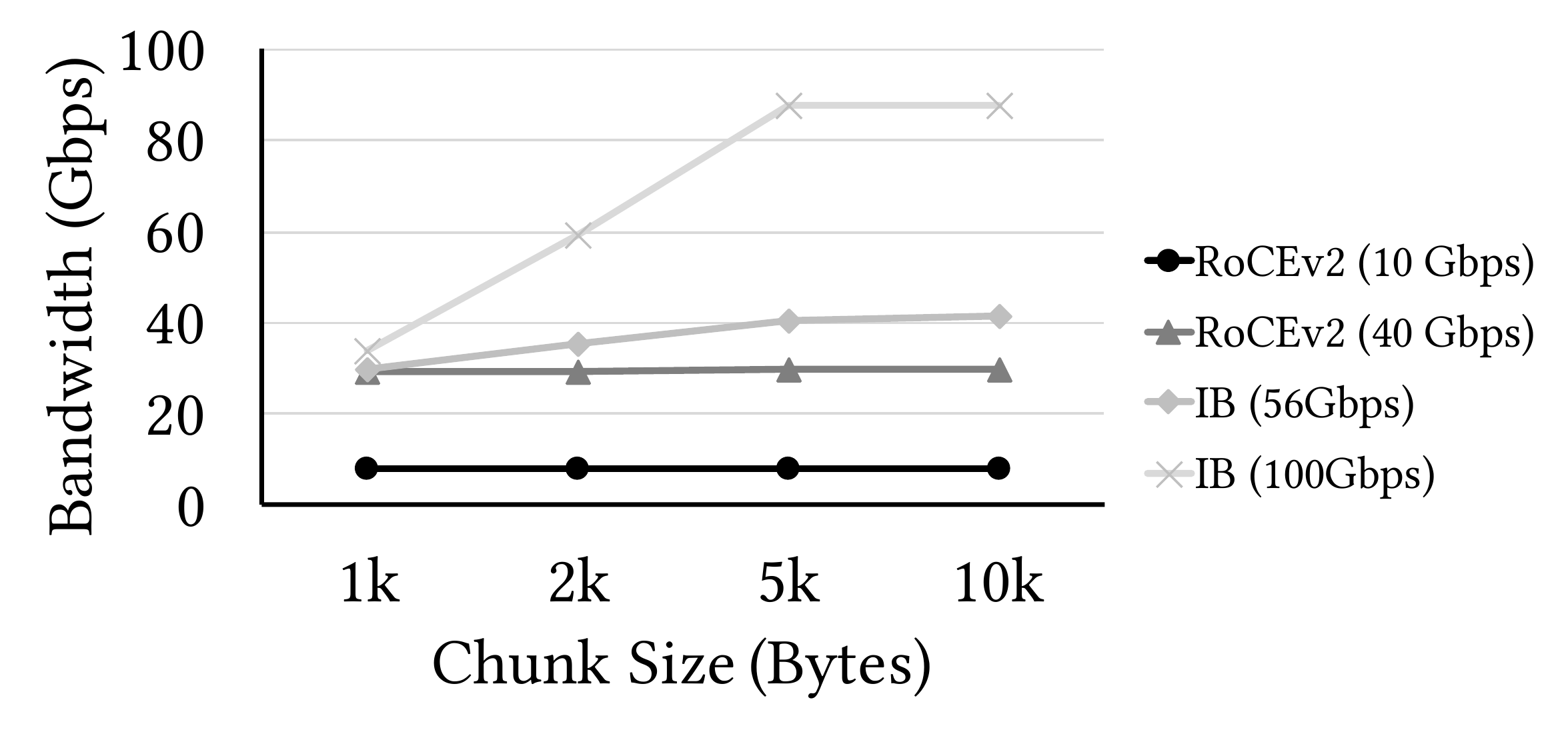}
    \caption{Bandwidth reachable using differnet chunk sizes with batch in various RNICs.}%
    \label{fig:pick-chunk-size}%
\end{figure}

\section{Characteristics of Latency- and \\ Throughput-Sensitive Flows in the Absence \\ of Bandwidth-Sensitive Flows}
\label{sec:lat-and-tput-anomalies}
Multiple latency-sensitive flows can coexist without affecting each other (Figure~\ref{fig:sec3-multi-M}). 
Although latencies increase, everyone suffers equally. 
All flows experience the same throughputs as well. 

Similarly, multiple throughput-sensitive flows receive almost equal throughputs when competing with each other,
as shown in Figure~\ref{fig:sec3-multi-tput}.

Finally, throughput-sensitive flows do not get affected by much when competing with latency-sensitive flows (Figure~\ref{fig:MOTI-TvL-ops}).
Nor do latency-sensitive flows experience noticeable latency degradations in the presence of throughput-sensitive flows except for iWARP (Figure~\ref{fig:MOTI-TvL-median} and Figure~\ref{fig:MOTI-TvL-tail}).

\section{100 Gbps Results With/Without {\name}}
\label{app:100gbps}

\begin{figure}[!t]
	\centering
		\subfloat[][Latency-sensitive Flow]{%
			\includegraphics[width=2.4in]{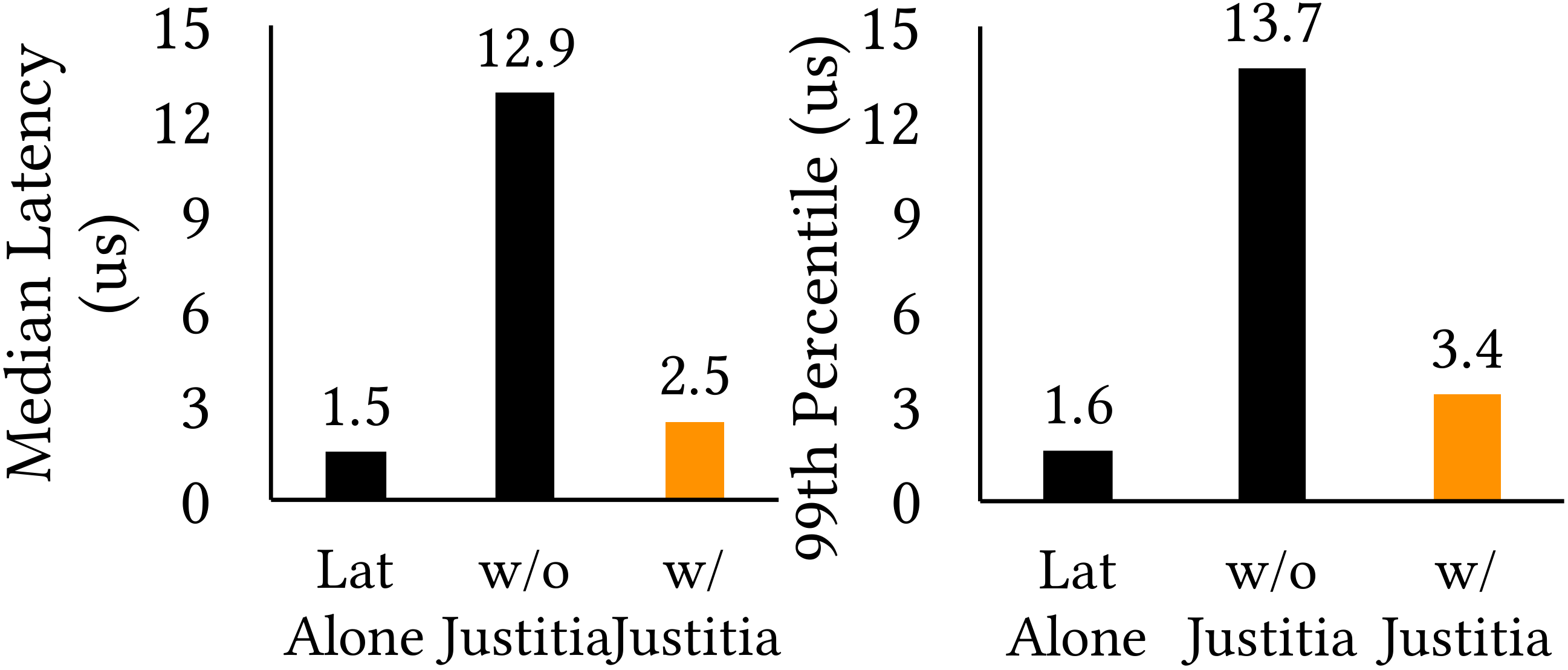}%
		}
		\hfill
		\subfloat[][{Bandwidth Flow}]{%
			\includegraphics[width=0.9in]{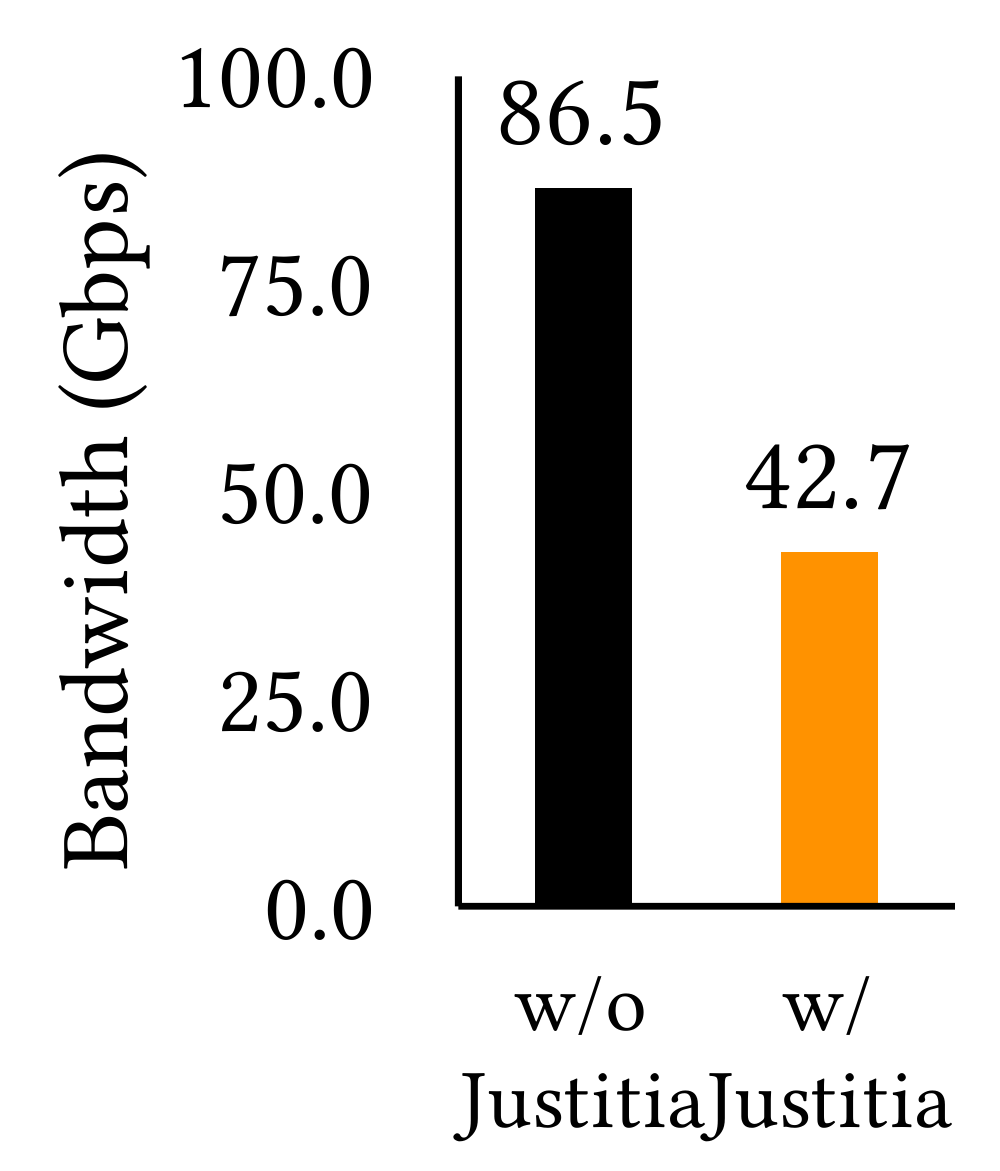}%
		}	
	\caption{[100 Gbps InfiniBand] Performance isolation of a latency-sensitive flow against a bandwidth-sensitive flow.}
  \label{fig:100gbps-lat-bw}
\end{figure}

\begin{figure}[!t]
	\centering
		\subfloat[][Throughput Flow]{%
			\includegraphics[width=1.1in]{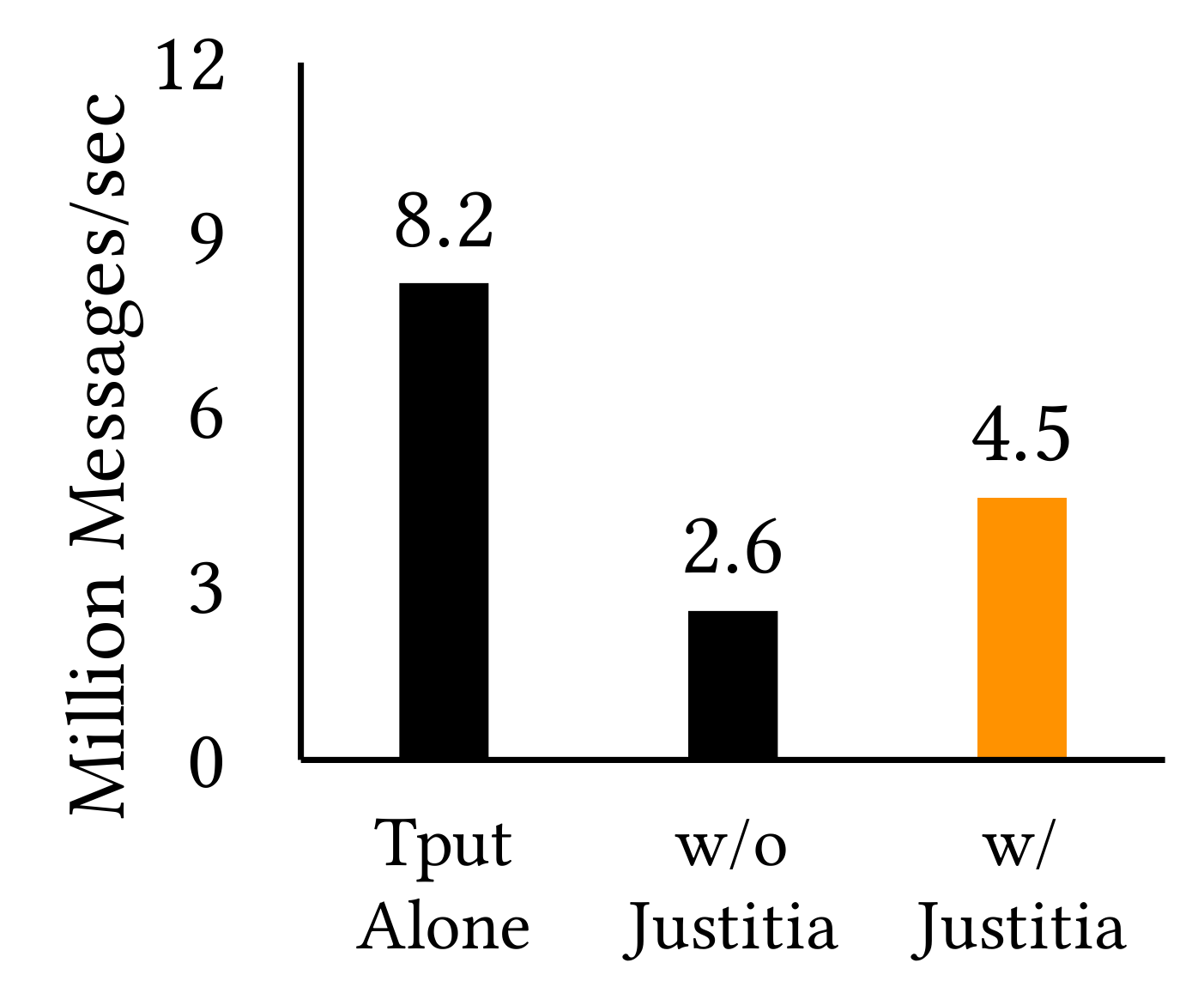}%
		}
		\hspace{1cm}
		\subfloat[][{B/w Flow}]{%
			\includegraphics[width=0.8in]{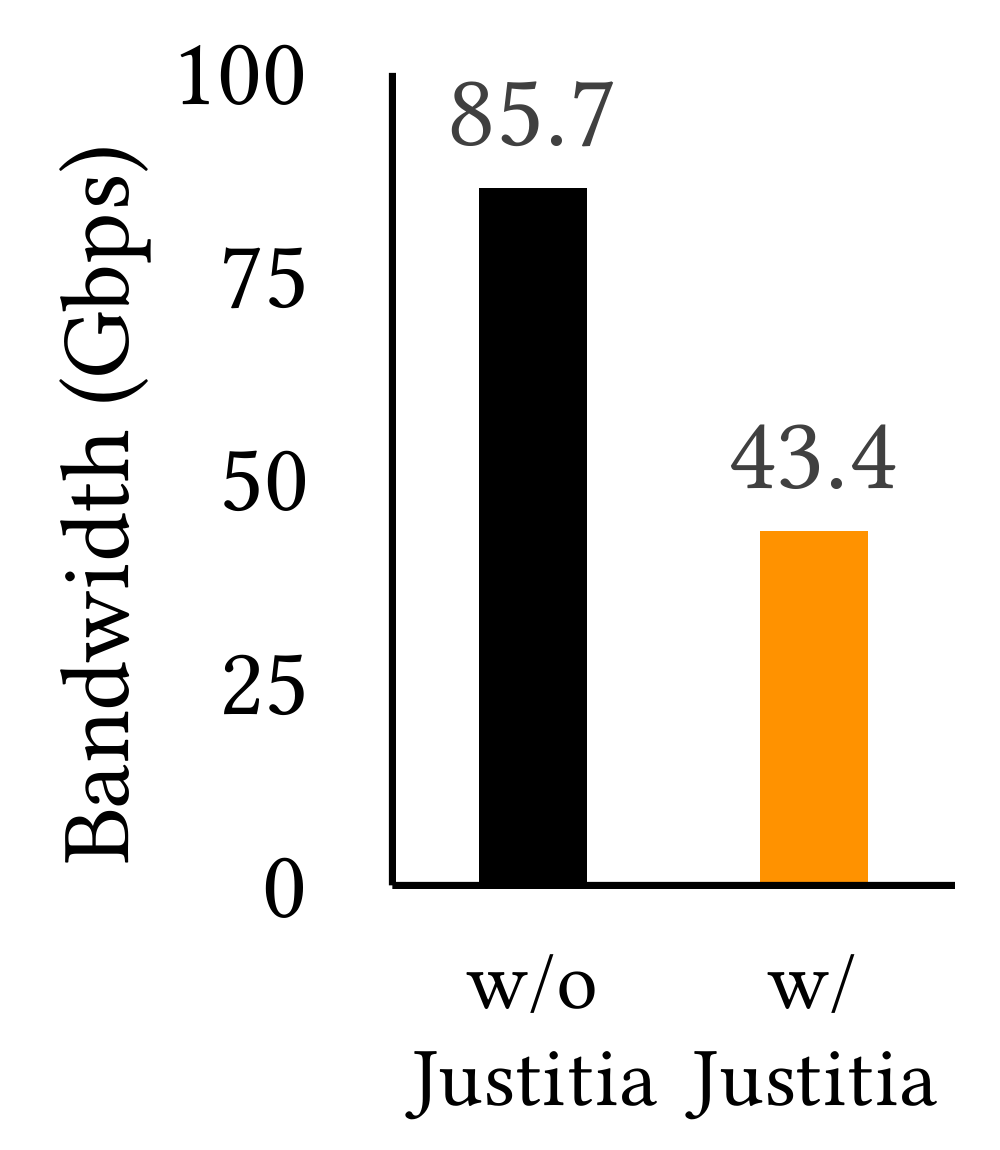}%
		}	
	\caption{[100 Gbps InfiniBand] Performance isolation of a throughput-sensitive flow against a bandwidth-sensitive flow.}
  \label{fig:100gbps-tput-bw}
\end{figure}

Similar to the anomalies observed for 10, 40, and 56 Gbps RDMA networks (\S\ref{sec:anomalies}), Figure~\ref{fig:100gbps-lat-bw} and Figure~\ref{fig:100gbps-tput-bw} show that latency- and throughput-sensitive flows are not isolated from bandwidth-sensitive flows even in 100 Gbps networks.
In these experiments, we use 5MB messages since 1MB messages are not large enough to saturate the 100 Gbps link.
{\name} can effectively mitigate the challenges by enforcing performance isolation.

\section{Reducing CPU Overhead of Using Small Tokens}
\label{sec:reduce-cpu-details}
Using small tokens lead to CPU overhead mainly from busy spinning to fetch tokens generated at a short period (around 1us) which precludes any context switches.
We solve this challenge by decoupling token generation from token enforcement (TE).

To preserve low CPU overhead, tokens are generated in {\name} daemon and distributed via IPC sockets using a large \tokenbytes whose \tokengentime is long.
Token enforcement happens in {\name} shapers: messages are split into smaller chunks, and a waiting interval is inserted before posting a work request for the next chunk.
The longer the waiting interval, the higher the CPU overhead caused by longer busy waiting, and the better isolation we achieve by allowing more small flows to sneak through during those intervals.
For example, if we set the waiting interval to be the time it takes to send out one small chunk at the current rate enforced by the pacer (\safeutil), the waiting intervals altogether will span the entire token generation time \tokengentime; this leads to 100\% CPU usage.
Any shorter interval leads to a lower CPU usage with a shorter interval, and any longer interval fails to maintain \safeutil.
If we denote the waiting interval by $t$, we get
\[t = \alpha \times t_{max} = \alpha \times \frac{ChunkSize}{SafeUtil}, \alpha \in \left[0,1\right]\]
\[CPU_{TE} = \alpha \times 100\%\]
where the shaper's CPU overhead can be easily controlled by periodically following hints provided by the pacer via shared memory.
The goal is to find the waiting interval that provides an acceptable isolation while minimizing CPU cost.
To dynamically adjust the waiting interval, {\name} increases waiting interval from 0 and stops when a significant improvement in the tail latency estimate can no longer be seen.

Note that the above CPU overhead is caused by pacing small chunks in bandwidth-sensitive applications only. 
{\name} currently minimizes CPU overhead to half of a core (50\%) per bandwidth-sensitive appications, and adds no CPU overhead to other types of applications.


\begin{figure}[!t]
	\centering
		\subfloat[][Latency-sensitive flow]{%
			\label{fig:LITE-E-vs-lat-latency}%
			\includegraphics[width=1.8in]{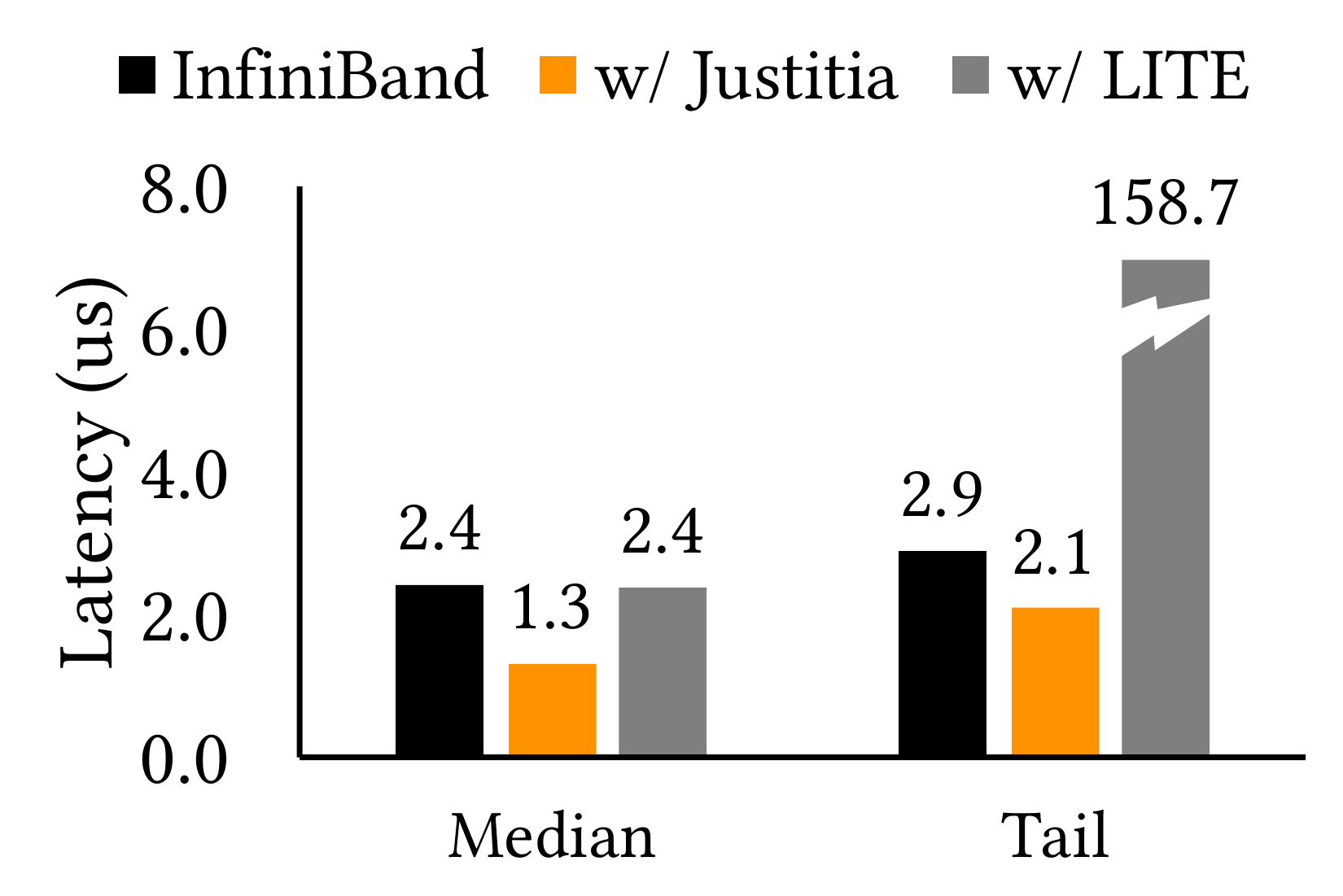}%
		}
		\hfill
		\subfloat[][{Bandwidth flow}]{%
			\label{fig:LITE-E-vs-lat-bw}%
			\includegraphics[width=1.5in]{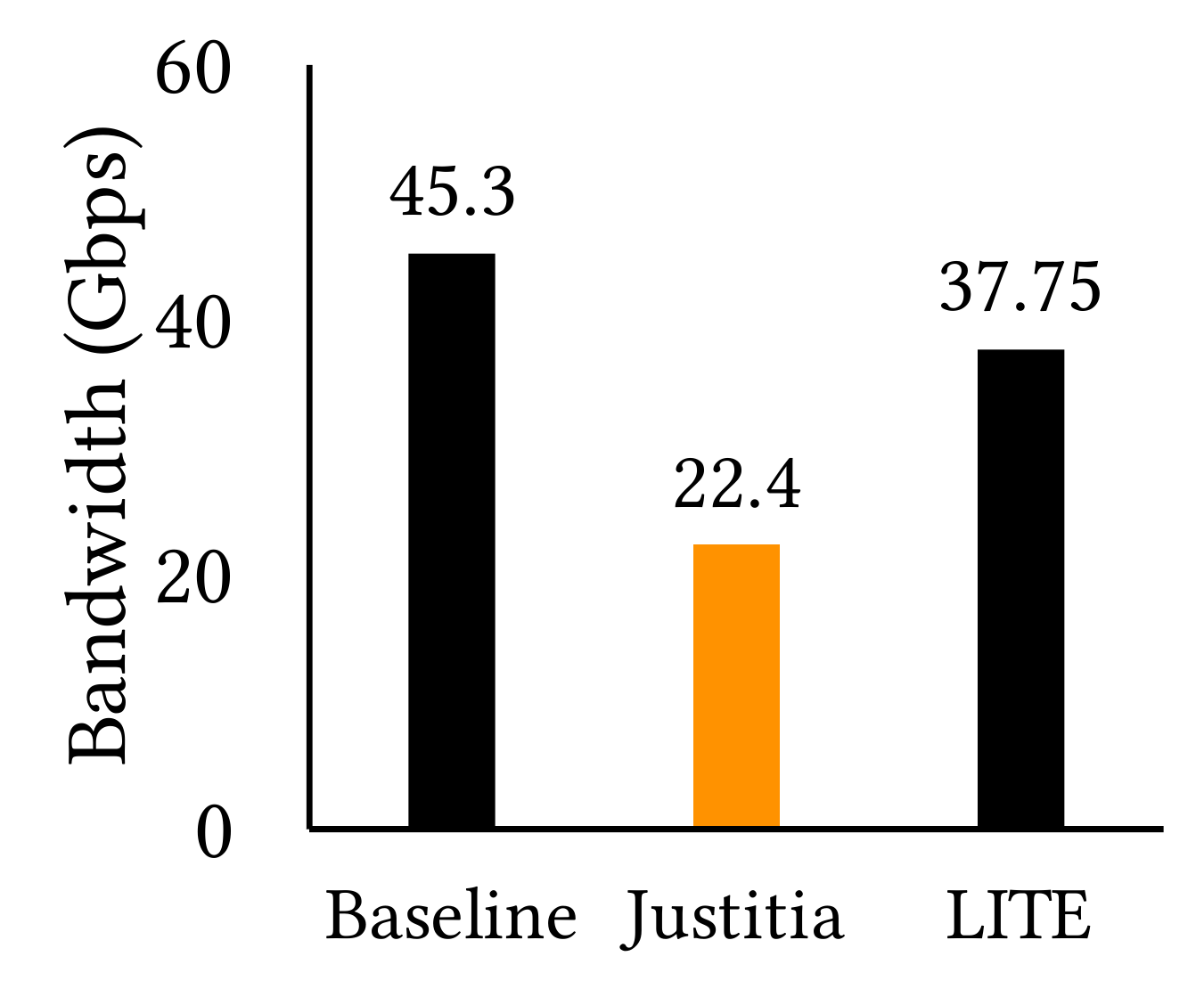}%
		}	
	\caption{[InfiniBand] Performance isolation of a latency-sensitive flow running against a 1MB background bandwidth-sensitive flow using {\name} and LITE.}
	\label{fig:sec7-LITE-E-vs-lat}%
\end{figure}

\begin{figure}[!t]
	\centering
	\includegraphics[width=1.9in]{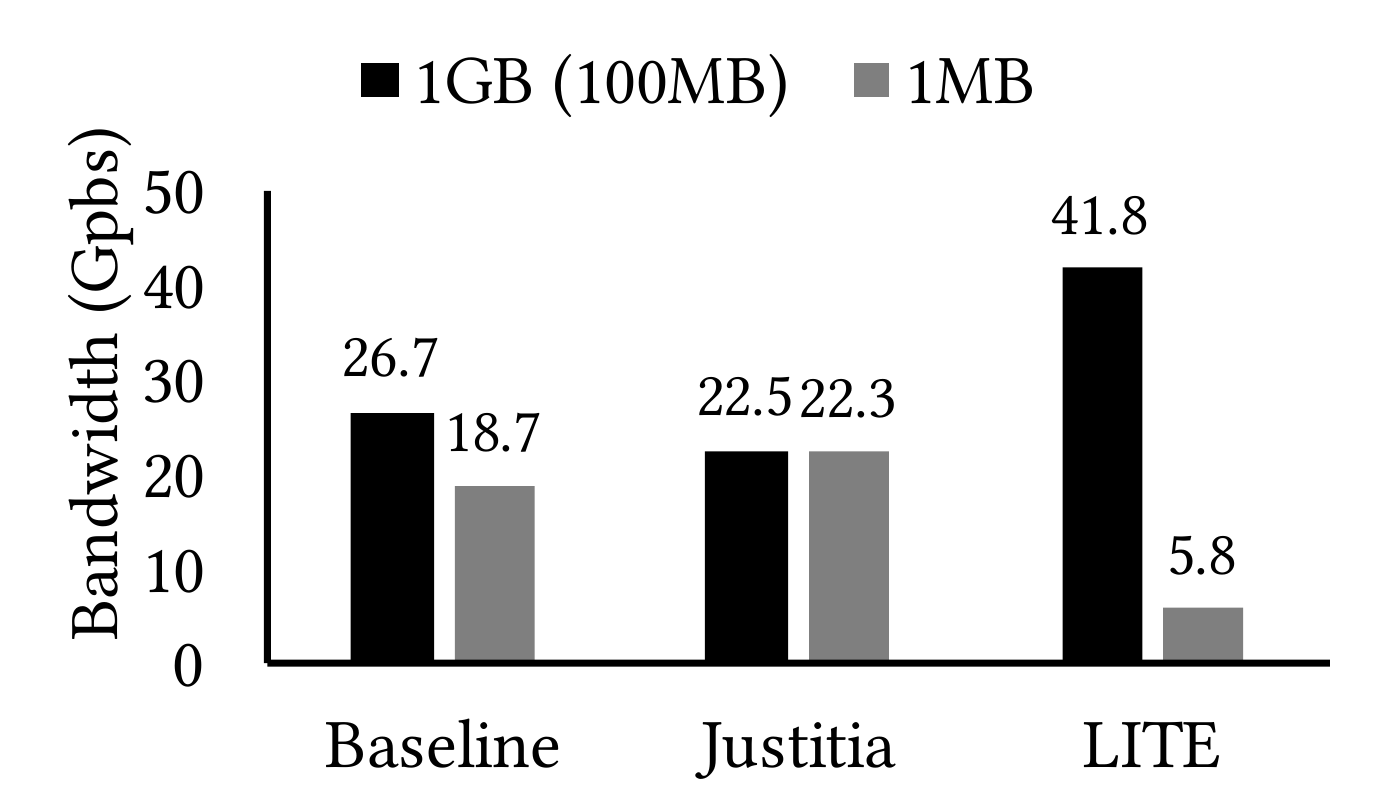}%
	\caption{[InfiniBand] Bandwidth allocations of two bandwidth-sensitive flows using {\name} and LITE. LITE uses 100MB messages instead of 1GB due to its own limitation.}%
  \label{fig:sec7-LITE-E-vs-E}%
\end{figure}

\section{{\name} vs. LITE}
\label{sec:lite-eval}
LITE \cite{lite} is a software-based RDMA implementation that adds a local indirection layer for RDMA in the Linux kernel to virtualize RDMA and enable resource sharing and performance isolation.
It can use hardware virtual lanes and also includes a software-based prioritization scheme.

We found that, in the absence of hardware virtual lanes, LITE does not perform well in isolating latency-sensitive flow from the bandwidth-sensitive one (Figure~\ref{fig:sec7-LITE-E-vs-lat}) -- $122\times$ worse 99th percentile latency than {\name}.
In terms of bandwidth-sensitive flows using different message sizes, LITE performs even worse than native InfiniBand (Figure~\ref{fig:sec7-LITE-E-vs-E}). 
{\name} outperforms LITE's software-level prioritization by being cognizant of the tradeoff between performance isolation and work conservation. 

%% file: open.tex
\section{Open Problems}

Interesting short- and long-term future directions of this work include, among others, dynamically determining a flow's performance requirements, handling multi-modal flows, handling in-network issues, extending to more complicated application- and/or tenant-level RDMA isolation issues, and implementing {\name} logic in programmable NICs.

We highlight two immediate next-steps in the following. 

\textbf{Co-Designing with Congestion Control.}
\label{sec:cc-ext}
Although {\name} effectively complements DCQCN (\S\ref{sec:dcqcn-exp}) in simple scenarios, DCQCN considers only bandwidth-sensitive flows. 
A key future work would be a ground-up co-design of {\name} with DCQCN \cite{msr-rdma-15} or TIMELY \cite{timely} to handle all three traffic types for the entire fabric with sender- and receiver-side contentions (\S\ref{sec:incast-eval}). 
While network calculus and service curves \cite{silo, hfsc, cruz1, cruz2} dealt with point-to-point bandwidth- and latency-sensitive flows, their straightforward applications can be limited by multi-resource RNICs and throughput-sensitive flows.
At the fabric level, exploring a Fastpass-style centralized solution \cite{fastpass} can be another future work.

\textbf{{\name} at Application and Tenant Levels.}
\label{sec:app-tenant}
Currently, {\name} isolates applications/tenants by treating all flows from the same originator as one logical flow with a single type.
This is an approximation of Seawall \cite{seawall}.
However, for an application with flows with different requirements, this straightforward approach is unlikely to work well.
 
A possible direction can be exploring Oktopus-style isolation schemes \cite{oktopus}, where we first isolate tenants and then apply {\name} inside each tenant.
Similar to hierarchical token bucket (HTB) \cite{htb}, a hierarchical instantiation of {\name} may be able to achieve this.
However, unlike HTB, we must deal with conflicting performance requirements and multi-resource RNICs.

Even in these scenarios, the same isolation-utilization tradeoff -- and more complicated variations \cite{faircloud, hug} -- will still apply.

\textbf{Strategyproof {\name}.}
\label{sec:sp-ext}
Applications may not always correctly or truthfully identify their flow types.
Augmenting {\name} with DRFQ \cite{drfq} while adding support for multiple parallel RNIC resources -- DRFQ considers multiple resources in sequence -- and all three traffic types can be interesting future work.